\documentstyle{elsart}

\begin{document}
\begin{frontmatter}

\title{Kaluza-Klein Gravity}

\author{J. M. Overduin}
\address{Department of Physics and Astronomy,
         University of Victoria, P.O. Box 3055, Victoria,
         British Columbia, Canada, V8W 3P6}
\and
\author{P. S. Wesson}
\address{Department of Physics,
         University of Waterloo, Ontario, Canada N2L 3G1}
\address{and}
\address{Gravity Probe-B,
         Hansen Physics Laboratories,
         Stanford University, Stanford, California, U.S.A. 94305}

\begin{abstract}
We review higher-dimensional unified theories from the general 
relativity, rather than the particle physics side.  
Three distinct approaches to the subject are identified 
and contrasted:  
compactified, projective and noncompactified.
We discuss the cosmological and astrophysical implications of extra
dimensions, and conclude that none of the three approaches can be 
ruled out on observational grounds at the present time.
\end{abstract}
\end{frontmatter}
\newpage

\section{Introduction} \label{INTRO}

Kaluza's \cite{Kal21} achievement was to show that five-dimensional
general relativity contains both Einstein's four-dimensional theory
of gravity and Maxwell's theory of electromagnetism.
He however imposed a somewhat artificial restriction (the cylinder 
condition) on the coordinates, essentially barring the fifth one 
{\em a priori\/} from making a direct appearance in the laws of 
physics.  Klein's \cite{Kle26a} contribution was to make this 
restriction less artificial by suggesting a plausible {\em physical\/} 
basis for it in compactification of the fifth dimension.  This idea
was enthusiastically received by unified-field theorists, 
and when the time came to include the strong and weak forces 
by extending Kaluza's mechanism to higher dimensions, 
it was assumed that these too would be compact.
This line of thinking has led through eleven-dimensional 
supergravity theories in the 1980s to the current favorite 
contenders for a possible ``theory of everything,'' 
ten-dimensional superstrings.

We review the field of Kaluza-Klein gravity, concentrating on the 
general relativity, rather than particle physics side of the subject.
(For the latter there are already many excellent 
books \cite{dSS83,Lee84,KFF84,PW86,ACF87,CMS89} and review 
articles \cite{Tom84,App84,Duf86,DNP86,BL87,Duf94} available.)
We also aim to {\em re-examine\/} the field to some extent,
as it seems to us that the cart of compactification
has in some ways gotten ahead of the horse of unification.
Kaluza unified not only gravity and electromagnetism, but also
{\em matter and geometry\/}, for the photon appeared in four
dimensions as a manifestation of {\em empty\/} five-dimensional 
spacetime.  Modern Kaluza-Klein theories, by contrast, 
routinely require the addition of explicit 
``higher-dimensional matter'' fields in order to achieve successful
compactification (among other things).  Are they necessary?  
Yes, if extra coordinates must be real, lengthlike and compact.  
There are, however, higher-dimensional unified field theories 
which require none of these things:
{\em projective\/} theories \cite{Les82,Sch83,Sch90b,Sch95}, in which
extra coordinates are not physically real; and {\em noncompactified\/} 
theories \cite{W90,W92a,PdL93,W93a,W95a,Rip95,W96a,Mas96},
in which they are not necessarily lengthlike or compact.
These theories receive special attention in our report.

We begin in \S~\ref{HISTORY} with a historical overview of 
higher-dimensional theories of gravity.  In \S~\ref{KALUZA} 
we review Kaluza's original mechanism, emphasizing what to us are 
its three principal features.
The three main approaches to higher-dimensional unification 
{\em since\/} Kaluza --- compactified, projective and 
noncompactified --- are reviewed in \S~\ref{COMPACT}, 
\S~\ref{PROJECTIVE} and \S~\ref{NONCOMPACT} respectively.  
We note that each one modifies or sacrifices at least one of 
the key features of Kaluza's theory, and discuss the implications.
The all-important question of experimental constraints is addressed 
in \S~\ref{COSMOLOGY} and \S~\ref{ASTROPHYSICS}, 
which deal respectively with cosmological 
and astrophysical effects of extra dimensions.
None of the three above-mentioned approaches can be ruled out 
on observational grounds at the present time.
Conclusions and prospects for further work are 
summarized in \S~\ref{CONC}.

\section{Historical Overview} \label{HISTORY}

\subsection{Higher Dimensions} \label{Minkowski}

The world of everyday experience is three-dimensional.
But why should this be so?
The question goes back at least to Kepler \cite{Kep81},
who speculated that the threefold nature of the Holy Trinity might be
responsible.  More recent arguments have involved the stability of
planetary orbits and atomic ground states, the use of wave propagation
for information transmission, the fundamental constants of nature,
and the anthropic principle \cite{Bar83}, as well as
wormhole effects \cite{Vol89},
the cosmological constant \cite{Gas89},
certain ``geometry-free'' considerations \cite{Lam93},
string theories \cite{Bra95}, 
and nucleation probabilities in quantum cosmology \cite{Emb95}.
All these lines of reasoning converge on the same conclusion:
that, in agreement with common intuition,
space is composed of three macroscopic spatial dimensions
$x^1,x^2$ and $x^3$.

Nevertheless, the temptation to tinker with the
dimensionality of nature has proved irresistible to physicists over
the years.  The main reason for this is that phenomena which
require very different explanations in three-dimensional space
can often be shown to be manifestations of simpler
theories in higher-dimensional manifolds.
But how can this idea be reconciled with the observed
three-dimensionality of space?  
If there are additional coordinates,
why does physics appear to be independent of them?

It is useful to keep in mind that the new coordinates need not
necessarily be lengthlike (in the sense of being measured in meters, 
say), or even spacelike (in regard to their metric signature).
A concrete example which violates both of these expectations 
was introduced in 1909 by Minkowski \cite{Min09},
who showed that the successes of Maxwell's unified electromagnetic
theory and Einstein's special relativity could be understood
geometrically if time, along with space, were considered part of a
four-dimensional spacetime manifold via $x^0=ict$.
Many of the abovementioned arguments against more than three dimensions
were circumvented by the fact that the fourth coordinate
did not mark {\em distance\/}.
And the reason that physics had appeared three-dimensional 
for so long was because of the large size of 
the dimension-transposing parameter $c$, which meant that
the effects of ``mixing'' space and time coordinates
(ie., length contraction, time dilation) appeared only at
very high speeds.

\subsection{Kaluza-Klein Theory} \label{KKtheory}

Inspired by the close ties between
Minkowski's four-dimensional spacetime
and Maxwell's unification of electricity and magnetism,
Nordstr\"om \cite{Nor14} in 1914 
and (independently) Kaluza \cite{Kal21} in 1921
were the first to try unifying {\em gravity\/}
with electromagnetism in a theory of {\em five\/} dimensions
($x^0$ through $x^4$).  
Both men then faced the question:
why had no fifth dimension been observed in nature?
In Minkowski's time, there had already been experimental phenomena
(namely, electromagnetic ones) whose invariance with respect to
Lorentz transformations could be interpreted as
four-dimensional coordinate invariance.
No such observations pointed to a fifth dimension.
Nordstr\"om and Kaluza therefore
avoided the question and simply demanded
that all derivatives with respect to $x^4$ vanish.
In other words, physics was to take place --- for as-yet unknown 
reasons --- on a four-dimensional
hypersurface in a five-dimensional universe
(Kaluza's ``cylinder condition'').

With this assumption, each was successful in
obtaining the field equations of both electromagnetism and
gravity from a single five-dimensional theory.
Nordstr\"om, working as he was before general relativity,
assumed a scalar gravitational potential;
while Kaluza used Einstein's tensor potential.
Specifically, Kaluza demonstrated that general relativity,
when interpreted as a five-dimensional theory in vacuum
(ie., $^{5}G_{AB} = 0$, with $A,B$ running over 0,1,2,3,4),
{\em contained\/} four-dimensional general relativity
in the presence of an electromagnetic field
(ie., $^{4}G_{\alpha\beta} = \, ^{4}T^{EM}_{\alpha\beta}$, 
with $\alpha,\beta$ running over $0,1,2,3$), together with
Maxwell's laws of electromagnetism.
(There was also a Klein-Gordon equation for a massless scalar
field, but this was not appreciated --- and was in fact suppressed ---
by Kaluza at the time.)
All subsequent attempts at higher-dimensional unification
spring from this remarkable result.

Various modifications of Kaluza's five-dimensional scheme, 
including Klein's idea \cite{Kle26a,Kle26b} of compactifying 
the extra dimension (which we will discuss in a moment) were 
suggested by Einstein, Jordan, Bergmann, and a few
others \cite{Ein38,Ein43,Jor47,Ber48,Thi48,Sou58,Sou63}
over the years, but it was not extended to {\em more\/} than five 
dimensions until theories of the strong and weak nuclear interactions 
were developed.  The obvious question was whether these 
new forces could be unified with gravity and electromagnetism 
by the same method.

The key to achieving this lay in the concept of gauge invariance, 
which was coming to be recognized as underlying all the 
interactions of physics.
Electrodynamics, for example, could be ``derived'' by imposing 
local $U(1)$ gauge-invariance on a free-particle Lagrangian.
From the gauge-invariant point of view, Kaluza's feat in extracting
electromagnetism from five-dimensional gravity was no longer 
so surprising:
it worked, in effect, because $U(1)$ {\em gauge-invariance\/} had been 
``added onto'' Einstein's equations in the guise of 
{\em invariance with respect to coordinate transformations\/} 
along the fifth dimension.  In other words, 
gauge symmetry had been ``explained'' as a geometric
symmetry of spacetime.  The electromagnetic field then appeared 
as a vector ``gauge field'' in four dimensions.
It was natural --- though not simple --- to extend this insight
to groups with more complicated symmetry.  
De Witt \cite{DeW63} in 1963 was the first to suggest 
incorporating the non-Abelian $SU(2)$ gauge group of Yang and Mills 
into a Kaluza-Klein theory of $(4+d)$ dimensions.
A minimum of three extra dimensions were required.  This problem was 
picked up by others \cite{Ray65,Ker68,Tra70} and solved completely 
by the time of Cho \& Freund \cite{Cho75a,Cho75b} in 1975.

\subsection{Approaches to Higher-Dimensional Unification} 
\label{Approaches}

We emphasize here three key features of all the models 
discussed so far:\\
({\em i\/}) They embody Einstein's vision \cite{Ein56,Whe68,Sal80} 
of nature as {\em pure geometry}.
(This idea can be traced in nonmathematical form at least to
Clifford in 1876 \cite{Cli76}, and there are hints of it as far back as
the Indian Vedas, according to Wheeler and others \cite{Whe62}.)
The electromagnetic and Yang-Mills fields, as well as the 
gravitational field,
are completely contained in the higher-dimensional Einstein tensor
$^{(4+d)}G_{AB}$; that is, in the metric and its derivatives.
No explicit energy-momentum tensor $^{(4+d)}T_{AB}$ is needed.\\
({\em ii\/}) They are {\em minimal\/} extensions of general relativity
in the sense that there is no modification to the mathematical
structure of Einstein's theory.  The only change is 
that tensor indices run over 0 to $(3+d)$ instead of 0 to 3.\\
({\em iii\/}) They are {\em a priori\/} cylindrical.
No mechanism is suggested to explain why physics depends on the 
first four coordinates, but not on the extra ones.

The first two of these are agreeable from the point of view of
elegance and simplicity.
The third, however, appears contrived to modern eyes.
In the effort to repair this defect,
higher-dimensional unified theory has evolved in three
more or less independent directions since the time of Kaluza.
Each one sacrifices or modifies one of the features
({\em i\/}) to ({\em iii\/}) above.

Firstly, it has been proposed that extra dimensions do not appear 
in physics because they are {\em compactified\/} and
unobservable on experimentally accessible energy scales.
This approach has been successful in many ways, and is
the dominant paradigm in higher-dimensional unification
(recent reviews include many excellent books
\cite{dSS83,Lee84,KFF84,PW86,ACF87,CMS89}
and articles \cite{Tom84,App84,Duf86,DNP86,BL87,Duf94}).
If one wants to unify more than just gravity and electromagnetism
in this way, however, it seems that one has in practice to abandon 
Einstein's goal of geometrizing physics, at least in the sense 
of ({\em i\/}) above.

A second way to sweep the extra dimensions out of sight
is to regard them as mathematical artifacts of a more complicated 
underlying theory, sacrificing ({\em ii\/}) above.  
This can be done, for example, if
one replaces the classical (affine) geometry underlying 
Einstein's general relativity with {\em projective} geometry 
(see for reviews \cite{Les82,Sch83,Sch90b,Sch95}).
``Extra dimensions'' then become visual aids which may or may not
help us understand the underlying mathematics of nature, 
but which do not correspond to physical coordinates.

The third approach to the problem of explaining exact cylindricity
is to consider the possibility that it {\em may not necessarily be\/} 
exact, relaxing ({\em iii\/}) above.
That is, one takes the new coordinates at face value,
allowing physics to depend on them in principle
\cite{W90,W92a,PdL93,W93a,W95a,Rip95,W96a,Mas96}.
This dependence presumably appears in regimes that have not yet been 
well-probed by experiment ---
much as the relevance of Minkowski's fourth dimension to mechanics was 
not apparent at non-relativistic speeds.  
When dependence on the extra dimensions is included, 
one finds that the five-dimensional Einstein equations $^{5}R_{AB}=0$
contain the four-dimensional ones 
$^{4}G_{\alpha\beta} = \, ^{4}T_{\alpha\beta}$
with a {\em general\/} energy-momentum tensor $^{4}T_{\alpha\beta}$
instead of just the electromagnetic one $^{4}T^{EM}_{\alpha\beta}$.

\subsection{The Compactified Approach} \label{Compact}

Klein showed in 1926 \cite{Kle26a,Kle26b}
that Kaluza's cylinder condition would arise naturally 
if the fifth coordinate had
(1) a circular topology,
in which case physical fields would depend on it only periodically,
and could be Fourier-expanded; and
(2) a small enough (``compactified'') scale,
in which case the energies of all Fourier modes above the ground state
could be made so high as to be unobservable
(except --- as we now add --- possibly in the very early universe).
Physics would thus be effectively independent of Kaluza's fifth 
dimension, as desired.  As a bonus, it seemed early on that the 
expansion of the electromagnetic field into Fourier modes 
could in principle explain the quantization of electric charge.
(This aspect of the theory has had to be abandoned, however,
as the charge-to-mass ratio of the higher modes did not match that
of any known particles.  Nowadays elementary charges are identified
with the ground state Fourier modes only, and their small mass is
attributed to spontaneous symmetry-breaking.)

The scheme was not perfect; one still needed to explain {\em why\/}
extra dimensions differed so markedly in topology and scale
from the familiar spacetime ones.  Their size in particular
had to be extremely small
(below the attometer (1 am = $10^{-18}$~m) scale,
according to current experiment \cite{Kos91}).
There was also the question of how to interpret a new scalar field
which appeared in the theory.  
These difficulties have, however, proved manageable.
Scalar fields are not as threatening as they once appeared;
one now just assumes that they are too massive to have been observed.
And an entire industry has grown up around the study of 
compactification mechanisms and the topology of compact spaces.

In fact, Klein's strategy of compactifying extra dimensions
has come to dominate higher-dimensional unified physics,
leading in recent years to new fields like eleven-dimensional
supergravity and ten-dimensional superstring theory.
We will survey these developments in this section,
and make contact with many of them throughout this report, 
but it is not our purpose to review them exhaustively. 
For this the reader is directed to the books
\cite{dSS83,Lee84,KFF84,PW86,ACF87,CMS89},
and review articles \cite{Tom84,App84,Duf86,DNP86,BL87,Duf94}
mentioned already.  Our goal here is to take a broad view,
comparing and contrasting the various approaches to higher-dimensional
gravity, and focusing in particular on those which have received 
less critical attention in the literature.
A semantic note:  while the term ``Kaluza-Klein theory'' ought,
strictly speaking, to apply only to models which assume both 
cylindricity and compactified dimensions, we follow popular usage 
and apply the term to any higher-dimensional unified theory of 
gravity in which the extra dimensions are regarded as real, 
whether compactified or not.  When distinguishing between these two, 
we will refer in the latter case to
``noncompactified Kaluza-Klein theories,'' though this is 
to some extent a contradiction in terms.

\subsection{Compactification Mechanisms} \label{Mechanisms}

A difficulty with compactification is that one cannot impose it
indiscriminately on whichever dimensions one likes --- 
the combination of macroscopic four-dimensional spacetime 
plus the compactified extra-dimensional space must be a 
solution of the higher-dimensional Einstein field equations.
In particular, one should be able to recover a ``ground state'' 
solution consisting of four-dimensional Minkowski space plus 
a $d$-dimensional compact manifold.
Although this is straightforward when $d=1$ (Klein's case),
the same thing is not true in higher-dimensional theories like that
of Cho \& Freund, where the compact spaces are in general curved
\cite{ACF87,Sal82,App83a,App83b,Dol84}.
The consequences of ignoring this inconsistency in 
``Kaluza-Klein ansatz'' have been emphasized by 
Duff {\em et al.} \cite{Duf86,DNP86,Duf94,Duf84}.
This and related problems have even led Cho
\cite{Cho87a,Cho87b,Cho89,Cho90} to call for the abandonment of
Klein's ``zero modes approximation'' as a means of 
dimensional reduction.

In general, however, spacetime can still be coaxed into compactifying 
in the desired manner --- at the cost of altering the 
higher-dimensional vacuum Einstein equations, either by incorporating 
torsion \cite{Thi72,Cha76,Dom78,Orz81},
adding higher-derivative terms (eg., $R^2$) onto the Einstein action
\cite{Wet82}, or --- last but not least --- adding an explicit
higher-dimensional energy-momentum tensor to the theory.
If chosen judiciously, this last will induce 
``spontaneous compactification'' of the extra dimensions, 
as first demonstrated by
Cremmer \& Scherk \cite{Cre76,Cre77}.
This approach, though, sacrifices Einstein and Kaluza's dream
\cite{Ein56,Whe68,Sal80,Cli76,Whe62}
of a purely {\em geometrical\/} unified theory of nature.
Rather than explaining the ``base wood'' of four-dimensional 
matter and forces as manifestations of the ``pure marble'' 
of geometry in higher dimensions, one has essentially been 
driven to invent {\em new\/} kinds of wood.
Weinberg \cite{Wei84a} has likened this situation
to the fable of ``stone soup,'' in which a miraculous stew, 
allegedly made out of rocks, turns out on deeper investigation 
to be made from rocks plus various kinds of vegetables, 
meat and spices.  

In spite of this aesthetic drawback, however,
the idea of spontaneous compactification gained rapid acceptance
\cite{Hor77,Luc78,Hor78,Ran82,Ran83,Tos84}
and has become the standard way to
reconcile extra dimensions with the observed four-dimensionality
of spacetime in Kaluza-Klein theory (see Bailin \& Love \cite{BL87} 
for a review).  An important variation
is that of Candelas \& Weinberg \cite{Wei83,Can84}, who showed
that the quantum Casimir energy of massless higher-dimensional fields,
when combined with a higher-dimensional cosmological constant, 
can also compactify the extra dimensions in a satisfactory way.  
Unfortunately, some $10^4 - 10^5$ matter fields are required.

\subsection{$D=11$ Supergravity} \label{SUGRA}

One way to make the addition ``by hand'' of extra matter fields
more natural was to make the theory {\em supersymmetric\/}
(ie., to match up every boson with an as-yet undetected 
fermionic ``superpartner'' and vice versa).  
The reason for this is that the (compactified) 
Kaluza-Klein programme of ``explaining'' gauge symmetries as 
(restricted) higher-dimensional spacetime symmetries can 
only give rise to four-dimensional gauge {\em bosons\/}.
If the theory is to include {\em fermionic\/} fields, as required 
by supersymmetry, then these fields at least must be put in by hand.
(This limitation may not necessarily apply to 
noncompactified Kaluza-Klein theories, in which the modest dependence 
on extra coordinates --- subject to experimental constraints ---
gives the Einstein equations a rich enough structure
that matter of a very {\em general\/} kind 
can be ``induced'' in the four-dimensional universe by 
pure geometry in higher dimensions.
In five dimensions, for example, one can obtain not only 
photons, the gauge bosons of electromagnetism, but also
dustlike, vacuum, or ``stiff'' matter.)

Supersymmetric gravity (``supergravity'') began life as a 
four-dimensional theory in 1976 \cite{Fre76,Des76}, but quickly 
made the jump to higher dimensions (``Kaluza-Klein supergravity'').
It was particularly successful in $D=11$, for three principal reasons.
First, Nahm \cite{Nah78} showed that
eleven was the {\em maximum\/} number of dimensions
consistent with a single graviton 
(and an upper limit of two on particle spin).
This was followed by Witten's proof \cite{Wit81} that eleven was also
the {\em minimum\/} number of dimensions required for a Kaluza-Klein 
theory to unify all the forces in the standard model of particle 
physics (ie., to contain the gauge groups of the strong ($SU(3)$) and
electroweak ($SU(2) \times U(1)$) interactions).
The combination of supersymmetry with Kaluza-Klein theory thus 
appeared to {\em uniquely fix\/} the dimensionality of spacetime.
Secondly, whereas in lower dimensions one had to choose between 
several possible configurations for the extra matter fields,
Cremmer, Julia \& Scherk \cite{Cre78} demonstrated in 1978 
that in $D=11$ exactly {\em one\/} choice was consistent with the 
requirements of supersymmetry (in particular, that there be 
equal numbers of Bose and Fermi degrees of freedom).  
In other words, while a higher-dimensional energy-momentum
tensor was still required, its form at least appeared less contrived.
Finally, Freund \& Rubin \cite{Fre80} showed in 1980 that 
compactification of the $D=11$ model could occur in only two ways:
to seven or four compact dimensions, leaving four 
(or seven, respectively) macroscopic ones.  
Not only did eleven-dimensional spacetime appear to 
be uniquely favoured for unification, but it also split perfectly
to produce the observed four-dimensional world.
(The other possibility, of a macroscopic {\em seven\/}-dimensional 
world, could unfortunately not be ruled out, and in fact at least 
one such model was explicitly constructed as well \cite{Pil84}.)
Buoyed by these successes, eleven-dimensional supergravity appeared 
set by the mid 1980s as a leading candidate for the hoped-for 
``theory of everything'' (see \cite{CMS89,DNP86,Wes86} for reviews,
and \cite{SS89} for an extensive collection of papers.
A nontechnical introduction is given in \cite{Fre85}.)

A number of blemishes, however --- one aesthetic and three more 
practical --- have dampened this initial enthusiasm.
Firstly, the compact manifolds originally envisioned by 
Witten \cite{Wit81} (those containing the standard model) 
turned out not to generate quarks or
leptons, and to be incompatible with supersymmetry \cite{CMS89,BL87}.
Their most successful replacements are the 7-sphere 
and the ``squashed'' 7-sphere \cite{DNP86}, described respectively 
by the symmetry groups $SO(8)$ and $SO(5) \times SU(2)$.
These groups, however, unfortunately do not contain the 
minimum symmetry requirements of the standard model 
($SU(3) \times SU(2) \times U(1)$).
This is commonly rectified by adding {\em more\/} matter fields, 
the ``composite gauge fields'' \cite{Cre79}, to the 
eleven-dimensional Lagrangian.
Secondly, it is very difficult to build chirality 
(necessary for a realistic fermion model) into an 
eleven-dimensional theory \cite{Wit81,Wet83}.
A variety of remedies have been proposed for this, 
including the ubiquitous additional higher-dimensional 
gauge fields \cite{Hor77,Ran83},
{\em noncompact\/} internal manifolds 
\cite{Wet84,Gel84,Wet85,Gel85,McI85,Ran86,Wet86},
and extensions of Riemannian geometry \cite{Wei84b,Wei84c,Wei86}.
Thirdly, $D=11$ supergravity theory is marred by a
large cosmological constant in four dimensions,
which is difficult to remove, even by fine-tuning \cite{ACF87,CMS89}.
Finally, quantization of the theory leads inevitably 
to anomalies \cite{SS89}.

Some of these difficulties can be eased by descending to 
ten dimensions: chirality is easier to obtain \cite{Wet83}, and 
many of the anomalies disappear \cite{Alv84}.
However, the introduction of chiral fermions leads to {\em new\/} 
kinds of anomalies.  
And the primary benefit of the $D=11$ theory --- its uniqueness --- 
is lost, since ten dimensions are not specially favoured, 
and the higher-energy theory does not break down naturally 
into four macroscopic and six compact dimensions.
(One can still find solutions in which this happens, 
but there is no reason why they should be preferred.)
In fact, most $D=10$ supergravity models not only 
require {\em ad hoc\/} higher-dimensional matter fields 
to ensure proper compactification, 
but entirely {\em ignore\/} gauge fields arising from the Kaluza-Klein 
mechanism (ie., from symmetries of the compact manifold),
so that {\em all\/} the gauge fields are effectively put into
the theory by hand \cite{CMS89}.
Kaluza's original aim of explaining forces in geometrical terms
is thus abandoned completely.

\subsection{$D=10$ Superstring Theory} \label{SUSY}

A breakthrough in solving the uniqueness and anomaly problems 
of $D=10$ theory occured when Green \& Schwarz \cite{Gre84} 
and Gross {\em et al.\/} \cite{Gro85} showed that there were two, 
and {\em only\/} two ten-dimensional supergravity models 
in which {\em all\/} anomalies could be made to vanish:  
those based on the groups $SO(32)$ and 
$E_8 \times E_8$ respectively.
Once again, extra terms (known as Chapline-Manton terms) 
had to be added to the higher-dimensional Lagrangian \cite{CMS89}.
This time, however, the addition was not completely arbitrary;
the extra terms were those which would appear anyway 
if the theory were a low-energy approximation to 
certain kinds of super{\em string} theory.

The state of the art in compactified Kaluza-Klein theory, then, 
has shifted from supergravity theories to
superstring theories, the main significance of the former now being
as low-energy limits of the latter \cite{SS89}.
Superstrings (supersymmetric generalizations of strings)
avoid the generic prediction of tachyons that plagued
the first string theories \cite{Sch74},
but retain their best features, especially the possibility of an
anomaly-free path to quantum gravity \cite{Gre82}.
In fact, their many virtues make them the current favorite
contender for a ``theory of everything'' \cite{Hor94}.  
Connections have recently been made between certain
superstring states and extreme black holes \cite{Duf94},
and it has even been argued that superstrings can help resolve
the long-standing black hole information paradox \cite{Sus94}.

Something of a uniqueness problem has persisted for $D=10$
superstrings in that the groups $SO(32)$ and $E_8 \times E_8$ admit
five different string theories between them.  But this difficulty
has recently been addressed by Witten \cite{Wit95}, who showed that it
is possible to view these five theories as aspects of a single
underlying theory, now known as M-theory (for ``Membrane'') \cite{Sch96a}.
The low-energy limit of this new theory, furthermore,
turns out to be $D=11$ supergravity!  So it appears that the
preferred dimensionality of spacetime in compactified Kaluza-Klein
theory may be switching back to eleven.

Perhaps the biggest obstacle to a wider acceptance of these
theories is the difficulty of extracting clear-cut physical
predictions from them.  String theory is ``promising ... ,''
one worker has said, ``... and promising, and promising'' \cite{Wil90}.
M-theory, which (unlike superstring theory) is not perturbative,
is even more opaque; Witten has suggested that the ``M'' might
equally well stand for ``Magic'' or ``Mystery'' at present \cite{Duf96}.
We will not consider these interesting developments further here,
directing the reader instead to the superstring reviews in
\cite{CMS89,Sch86,Gre87} 
(a nontechnical account may be found in \cite{Gre86}),
or the recent reviews of M-theory in \cite{Duf96,Sch96b}.

\subsection{The Projective Approach} \label{Projective}

Compactification of extra dimensions is not the only way to explain
Kaluza's cylinder condition.
Another, less well-known approach goes back
to Veblen \& Hoffmann \cite{Veb31} in 1931.
These authors showed that the fifth dimension 
could be ``absorbed into'' ordinary
four-dimensional spacetime if one replaced the {\em classical\/}
(affine) tensors of general relativity with {\em projective\/} ones.
Rather than being regarded as new coordinates, the extra dimensions
were effectively demoted to visual aids.
Because they were not physically real, 
there was no need to explain why they were not observed.  
The price for this resolution of the problem was that one had to 
alter the geometrical foundation of Einstein's theory.
This idea received attention from 
Jordan, Pauli and several others over the 
years \cite{Jor47,Pau33,Pai41,Jor48,Lud51,Jor55,Sch57,Sch68,Les74}.
Early versions of the theory ran afoul of experimental constraints 
on the Brans-Dicke parameter $\omega$
and had to be ruled out as untenable \cite{Les82}.
However the projective approach has been revived in at least two
new formulations.  That of Lessner \cite{Les82} assigns the
scalar field a purely {\em microscopic\/} meaning; this 
has interesting consequences for elementary particles
\cite{Les86,Les87,Les90,Les91}.
The other, due to Schmutzer \cite{Sch83,Sch90b,Sch95},
endows the vacuum with a special kind of higher-dimensional matter, 
the ``non-geometrizable substrate,''
thereby sacrificing Einstein's dream of reducing physics to geometry.
This theory, however, does make a number of testable 
predictions \cite{Sch90b,Sch88a,Sch88b,Sch90a}
which are so far compatible with observation.

\subsection{The Noncompactified Approach} \label{Noncompact}

An alternative to both the compactified and projective approaches
is to take the extra dimensions at face value,
without necessarily compactifying them, 
and assume that nature is only {\em approximately\/}
independent of them --- much as it was on
Minkowski's fourth coordinate at nonrelativistic speeds.
In other words, one avoids having to explain why cylindricity
should be exact by {\em relaxing\/} it in principle.
Of course, the question remains as to why nature should be
so {\em nearly\/} cylindrical in practice.
If the extra dimensions are lengthlike, 
then one might try to answer this by supposing that particles
are trapped near a four-dimensional hypersurface
by a potential well.
Ideas of this kind have been around since at least 1962 \cite{Jos62};
for recent discussions see \cite{Rub83,Vis85,Lag87,Gib87a}.

Confining potentials are not, however, 
an obvious improvement over compactification mechanisms
in terms of economy of thought.
An alternative is to take Minkowski's example more literally and
entertain the idea that extra dimensions, like time, 
might not necessarily be lengthlike.
In this case the explanation for the near-cylindricity of
nature is to be found in the physical interpretation of the extra
coordinates; ie., in the values of the dimension-transposing
parameters (like $c$) needed to give them units of length.
The first such proposal of which we are aware is the 1983
``space-time-mass'' theory of Wesson \cite{W83},
who suggested that a fifth dimension might be associated
with {\em rest mass\/} via $x^4 = Gm/c^2$.
The chief effect of this new coordinate on four-dimensional physics
was that particle rest mass, usually assumed to be constant,
varied with time.  The variation was, however, 
small and quite consistent with experiment.
This model has been studied in some detail, particularly
with regard to its consequences for astrophysics and cosmology,
by Wesson \cite{W90,W84,W85,W86,W88} and others
\cite{Cha86a,Cha87,Cha90a,Cha90b,Ban90a,PdL88,Gro88a,Gro88b,Bao89},
\cite{Ma90a,Ma90b,Ma91,Col90,Chi90,Roq91,Car91,Car92,Wag92,Mac93},
\cite{Fuk87,Fuk88a,Fuk88b,Fuk92a,Fuk93},
and has been extended to {\em more\/} than
five dimensions by Fukui \cite{Fuk88c,Fuk92b},
with the constants $\hbar$ and $e$ playing roles 
analogous to $c$ and $G$.

Variable gravity theories are, of course, not new.
What {\em is\/} new in the models just described ---
and what is important about noncompactified Kaluza-Klein theory
in principle --- is not so much the particular physical 
interpretation one attaches to the new coordinates, 
but the bare fact that 
{\em physics is allowed to depend on them at all}.
It is clearly of interest to study the higher-dimensional 
Einstein equations with a {\em general\/} dependence 
on the extra coordinates; ie., 
without any preconceived notions as to their physical meaning.
A pioneering effort in this direction was made 
in 1986 by Yoshimura \cite{Yos86},
who however considered only the case where the $d$-dimensional
part of the $(4+d)$-space could depend on the new coordinates.
The general theory, in which {\em any\/} part of the metric can
depend on the fifth coordinate, has been explored recently by Wesson
and others \cite{W90,W92a,PdL93,W93a,W95a,Rip95,W96a,Mas96},
and its implications for cosmology \cite{W92b,W92c,W94b,W95c},
\cite{Bil96a,W95b,Liu94,McM94,Col94,Col95,W94a,Mas94,Liu95b}
and astrophysics \cite{W92d,Liu92,W94c,Liu96a,W92e,W94d},
\cite{Bil96b,Lim92,Kal95,Lim95,W96b,W93b,Liu93,Liu96b,Liu96c}
have become the focus of a growing research effort.
As this branch of Kaluza-Klein theory has not yet been reviewed 
in a comprehensive manner,
we propose to devote special attention to it in this report.
Our intention is to compare
and contrast this branch of the subject with other ones, however,
so we will make frequent contact with the compactified and
(to a lesser extent) projective theories.

\section{The Kaluza Mechanism} \label{KALUZA}

Kaluza unified electromagnetism with gravity by
applying Einstein's general theory of relativity to a {\em five-},
rather than four-dimensional spacetime manifold.  
In what follows, 
we consider generalizations of his procedure that may
be new to some readers, so it will be advantageous to briefly
review the mathematics and underlying assumptions here.

\subsection{Matter from Geometry} \label{Geometry}

The Einstein equations in five dimensions with 
{\em no five-dimensional energy-momentum tensor\/} are:
\begin{equation}
\hat{G}_{AB} = 0 \; \; \; ,
\label{5dEFE1}
\end{equation}
or, equivalently:
\begin{equation}
\hat{R}_{AB} = 0 \; \; \; ,
\label{5dEFE2}
\end{equation}
where $\hat{G}_{AB} \equiv \hat{R}_{AB} - \hat{R} \, \hat{g}_{AB} / 2$
is the Einstein tensor, $\hat{R}_{AB}$ and
$\hat{R} = \hat{g}_{AB} \hat{R}^{AB}$ are 
the five-dimensional Ricci tensor and scalar respectively,
and $\hat{g}_{AB}$ is the five-dimensional metric tensor.
(Throughout this report
capital Latin indices $A,B, ...$ run over $0,1,2,3,4$,
and five-dimensional quantities are denoted by hats.)
These equations can be derived by varying a five-dimensional
version of the usual Einstein action:
\begin{equation}
S = -\frac{1}{16 \pi \hat{G}} \int \hat{R} \sqrt{-\hat{g}} 
     \; d^4 x \, dy \; \; \; ,
\label{5dAction}
\end{equation}
with respect to the five-dimensional metric,
where $y=x^4$ represents the new (fifth) coordinate and
$\hat{G}$ is a ``five-dimensional gravitational constant.''

The absence of matter sources in these equations reflects
what we have emphasized as Kaluza's first key assumption ({\em i\/}),
inspired by Einstein:  that 
{\em the universe in higher dimensions is empty}.
The idea is to explain matter (in four dimensions) as a 
manifestation of pure geometry (in higher ones).  
If, instead, one introduced new kinds of higher-dimensional matter, 
then one would have gained little in economy of thought.  
One would, so to speak, be getting Weinberg's 
``stone soup'' \cite{Wei84a} from a can.

\subsection{A Minimal Extension of General Relativity} \label{Minimal}

The five-dimensional Ricci tensor and Christoffel symbols are defined
in terms of the metric exactly as in four dimensions:
\begin{eqnarray}
\hat{R}_{AB}        & = & \partial_C \hat{\Gamma}^C_{AB} -
                          \partial_B \hat{\Gamma}^C_{AC} +
                          \hat{\Gamma}^C_{AB} \hat{\Gamma}^D_{CD} -
                          \hat{\Gamma}^C_{AD} \hat{\Gamma}^D_{BC} 
                          \; \; \; , \nonumber \\
\hat{\Gamma}^C_{AB} & = & \frac{1}{2} \hat{g}^{CD} \left( 
                          \partial_A \hat{g}_{DB} +
                          \partial_B \hat{g}_{DA} -
                          \partial_D \hat{g}_{AB} \right) \; \; \; .
\label{5dChristRicci}
\end{eqnarray}
Note that, aside from the fact that tensor indices run over $0 - 4$ 
instead of $0 - 3$, all is exactly as it was in Einstein's theory.  
We have emphasized this as the second key feature ({\em ii\/}) of 
Kaluza's approach to unification.

Everything now depends on one's choice for the form of the 
five-dimensional metric.  In general, one identifies the 
$\alpha\beta$-part of $\hat{g}_{AB}$ with $g_{\alpha\beta}$ 
(the four-dimensional metric tensor),
the $\alpha 4$-part with $A_{\alpha}$ (the electromagnetic potential),
and the $44$-part with $\phi$  (a scalar field).
A convenient way to parametrize things is as follows:
\begin{equation}
\left( \hat{g}_{AB} \right) = \left( \begin{array}{cc}
   g_{\alpha\beta} + \kappa^2 \phi^2 A_{\alpha} A_{\beta} \; \; & \; \; 
      \kappa \phi^2 A_{\alpha} \\
   \kappa \phi^2 A_{\beta} \; \;                                & \; \; 
      \phi^2
   \end{array} \right) \; \; \; ,
\label{5dMetric}
\end{equation}
where we have scaled the electromagnetic potential $A_{\alpha}$ 
by a constant $\kappa$ in order to get the right 
multiplicative factors in the action later on.
(Throughout this report, Greek indices $\alpha,\beta, ...$ run 
over $0,1,2,3$, and small Latin indices $a,b, ...$ run over $1,2,3$.
The four-dimensional metric signature is taken 
to be $(+ \, - \, - \, -)$, and we work in units such that $c=1$.  
In addition, for convenience and accord with other work, 
we set $\hbar=1$ in \S~\ref{KALUZA}, 
and $G=1$ in in \S~\ref{COSMOLOGY} and \S~\ref{ASTROPHYSICS}.)

\subsection{The Cylinder Condition} \label{Cylinder}

If one then applies the third key feature ({\em iii\/}) of Kaluza's 
theory (the cylinder condition), which means dropping all derivatives 
with respect to the fifth coordinate, then one finds, using the 
metric~(\ref{5dMetric}) and the definitions~(\ref{5dChristRicci}), 
that the $\alpha\beta$-, $\alpha 4$-, and $44$-components of 
the five-dimensional field equation~(\ref{5dEFE2}) reduce 
respectively to the following field equations \cite{Les82,Thi48}
in four dimensions:
\begin{eqnarray}
G_{\alpha\beta}                    & = & \frac{\kappa^2 \phi^2}{2} 
   T_{\alpha\beta}^{EM} - \frac{1}{\phi} \left[ \nabla_{\alpha} 
   (\partial_{\beta} \phi) - g_{\alpha\beta} \Box \phi \right] 
   \; \; \; , \nonumber \\
\nabla^{\alpha} \, F_{\alpha\beta} & = & -3 \, \frac{\partial^{\alpha} 
   \phi}{\phi} \, F_{\alpha\beta} \; \; \; , \; \; \;
\Box \phi = \frac{\kappa^2 \phi^3}{4} \, F_{\alpha\beta} 
   F^{\alpha\beta} \; \; \; ,
\label{4dFieldEquns}
\end{eqnarray}
where $G_{\alpha\beta} \equiv R_{\alpha\beta} - R g_{\alpha\beta} / 2$ 
is the Einstein tensor,
$T_{\alpha\beta}^{EM} \equiv g_{\alpha\beta} F_{\gamma\delta} 
F^{\gamma\delta}/4 - F_{\alpha}^{\gamma} F_{\beta\gamma}$
is the electromagnetic energy-momentum tensor, and
$F_{\alpha\beta} \equiv \partial_{\alpha} A_{\beta} - 
\partial_{\beta} A_{\alpha}$.
There are a total of $10+4+1=15$ equations, as expected since 
there are fifteen independent elements in
the five-dimensional metric~(\ref{5dMetric}).

\subsection{The Case $\phi=$ constant} \label{GR+EM}

If the scalar field $\phi$ is constant throughout spacetime, 
then the first two of eqs.~(\ref{4dFieldEquns}) are just the 
Einstein and Maxwell equations:
\begin{equation}
G_{\alpha\beta} = 8 \pi G \phi^2 T_{\alpha\beta}^{EM} \; \; \; , 
   \; \; \;
\nabla^{\alpha} \, F_{\alpha\beta} = 0 \; \; \; ,
\label{EinsteinMaxwell}
\end{equation}
where we have identified the scaling parameter $\kappa$ in terms of 
the gravitational constant $G$ (in four dimensions) by:
\begin{equation}
\kappa \equiv 4 \sqrt{\pi G} \; \; \; .
\label{DefnKappa}
\end{equation}
This is the result originally obtained by Kaluza and Klein, 
who set $\phi=1$.  
(The same thing has done by some subsequent authors 
employing ``special coordinate systems'' \cite{Tom84,Ber76}.)
The condition $\phi=$ constant is, however, 
only consistent with the the {\em third\/} of the field
equations~(\ref{4dFieldEquns}) 
when $F_{\alpha\beta} F^{\alpha\beta} = 0$, 
as was first pointed out by Jordan \cite{Jor47,Ber48} 
and Thiry \cite{Thi48}.
The fact that this took twenty years to be acknowledged 
is a measure of the deep suspicion with which
scalar fields were viewed in the first half of this century.

Nowadays the same derivation is usually written in variational 
language.  Using the metric~(\ref{5dMetric}) and the 
definitions~(\ref{5dChristRicci}), and invoking the 
cylinder condition not only to drop derivatives with respect to $y$, 
but also to pull $\int dy$ out of the action integral,
one finds that eq.~(\ref{5dAction}) contains
three components \cite{App84}:
\begin{equation}
S = -\int d^4x \, \sqrt{-g} \, \phi \, \left( \frac{R}{16 \pi G} +
     \frac{1}{4} \phi^2 F_{\alpha\beta} F^{\alpha\beta} +
     \frac{2}{3 \kappa^2} \, \frac{\partial^{\alpha} \phi 
     \; \partial_{\alpha} \phi} {\phi^2} \right) \; \; \; ,
\label{4dAction}
\end{equation}
where $G$ is defined in terms of its five-dimensional 
counterpart $\hat{G}$ by:
\begin{equation}
G \equiv \hat{G} / \int dy \; \; \; ,
\label{DefnG}
\end{equation}
and where we have used equation~(\ref{DefnKappa}) to bring the
factor of $16\pi G$ inside the integral.
As before, if one takes $\phi=$ constant, 
then the first two components of this action are just the 
Einstein-Maxwell action for gravity and electromagnetic radiation 
(scaled by factors of $\phi$).  The third component is the action 
for a massless Klein-Gordon scalar field.

The fact that the action~(\ref{5dAction}) leads to~(\ref{4dAction}),
or --- equivalently --- that 
the sourceless field equations~(\ref{5dEFE2})
lead to~(\ref{4dFieldEquns}) {\em with} source matter, 
constitutes the central miracle of Kaluza-Klein theory.
Four-dimensional matter (electromagnetic radiation, at least)
has been shown to arise purely from the geometry of empty
five-dimensional spacetime.  The goal of all subsequent 
Kaluza-Klein theories has been to extend this success to 
{\em other\/} kinds of matter.

\subsection{The Case $A_{\alpha}=0$:  Brans-Dicke Theory} 
\label{BransDicke}

If one does {\em not\/} set $\phi=$ constant, then Kaluza's 
five-dimensional theory contains besides electromagnetic effects 
a Brans-Dicke-type scalar field theory, 
as becomes clear when one considers the case in
which the electromagnetic potentials vanish, $A_{\alpha}=0$.  
Without the cylinder condition, this would be no more than a choice
of coordinates, and would not entail any loss of algebraic generality.
(It would be exactly analogous to the common procedure in ordinary
electrodynamics of choosing four-space coordinates in which either 
the electric or magnetic field disappears.)
With the cylinder condition, however, we are effectively working
in a special set of coordinates, so that the theory is no
longer invariant with respect to general (ie., five-dimensional)
coordinate transformations.
The restriction $A_{\alpha}=0$ is, therefore, a physical 
and not merely mathematical one, 
and restricts us to the ``graviton-scalar sector'' of the theory.

This is acceptable in some contexts ---
in homogeneous and isotropic situations, for example,
where off-diagonal metric coefficients 
would pick out preferred directions; or
in early-universe models which are dynamically dominated
by scalar fields.  Neglecting the $A_{\alpha}$-fields, then, 
eq.~(\ref{5dMetric}) becomes:
\begin{equation}
\left( \hat{g}_{AB} \right) = \left( \begin{array}{cc}
                              g_{\alpha\beta} \; \; & \; \; 0 \\
                              0                     & \; \; \phi^2
                              \end{array} \right) \; \; \; .
\label{New5dMetric}
\end{equation}
With this metric, the field equations~(\ref{5dEFE2}), and Kaluza's
assumptions ({\em i\/}) -- ({\em iii\/}) as before, the
action~(\ref{5dAction}) reduces to:
\begin{equation}
S = -\frac{1}{16 \pi G} \int d^4x \, \sqrt{-g} \; R \phi \; \; \; .
\label{4dAction2}
\end{equation}
This is the special case $\omega = 0$ of
the Brans-Dicke action \cite{Bra61}:
\begin{equation}
S_{BD} = -\int d^4x \, \sqrt{-g} \; \left( \frac{R \phi}{16 \pi G} +
         \omega \, \frac{\partial^{\alpha} \phi \,
         \partial_{\alpha} \phi} {\phi} \right) 
         + S_m \; \; \; ,
\label{BDaction}
\end{equation}
where $\omega$ is the dimensionless Brans-Dicke constant and the
term $S_m$ refers to the action associated with any
matter fields which may be coupled to the metric or scalar field.

The value of $\omega$ is of course constrained to be greater than 
$\sim \! \! 500$ by observation \cite{Wil81}, so this simple model
is certainly not viable, in the present era at least.
One can however evade this limit by adding a nonzero potential 
$V(\phi)$ to the above action, as in extended inflation \cite{LaS89} 
and other theories \cite{Wet88,Sol91}; 
or by allowing the Brans-Dicke parameter $\omega$ to vary 
as a function of $\phi$, as in hyperextended \cite{Ste90} 
and other inflationary models \cite{McD93}.

\subsection{Conformal Rescaling} \label{Rescaling}

One can also re-formulate the problem by carrying out a Weyl, 
or {\em conformal rescaling\/} of the metric tensor.
Conformal factors have begun to appear frequently in papers on
Kaluza-Klein theory, but they have as yet received little attention 
in reviews of the subject, so we will discuss them briefly here, 
referring the reader to the literature for details.

The extra factor of $\phi$ in the action~(\ref{4dAction})
above implies that, strictly speaking, the scalar field would have
to be constant throughout spacetime \cite{App84,BL87} in order for 
the gravitational part of the action to be in canonical form.
Some authors \cite{Tom84,Ber76} have in fact set it equal to one 
by definition, though this is of course not a generally covariant 
procedure.  The offending factor can however be removed by
conformally rescaling the five-dimensional metric:
\begin{equation}
\hat{g}_{AB} \rightarrow \hat{g}^{\prime}_{AB} =
   \Omega^2 \hat{g}_{AB} \; \; \; ,
\label{ConfRescaling}
\end{equation}
where $\Omega^2 > 0$ is the conformal (or Weyl) factor,
a function of the first four coordinates only
(assuming Kaluza's cylinder condition).
This is one step removed from the simplest possible realization
of Kaluza's idea.  In compactified and projective theories, however,
there can be no physical objection to such a procedure since
it takes place ``in higher dimensions'' which are not accessible
to observation.  Questions only arise in the process of
dimensional reduction; ie., in interpreting the ``real,''
four-dimensional quantities in terms of the 
rescaled five-dimensional ones.

The four-dimensional metric tensor is rescaled by 
the same factor as the five-dimensional one
($g_{\alpha\beta} \rightarrow g^{\prime}_{\alpha\beta}
= \Omega^2 g_{\alpha\beta}$), and this has the following
effect on the four-dimensional Ricci scalar \cite{Wal84}:
\begin{equation}
R \rightarrow R^{\prime} = \Omega^{-2} \left( R +
   6 \frac{\Box \Omega}{\Omega} \right) \; \; \; .
\end{equation}
A convenient parametrization is obtained by making the trivial 
redefinition $\phi^2 \rightarrow \phi$ and then introducing the
conformal factor $\Omega^2 = \phi^{-1/3}$, so that 
the five-dimensional metric reads:
\begin{equation}
\left( \hat{g}_{AB}^{\prime} \right) = 
\phi^{-1/3} \; \left( \begin{array}{cc}
   g_{\alpha\beta} + \kappa^2 \phi A_{\alpha} A_{\beta} \; \; & \; \; 
      \kappa \phi A_{\alpha} \\
   \kappa \phi A_{\beta} \; \;                                & \; \; 
      \phi
   \end{array} \right) \; \; \; ,
\label{Conf5dMetric}
\end{equation}
The same procedure as before then leads \cite{CMS89,DNP86,BL87}
to the following {\em conformally rescaled\/} action
instead of eq.~(\ref{4dAction}) above:
\begin{equation}
S^{\prime} = -\int d^4x \, \sqrt{-g^{\prime}} \left( \frac{R^{\prime}}
             {16 \pi G} + \frac{1}{4} \phi F^{\prime}_{\alpha\beta}
             F^{\prime \, \alpha\beta} + \frac{1}{6 \kappa^2} \,
             \frac{\partial^{\prime \, \alpha} \phi \; 
             \partial^{\prime}_{\alpha} \phi}{\phi^2} 
             \right) \; \; \; ,
\label{Conf4dAction}
\end{equation}
where primed quantities refer to the rescaled metric
(ie., $\partial^{\prime \, \alpha} \phi = g^{\prime \, \alpha\beta}
\partial_{\beta} \phi$), and $G$ and $\kappa$ are defined as before.
The gravitational part of the action then has the conventional form,
as desired.

The Brans-Dicke case, obtained by putting $A_{\alpha}=0$ 
in the metric, is also modified by the presence of 
the conformal factor.  One finds (again making the redefinition
$\phi^2 \rightarrow \phi$ and using $\Omega^2=\phi^{-1/3}$)
that the action~(\ref{4dAction2}) becomes \cite{BL87}:
\begin{equation}
S^{\prime} = -\int d^4x \, \sqrt{-g^{\prime}} \left( 
             \frac{R^{\prime}} {16 \pi G} + \frac{1}{6 \kappa^2} \,
             \frac{\partial^{\prime \, \alpha} \phi \; 
             \partial^{\prime}_{\alpha} \phi}{\phi^2} 
             \right) \; \; \; .
\label{Conf4dAction2}
\end{equation}
In terms of the ``dilaton'' field 
$\sigma \equiv \ln \phi / (\sqrt{3} \, \kappa)$,
this action can be written:
\begin{equation}
S^{\prime} = -\int d^4x \, \sqrt{-g^{\prime}} \left( 
             \frac{R^{\prime}} {16 \pi G} + \frac{1}{2} \,
             \partial^{\prime \, \alpha} \sigma \; 
             \partial^{\prime}_{\alpha} \sigma \right) \; \; \; ,
\label{Conf4dAction3}
\end{equation}
which is the canonical action for a minimally coupled scalar field 
with no potential \cite{Mad88}.

\subsection{Conformal Ambiguity} \label{Ambiguity}

The question of conformal ambiguity arises when we ask,
``Which is the {\em real\/} four-dimensional metric 
(ie., the one responsible for Einstein's gravity) --- 
the original $g_{\alpha\beta}$,
or the rescaled $g^{\prime}_{\alpha\beta}$?''
The issue was already raised at least as far back as 1955 by 
Pauli \cite{Jor55}.  (The rescaled metric is sometimes referred to in
the literature as the ``Pauli metric,'' as opposed to the unrescaled
``Jordan metric.'')  Most authors have worked with the traditional
(unrescaled) metric, if indeed they have troubled themselves 
over the matter at all \cite{Sok89}.  Others \cite{Dam90,Cho92} 
have considered the interesting idea of coupling visible matter 
(including that involved in the classical tests of general relativity) 
to the Jordan metric, but allowing {\em dark matter\/} to couple to a 
rescaled Pauli metric.  In recent years, a variety of new arguments 
have been advanced in favor of regarding the rescaled metric as the 
true ``Einstein metric'' for {\em all\/} types of matter in 
compactified Kaluza-Klein theory.
The following paragraph is intended as a brief review of these;
many are discussed more thoroughly in \cite{Sok89}.

The first use of conformal rescaling to pick out ``physical fields'' 
was in certain ten-dimensional supergravity \cite{Cha81} 
and superstring \cite{Din85} models of the early 1980s.  
It then appeared in work on the quantum aspects 
of Kaluza-Klein theory \cite{Kun86}, and on the stability of
compactified Kaluza-Klein cosmologies \cite{Mae86,Mae87}.
In these papers it was asserted that conformal ambiguity affected 
the physics at the quantum but not the classical level.  
This was supported by a demonstration \cite{Bom87} 
that the mass of a five-dimensional Kaluza-Klein monopole 
was invariant with respect to conformal rescaling, 
although it was speculated in this paper that the addition 
of matter fields would complicate the situation.  
Cho \cite{Cho92} confirmed this suspicion
by showing explicitly that the conformal invariance of the Brans-Dicke 
action~(\ref{BDaction}) would be broken for $S_m \ne 0$.
This resulted in different matter couplings to the metric 
for different conformal factors, which would manifest themselves as
``fifth force''-type violations of the weak equivalence 
principle \cite{Cho91}.
He argued in addition that only one conformal factor --- 
the factor $\phi^{-1/3}$ used above --- could allow one 
to properly interpret the metric as a massless spin-two graviton
\cite{Cho92}; and moreover that without this factor the
kinetic energy of the scalar field would be unbounded from below, 
making the theory unstable \cite{Cho87a}.  
This last point has also been emphasized
by Soko{\l}owski and others \cite{Sok89,Sok86,Sok87}.
(Note that the conformal factor $\sqrt{\varphi}$ used by these authors 
is the same as the one discussed above; the exponent depends on 
whether one rescales the five-, or only the four-dimensional metric.
The scalar $\varphi$ is related to $\phi$ simply 
by $\varphi=\phi^{-3/2}$.)
It has also been claimed that conformal rescaling is necessary 
in scale-invariant Kaluza-Klein cosmology \cite{Gib87b} if one is to 
properly interpret the effective four-dimensional 
Friedmann-Robertson-Walker scale factor.
A recent discussion of conformal ambiguity in compactified 
Kaluza-Klein theory is found in \cite{Mag94}.

There is also something much like a conformal rescaling of 
coordinates in projective Kaluza-Klein theory, notably in 
the work of Schmutzer after 1980 \cite{Sch83,Sch90b,Sch88a,Sch88b}, 
where it is introduced in order to eliminate unwanted second-order 
scalar field terms from the generalized gravitational field equations.
In {\em noncompactified\/} Kaluza-Klein theory, by contrast, 
there has been no discussion of conformal rescaling.  
This is largely because the extra dimensions are regarded 
as {\em physical\/} (if not necessarily lengthlike or timelike).
The five-dimensional metric, in effect, becomes accessible 
(in principle) to observation, and conformally transforming it at will
may no longer be so innocuous.  
We will not consider the issue further in this report;
interested readers are directed to Soko{\l}owski's paper \cite{Sok89}.

\section{Compactified Theories} \label{COMPACT}

So far we have introduced Kaluza's theory, with its cylinder condition,
but have deliberately postponed discussion of compactification 
because we wish to emphasize that it is logically distinct 
from cylindricity,
and in particular that it is only one mechanism by which to
explain the apparently four-dimensional nature of the world.
We now turn to compactified Kaluza-Klein theory, 
but keep our discussion short as this subject has been 
thoroughly reviewed elsewhere (\cite{dSS83,Lee84,KFF84,PW86,ACF87},
\cite{CMS89,Tom84,App84,Duf86,DNP86,BL87,Duf94,Ber76}).

\subsection{Klein's Compactification Mechanism} \label{Klein}

The somewhat contrived nature of Kaluza's assumption, that a fifth 
dimension exists but that no physical quantities depend upon it,
has struck generations of unified field theorists as inadequate.
Klein arrived on the scene during the tremendous excitement surrounding
the birth of quantum theory, and perhaps not surprisingly had the 
idea \cite{Kle26a,Kle26b} of explaining the lack of dependence by 
making the extra dimension very small.  
(The story that this was suggested to him on hearing a colleague 
address him by his last name has, so far as we know, 
no basis in historical fact.)

Klein assumed that the fifth coordinate was to be a lengthlike one
(like the first three), and assigned it two properties:
(1) a {\em circular topology} ($S^1$); and
(2) a {\em small scale}.
Under property (1), any quantity $f(x,y)$ 
(where $x=(x^0,x^1,x^2,x^3)$ and $y=x^4$) becomes {\em periodic\/}; 
$f(x,y) = f(x,y + 2 \pi r)$ where $r$ is 
the scale parameter or ``radius'' of the fifth dimension.
Therefore all the fields can be Fourier-expanded:
\begin{eqnarray}
g_{\alpha\beta}(x,y) & = & \sum_{n=-\infty}^{n=\infty} 
   g_{\alpha\beta}^{(n)}(x) e^{iny/r} \; \; \; , \; \; \;
A_{\alpha}(x,y) = \sum_{n=-\infty}^{n=\infty} A_{\alpha}^{(n)}(x)
   e^{iny/r} \; \; \; , \nonumber \\
\phi(x,y)            & = & \sum_{n=-\infty}^{n=\infty} \phi^{(n)}
   e^{iny/r} \; \; \; ,
\label{Fourier}
\end{eqnarray}
where the superscript $^{(n)}$ refers to the {\em n\/}th Fourier mode.
Thanks to quantum theory, these modes carry a momentum 
in the $y$-direction of the order $|n| / r$.  
This is where property (2) comes in:
if $r$ is small enough, then the $y$-momenta of even the $n=1$ modes 
will be so large as to put them beyond the reach of experiment.  
Hence only the $n=0$ modes, which are independent of $y$, 
will be observable, as required in Kaluza's theory.

How big could the scale size $r$ of a fourth spatial dimension be?
The strongest constraints have come from high-energy particle physics, 
which probes increasingly higher mass scales and correspondingly 
smaller length scales (the Compton wavelength of massive modes 
is of the order $M^{-1}$).
Experiments of this kind \cite{Kos91} presently 
constrain $r$ to be less than an attometer in size 
(1 am = $10^{-18}$~m).
Theorists often set $r$ equal to the Planck length 
$\ell_{pl} \sim 10^{-35}$~m, which is both a natural value and small 
enough to guarantee that the mass of any $n \ne 0$ Fourier modes 
lies beyond the Planck mass $m_{pl} \sim 10^{19}$~GeV.

In general, one identifies Kaluza's five-dimensional 
metric~(\ref{5dMetric}) with the full (Fourier-expanded) 
metric $\hat{g}_{AB}(x,y)$, higher modes and all.  
One then makes what is known in compactified theory as the 
``Kaluza-Klein ansatz,'' which consists in discarding all massive
($n \ne 0$) Fourier modes, as justified above.
In the five-dimensional case, the Kaluza-Klein ansatz amounts to 
simply dropping the $y$-dependency of $g_{\alpha\beta}, A_{\alpha}$, 
and $\phi$, giving the effective four-dimensional ``low-energy'' 
theory of the graviton $g_{\alpha\beta}^{(0)}$, 
photon $A_{\alpha}^{(0)}$ and scalar $\phi^{(0)}$.  
For higher dimensions, though, the relationship between the full 
metric and the metric obtained with the ``Kaluza-Klein ansatz''
is more complicated, as has been emphasized by 
Duff et al. \cite{Duf86,DNP86}.
These authors also stress the difference between these two metrics 
and a third important metric in Kaluza-Klein theory,
the ground state metric $\langle \hat{g}_{AB} \rangle$ which is 
the vacuum expectation value of the full metric $\hat{g}_{AB}(x,y)$, 
and determines the topology of the compact space.
In the five-dimensional case described above, which is topologically 
$M^4 \times S^1$, this looks like:
\begin{equation}
\left( \langle \hat{g}_{AB} \rangle \right) = \left( \begin{array}{cc}
   \eta_{\alpha\beta} \; \; & \; \; 0 \\
                    0 \; \; & \; \; -1
   \end{array} \right) \; \; \; ,
\label{5dGround}
\end{equation}
where $\eta_{\alpha\beta}$ is the four-dimensional 
Minkowski space metric.

\subsection{Quantization of Charge} \label{Quantized}

The expansion of fields into Fourier modes suggests a possible
mechanism to explain {\em charge quantization}, and it is interesting
to see what became of this idea \cite{CMS89}.
One begins by introducing five-dimensional matter into the theory, 
leaving aside for the moment questions as to what this would 
correspond to physically.
The simplest kind of matter is a massless five-dimensional 
scalar field $\hat{\psi}(x,y)$.  
Its action would have a kinetic part only:
\begin{equation}
S_{\hat{\psi}} = -\int d^4x dy \sqrt{-\hat{g}} \;
                  \partial^A \hat{\psi} \, \partial_A \hat{\psi} 
                  \; \; \; .
\label{5dPsiAction}
\end{equation}
The field can be expanded like those in eq.~(\ref{Fourier}):
\begin{equation}
\hat{\psi}(x,y) = \sum_{n=-\infty}^{n=\infty} \hat{\psi}^{(n)}
                  e^{iny/r} \; \; \; .
\end{equation}
When this expansion is put into the action~(\ref{5dPsiAction}),
one finds (using eq.~(\ref{Conf5dMetric})) the following result
\cite{CMS89}, analogous to eq.~(\ref{4dAction}):
\begin{eqnarray}
S_{\hat{\psi}} & = & -\left( \int dy \right) \sum_n
                      \int d^4x \sqrt{-g} \; \left[ \left(
                      \partial^{\alpha} + \frac{i n \kappa 
                      A^{\alpha}}{r} \right) \hat{\psi}^{(n)} \, 
                      \left( \partial_{\alpha} + \frac{i n \kappa 
                      A_{\alpha}}{r} \right) \hat{\psi}^{(n)}
                      \right. \nonumber \\
               &   & \hspace{5cm} -\left. 
                     \frac{n^2}{\phi \, r^2} \, \hat{\psi}^{(n) \, 2} 
                     \right] \; \; \; ,
\label{4dPsiAction}
\end{eqnarray}
From this action one can read off both the charge and mass of
the scalar modes $\hat{\psi}^{(n)}$.
Comparison with the minimal coupling rule 
$\partial_{\alpha} \rightarrow \partial_{\alpha} + i e A_{\alpha}$
of quantum electrodynamics (where $e$ is the electron charge)
shows that in this theory the $nth$ Fourier mode of the scalar 
field $\hat{\psi}$ also carries a {\em quantized charge\/}:
\begin{equation}
q_n = \frac{n \kappa}{r} \left( \phi \int dy \right)^{-1/2} 
    = \frac{n \sqrt{16 \pi G}}{r \sqrt{\phi}} \; \; \; ,
\label{ScalarCharge}
\end{equation}
where we have normalized the definition of $A_{\alpha}$ in 
the action~(\ref{Conf4dAction}) by dividing out the 
factor $(\phi \int dy)^{1/2}$, and made use of
the definitions~(\ref{DefnKappa}) and (\ref{DefnG})
for $\kappa$ and $G$ respectively.
As a corollary to this result one can also come close to 
{\em predicting\/} the value of the fine structure constant,
simply by identifying the charge $q_1$ of the first Fourier
mode with the electron charge $e$.
Taking $r \sqrt{\phi}$ to be on the order of the Planck length 
$\ell_{pl} = \sqrt{G}$, one has:
\begin{equation}
\alpha \equiv \frac{q_1^2}{4 \pi} 
         \sim \frac{(\sqrt{16\pi G}/\sqrt{G})^2}{4\pi}
            = 4 \; \; \; .
\end{equation}
(An improved determination of $r \sqrt{\phi}$ would presumably hit 
closer to the mark.)  The possibility of thus explaining 
an otherwise ``fundamental constant'' would have made compactified 
five-dimensional Kaluza-Klein theory very attractive.

However, the {\em masses\/} of the scalar modes are not at all 
compatible with these ideas.  These are given by the square root 
of the coefficient of the $\hat{\psi}^{(n) \, 2}$-term:
\begin{equation}
m_n = \frac{|n|}{r \sqrt{\phi}} \; \; \; .
\label{ScalarMass}
\end{equation}
If $r \sqrt{\phi} \sim \ell_{pl}$ as we assumed,
then the electron mass $m_1$ (corresponding to the first Fourier mode)
would be $\ell_{pl}^{-1}$; ie., the Planck mass 
$m_{pl} \sim 10^{19}$ GeV, rather than 0.5 MeV.
This discrepancy of some twenty-two orders of magnitude between 
theory and observation played a large role in the abandonment 
of five-dimensional Kaluza-Klein theory.  

In modern compactified theories, one avoids this problem 
by doing three things \cite{CMS89}:
(1) identifying observed (light) particles like the electron with the 
$n=0$, rather than the higher modes of the Fourier expansion above.
From eq.~(\ref{ScalarMass}), these particles therefore have zero mass 
at the level of the field equations.  However, one then invokes:
(2) the mechanism of spontaneous symmetry-breaking to bestow on them
them the modest masses required by observation.  
From eq.~(\ref{ScalarCharge}) above, there is also the problem of 
explaining how the $n=0$ modes can have nonzero {\em charge\/} 
(or, more generally, nonzero couplings to the gauge fields).  
This is solved by:
(3) going to higher dimensions, where massless particles are no longer
``singlets of the gauge group'' corresponding to the ground state
(eg., the $44$-part of the metric~(\ref{5dGround}) above).
We look at this procedure briefly in the next section.

The other way to avoid the problems of compactified five-dimensional 
Kaluza-Klein theory is, of course, to look at projective theories, 
or indeed to loosen the restriction of compactification on the 
fifth dimension altogether. 
These approaches probably mean giving up the ready-made 
explanation for charge quantization described above.

\subsection{Extension to Higher Dimensions} \label{HigherD}

The key to extending the Kaluza-Klein formalism to strong and 
weak nuclear interactions lies in recognizing that electromagnetism 
has been effectively incorporated into general relativity 
by adding $U(1)$ local gauge invariance to the theory,
in the form of local coordinate invariance with respect to $y=x^4$.
Assuming the extra coordinate has a circular topology and 
a small scale, the theory is invariant under transformations:
\begin{equation}
y \rightarrow y^{\prime} = y + f(x) \; \; \; ,
\label{5dTrans}
\end{equation}
where $x$ stands for the four-space coordinates $x^0,x^1,x^2,x^3$.
With the aid of the usual tensor transformation law 
(in {\em five\/} dimensions):
\begin{equation}
\hat{g}_{AB} \rightarrow
\hat{g}^{\prime}_{AB} = \frac{\partial x^C}{\partial x^{\prime \, A}}
                        \frac{\partial x^D}{\partial x^{\prime \, B}} 
                        \, \hat{g}_{CD} \; \; \; ,
\label{TensorTrans}
\end{equation}
one then finds that the only change to the metric~(\ref{5dMetric}) is:
\begin{equation}
A_{\alpha} \rightarrow
A^{\prime}_{\alpha} = A_{\alpha} + \partial_{\alpha} f(x) \; \; \; ,
\label{5dGauge}
\end{equation}
which is just a $U(1)$ local gauge transformation.
In other words the theory is locally $U(1)$ gauge invariant.
It is thus not surprising that electromagnetism
could be contained in five-dimensional general relativity.

To extend the same approach to more complicated symmetry groups,
one goes to higher dimensions \cite{CMS89}.  The metric corresponding to
the ``Kaluza-Klein ansatz'' ($n=0$ modes only) can be written
({\em cf.\/} eq.~(\ref{5dMetric})):
\begin{equation}
\left( \hat{g}_{AB}^{(0)} \right) = \left( \begin{array}{cc}
   g_{\alpha\beta} + \tilde{g}_{\mu\nu} K_i^{\mu} A^i_{\alpha}
      K_j^{\nu} A^j_{\beta} \; \; & \; \; 
         \tilde{g}_{\mu\nu} K_i^{\mu} A^i_{\alpha} \\
      \tilde{g}_{\mu\nu} K_i^{\nu} A^i_{\beta} \; \; & \; \; 
         \tilde{g}_{\mu\nu}
   \end{array} \right) \; \; \; ,
\label{DdMetric}
\end{equation}
where $\tilde{g}_{\mu\nu}$ is the metric of the $d$-dimensional
{\em compact\/} space.
Indices $\mu,\nu,...$ run from 1 to $d$, while $A,B,...$ run 
from 0 to $(3+d)$, and $\alpha,\beta,...$ run from 0 to 3 as usual.
The $K_i^{\mu}$ are a set of $n$ linearly independent 
Killing vectors for the compact manifold ($i=1,...,n$).
Analogously to eq.~(\ref{5dTrans}) one then assumes that
the theory is locally invariant under transformations:
\begin{equation}
y^{\mu} \rightarrow y^{\prime \, \mu} = 
y^{\mu} + \sum_{i=1}^n \varepsilon^i(x) K_i^{\mu} \; \; \; ,
\end{equation}
where the $\varepsilon^i(x)$ are a set of $n$ infinitesimal parameters.
Because Killing vectors by definition satisfy:
\begin{equation}
\frac{\partial K_i^{\lambda}}{\partial y^{\mu}} \, 
   \tilde{g}_{\lambda\nu} + 
   \frac{\partial K_i^{\lambda}}{\partial y^{\nu}} \, 
   \tilde{g}_{\lambda\mu} + K_i^{\lambda} \, \frac{\partial 
   \tilde{g}_{\mu\nu}}{\partial y^{\lambda}} = 0 \; \; \; ,
\end{equation}
the transformation law~(\ref{TensorTrans}) leaves the
$\tilde{g}_{\mu\nu}$-part of the metric untouched,
and the only effect on eq.~(\ref{DdMetric}) is:
\begin{equation}
A^i_{\alpha} \rightarrow A^{i \, \prime}_{\alpha} = 
   A^i_{\alpha} + \partial_{\alpha} \varepsilon^i(x) \; \; \; ,
\label{DdGauge}
\end{equation}
which is a local gauge transformation whose gauge group is 
the {\em isometry group\/} ($G$, say) of the compact manifold.
Thus one might hope that higher-dimensional general relativity 
could contain {\em any\/} gauge theory.

The larger symmetry of the higher-dimensional mechanism also allows 
for nonzero couplings of the $n=0$ modes to the gauge fields; 
ie., for ``charged'' massless particles (which, as we saw, was 
impossible in the five-dimensional case).  
Massless scalar fields $\phi_a(x)$ in the 
adjoint representation of the gauge group,
for example, can be introduced \cite{CMS89} via:
\begin{equation}
\Phi_a^{\mu} = \phi_a(x) K_a^{\mu}(y) \; \; \; ,
\end{equation}
and these in general have nonzero couplings to the gauge fields 
because the $K_a^{\mu}(y)$ are not covariantly constant.

\subsection{Higher-Dimensional Matter} \label{Matter}

It is crucial to realize, however, that the above ``ansatz'' 
metric~(\ref{DdMetric}) does {\em not\/} satisfy Einstein's equations 
in $4+d$ dimensions unless the Killing vectors are independent 
of $\{ y \}$, the extra coordinates \cite{CMS89} ---
ie., unless the compact manifold is {\em flat\/} \cite{ACF87}.
The ground state metric ({\em cf.\/} eq.~(\ref{5dGround})) is:
\begin{equation}
\left( \langle \hat{g}_{AB} \rangle \right) = \left( \begin{array}{cc}
         \eta_{\alpha\beta} \; \; & \; \; 0 \\
                          0 \; \; & \; \; \tilde{g}_{\mu\nu}(y)
         \end{array} \right) \; \; \; .
\label{DdGround}
\end{equation}
The vacuum Einstein equations are 
$\hat{R}_{AB}-\hat{R}\hat{g}_{AB}/2=0$.
Since $\langle \hat{g}_{\alpha\beta} \rangle = \eta_{\alpha\beta}$ 
is flat, $\hat{R}_{\alpha\beta}=0$.
Therefore, from the $\alpha\beta$-components of the field equations,
$\hat{R}$ must vanish.
But then the $\mu\nu$-parts of the same equations imply that
$\hat{R}_{\mu\nu}=0$; ie., that $\tilde{g}_{\mu\nu}$ must also be flat.
In what is perhaps a symptom of the split that has developed since 
Klein between the particle physics and general relativity sides of 
higher-dimensional unification research, early workers tended to 
ignore this ``consistency problem'' \cite{Duf86,DNP86},
and placed no restrictions on the compact manifold while continuing 
to use the metric~(\ref{DdMetric}).
Recently Cho \cite{Cho87a,Cho87b,Cho89,Cho90} has raised related 
questions about whether the ``zero modes'' might not become massive
(and $\{ y \}$-dependent) in the event of spontaneous 
symmetry-breaking, and has even suggested ``kicking away the ladder'' 
of Klein's Fourier modes entirely, basing dimensional reduction
{\em a priori\/} on isometry instead.

It is now widely recognized \cite{BL87} that conventional 
compactification of $d$ extra spatial dimensions
(where $d>1$) requires either (1) explicit higher-dimensional
matter terms, which can induce ``spontaneous compactification'' 
by imposing constant curvature on the compact 
manifold \cite{Cre76,Cre77}; or
(2) other modifications of the higher-dimensional theory, 
such as the inclusion of torsion \cite{Thi72,Cha76,Dom78,Orz81} 
or higher-derivative (eg. $R^2$) terms \cite{Wet82}.
Most higher-dimensional compactified Kaluza-Klein theories rely on
higher-dimensional matter of one kind or another.
For example, in Freund-Rubin compactification \cite{Fre80}, 
which is the basis of eleven-dimensional supergravity, 
one introduces a third-rank antisymmetric tensor field $\hat{A}_{BCD}$ 
with field strength:
\begin{equation}
\hat{F}_{ABCD} \equiv \partial_A \hat{A}_{BCD} -
                      \partial_B \hat{A}_{ACD} +
                      \partial_C \hat{A}_{ABD} -
                      \partial_D \hat{A}_{ABC} \; \; \; ,
\end{equation}
and free action given by:
\begin{equation}
S_{\hat{A}} = -\frac{1}{384\pi G} \int d^{(4+d)}x \sqrt{-\hat{g}} \, 
               \hat{F}_{ABCD} \hat{F}^{ABCD} \; \; \; .
\end{equation}
The effect of this \cite{BL87} is to add an explicit energy-momentum 
tensor to the right-hand side of the higher-dimensional 
Einstein equations~(\ref{5dEFE1}):
\begin{equation}
\hat{T}_{AB} = -\frac{1}{48\pi G} 
                \left( \hat{F}_{CDEA} \hat{F}^{CDE}_{B} - 
                \frac{1}{8} \hat{F}_{CDEF} \hat{F}^{CDEF} \, 
                \hat{g}_{AB} \right) \; \; \; ,
\end{equation}
The matter fields required to achieve compactification are not the 
end of the story, however.  Others are in general needed if the 
theory is to contain the full gauge group of the standard model 
(including strong and weak interactions).
Witten \cite{Wit81} has shown that this requires a theory of at 
least eleven dimensions (including the four macroscopic ones).
While there are an infinite number of compact seven-dimensional 
manifolds whose isometry groups 
$G \supset SU(3) \times SU(2) \times U(1)$, 
none of them give rise to realistic quark and 
lepton representations \cite{CMS89,BL87}.
It is possible to obtain quarks and leptons from other manifolds 
such as the 7-sphere and the ``squashed'' 7-sphere \cite{DNP86}.  
The symmetry groups of these manifolds
($SO(8)$ and $SO(5) \times SU(2)$ respectively) are, however, 
not large enough to contain the standard model, 
and additional ``composite'' matter fields \cite{BL87,Cre79} 
are therefore required.  

Explicit higher-dimensional fields may also be required to incorporate 
{\em chirality\/} into eleven-dimensional compactified theory 
\cite{ACF87,Hor77,Ran83} (this is difficult in an odd number of 
dimensions).  Two other schemes by which this might be accomplished 
are modifications of Riemannian geometry \cite{Wei84b,Wei84c,Wei86} 
and {\em non\/}compact internal 
manifolds \cite{Wet84,Gel84,Wet85,Gel85,McI85,Ran86,Wet86}.  
Thus in the $D=11$ ``chirality problem'' one finds again a choice 
between sacrificing either 
({\em i\/}) the equation ``Matter=Geometry;'' 
({\em ii\/}) the geometrical basis of Einstein's theory; or 
({\em iii\/}) cylindricity.
Compactified theory has in general been characterized by a readiness
to drop ({\em i\/}).

We conclude this section by noting that the situation 
(with regard to non-geometrized matter) does not improve in 
ten-dimensional compactified theory; in fact,
in many cases a six-dimensional internal manifold with {\em no\/} 
isometries is used \cite{CMS89}, which means that {\em all\/} 
the matter is effectively put in by hand, 
marking a complete abandonment of the original Kaluza programme.
Besides ensuring compactification and making room for fermions, 
extra terms in the ten-dimensional Lagrangian also play a role in 
suppressing {\em anomalies\/}.
In the two most popular $D=10$ theories, for example 
(those based on the symmetry groups $SO(32)$ \cite{Gre84} 
and $E_8 \times E_8$ \cite{Gro85}),
this is accomplished by Chapline-Manton terms \cite{CMS89}.
To the extent that these terms arise naturally in the low-energy 
limit of ten-dimensional {\em superstring\/} theory, however, 
they are less arbitrary than some of the others we have mentioned.  
There is no doubt that superstrings currently offer, 
within the context of compactified Kaluza-Klein theory, the best
hope for a unified ``theory of everything'' \cite{SS89,Hor94}.
Whether the compactified approach is the best one remains ---
as we hope to show in the rest of this report --- an open question.

\section{Projective Theories} \label{PROJECTIVE}

Projective theories were designed to emulate the successes of Kaluza's
five-dimensional theory without the epistemological burden of a real 
fifth dimension.  Early models did this too well:  like Kaluza's
(with no dependence on the fifth coordinate and no added 
``higher-dimensional matter'' fields) 
they gave back $\omega=0$ Brans-Dicke theory 
when the electromagnetic potentials were switched off.
This contradicted time-delay measurements like that of
the Martian Viking lander \cite{Wil81}.
There were other problems as well \cite{Sch83}.
Modern projective theories \cite{Les82,Sch83,Sch90b,Sch95} attempt to
overcome these shortcomings in at least two different ways.

\subsection{A Theory of Elementary Particles} \label{Lessner}

Lessner \cite{Les82} has suggested that, although experiments rule 
out a macroscopic Brans-Dicke-type scalar field, the theory might 
still be applicable on {\em microscopic\/} scales, and could be used 
to describe the internal structure of elementary particles.
He begins with the same five-dimensional field equations~(\ref{5dEFE2})
(now interpreted as {\em projector equations}), and obtains the 
same four-dimensional field equations~(\ref{4dFieldEquns}), 
except that the constant $G$ is replaced by ``$B$,'' 
which becomes essentially a free parameter of the theory.  
A solution $(g_{\alpha\beta},F_{\alpha\beta},\phi)$
of the field equations is called a ``particle'' if it satisfies 
certain conditions on symmetry, positivity and asymptotic 
behaviour \cite{Les82}.  Some of the properties of these particles 
are explored in \cite{Les86,Les87,Les90,Les91}.
The theory is only applicable to {\em macroscopic\/} phenomena 
when $\phi=1$ and the third of the field 
equations~(\ref{4dFieldEquns}) is omitted.

\subsection{Projective Unified Field Theory} \label{Schmutzer}

Schmutzer has taken an alternate approach since 1980 in his
``projective unified field theory'' or PUFT \cite{Sch83,Sch90b,Sch95}
by explicitly introducing ``non-geometrizable matter'' 
(the so-called ``substrate'').
In ordinary Kaluza-Klein theory this would correspond to 
higher-dimensional matter and be represented in the five-dimensional 
Einstein equations~(\ref{5dEFE1}) by a nonzero 
energy-momentum tensor $\hat{T}_{AB}$; in the projective theory one 
has instead an {\em energy projector} $\hat{\theta}_{AB}$:
\begin{equation}
\hat{G}_{AB} = 8 \pi \hat{G} \; \hat{\theta}_{AB} \; \; \; .
\label{ProjEFE}
\end{equation}
There is also a conformal rescaling of the four-dimensional metric,
as mentioned earlier:
\begin{equation}
g_{\alpha\beta} \rightarrow g^{\prime}_{\alpha\beta} =
   e^{-\sigma} \, g_{\alpha\beta} \; \; \; ,
\end{equation}
where $\sigma$ is a new scalar field.
Eqs.~(\ref{ProjEFE}) break down, analogously to the five-dimensional 
ones~(\ref{5dEFE2}), to the following set of equations 
in four dimensions:
\begin{eqnarray}
G_{\alpha\beta} & = & 8 \pi G \left( T_{\alpha\beta}^{EM} + 
   \Sigma_{\alpha\beta} + \theta_{\alpha\beta} \right) \; \; \; , 
   \; \; \;
\nabla_{\alpha} H^{\alpha\beta} = J^{\beta} \; \; \; , 
   \nonumber \\
\Box \sigma     & = & 8 \pi G \left( \frac{2}{3} \vartheta +
   \frac{1}{2} B_{\alpha\beta} H^{\alpha\beta} \right) \; \; \; ,
\label{EffEFE}
\end{eqnarray}
where $T_{\alpha\beta}^{EM}$ is the electromagnetic energy-momentum 
tensor as before, and where there are also two new energy-momentum 
tensors:  the {\em substrate energy tensor\/} 
$\theta_{\alpha\beta} = \hat{\theta}_{\alpha\beta}$ 
and the {\em scalaric energy tensor} 
$\Sigma_{\alpha\beta}$ defined by:
\begin{equation}
\Sigma_{\alpha\beta} = -\frac{3}{16\pi G} \left( \partial_{\alpha} 
   \sigma \, \partial_{\beta} \sigma - \frac{1}{2} \, 
   g_{\alpha\beta} \, \partial^{\gamma} \sigma \, 
   \partial_{\gamma} \sigma \right) \; \; \; .
\end{equation}
The other terms in eqs.~(\ref{EffEFE}) are the electric four-current 
density $J^{\alpha}$, 
the electromagnetic field strength tensor $B_{\alpha\beta}$,
the induction tensor $H_{\alpha\beta} = e^{3\sigma} B_{\alpha\beta}$
(the factor $e^{3\sigma}$ acts here as a kind of 
``scalaric dielectricity''), and one more new quantity, 
the {\em scalaric substrate density}:
\begin{equation}
\vartheta = e^{-\sigma} \hat{\theta}^{A}_{A} - 
            \frac{3}{2} \theta^{\alpha}_{\alpha} \; \; \; .
\end{equation}
The conservation of energy $\hat{\nabla}^{A} \hat{\theta}_{AB} = 0$
implies not only conservation of four-current 
($\nabla_{\alpha} J^{\alpha}=0$)
but also conservation of {\em substrate energy}:
\begin{equation}
\nabla_{\beta} \theta^{\alpha\beta} = -B^{\alpha}_{\beta} J^{\beta} +
                                      \vartheta \nabla^{\alpha} 
                                      \sigma \; \; \; .
\end{equation}
The existence of substrate and scalaric matter in PUFT gives rise to
phenomena such as ``scalaric polarization'' of the vacuum, 
and violations of the weak equivalence principle for time-dependent 
scalaric fields.
These can be quantified in terms of a ``scalarism parameter'' $\gamma$,
defined as the ratio of scalaric substrate density to 
the density $\rho$ of ordinary matter:
\begin{equation}
\gamma \equiv \vartheta/\rho \; \; \; .
\end{equation}
This number becomes in practice the primary free parameter of the 
theory, showing up in PUFT-based cosmological 
models \cite{Sch88a,Sch90a},
equivalence principle-type experiments \cite{Sch88b},
and Solar System tests (perihelion shift, 
light deflection and time delay) \cite{Sch90b}.  
Experimental constraints on the theory take the form of
upper limits on the size of $\gamma$.

In comparing these projective theories to the compactified 
Kaluza-Klein theories of the last section, 
one could perhaps summarize as follows:
Kaluza's unified theory as it stands is an elegant 
(no higher-dimensional matter)
and minimal extension of general relativity, but suffers from 
the defect of a very contrived-looking cylinder condition.  
Five-dimensional compactified theory, beginning with Klein, 
repairs this flaw (and even offers the possibility of explaining 
charge quantization) but turns out to disagree radically with 
observation.
To overcome this problem within the context of compactified theory, 
one has to go to higher dimensions and either introduce
higher-dimensional matter or higher-derivative terms to 
the Einstein action,
if one wishes to obtain satisfactory compactification.  
Projective theory offers an alternative way to ``explain'' the 
cylinder condition, and can (unlike compactified theory) 
be formulated in a way that is compatible with experiment using
only one extra ``dimension.''  This comes at the price, however,
not only of modifying the geometrical foundation of Einstein's theory,
but (in Schmutzer's case) of introducing a 
``non-geometrizable substrate,'' or (in Lessner's case) 
of limiting one's ambitions to microscopic phenomena.  
Overall, the projective approach does not appear to us to be an
improvement over compactified theory.

\section{Noncompactified Theories} \label{NONCOMPACT}

An alternative is to stay with the idea that the new coordinates are 
physical, but to {\em generalize\/} the compactified approach by
relaxing the cylinder condition 
\cite{W90,W92a,PdL93,W93a,W95a,Rip95,W96a,Mas96},
instead of restricting the topology and scale of the fifth dimension
in an attempt to satisfy it exactly.
This means that physical quantities, including in particular those 
derived from the metric tensor, will depend on the fifth coordinate.
In fact it is precisely this dependence which allows one to obtain 
not only electromagnetic radiation, but matter of a very 
{\em general\/} kind from geometry via the higher-dimensional 
field equations.  The equations of motion, too, are modified by 
dependence on extra coordinates.
We review these facts in the next few sections.

Of course, the fifth dimension might also be expected to appear 
elsewhere in physics, and one of the primary challenges of 
noncompactified theory is to explain why its effects have not 
been noticed so far.  Why, for example, have experiments
such as those mentioned earlier \cite{Kos91} been able to restrict the
size of any extra dimensions to below the attometer scale?  
In noncompactified theory, the answer is that extra coordinates 
are not necessarily {\em lengthlike}, as these experiments assume.  
Following Minkowski's example, one can imagine coordinates of 
other kinds, scaled by appropriate dimension-transposing parameters 
(like $c$) to give them units of length.
We review this important issue, and the evidence for the hypothesis
that a fifth dimension might be physically related to {\em rest mass}, 
at the end of \S~\ref{NONCOMPACT}.  For the moment, however, we put 
off questions of interpretation and begin by simply seeing how far 
Kaluza's five-dimensional unified field theory can be taken when
the cylinder condition is dropped.

\subsection{The Metric} \label{Metric}

Without cylindricity, there is no reason to compactify the fifth 
dimension, so this approach is properly called ``noncompactified.''
Noncompact extra dimensions have also been considered in
{\em compactified\/} Kaluza-Klein theory by Wetterich and others
\cite{Wet84,Gel84,Wet85,Gel85,McI85,Ran86,Wet86} as a way to bring
chiral fermions into the theory and arrange for a vanishing
four-dimensional cosmological constant.
These authors, however, retain Klein's mechanism of harmonic expansion, 
which in turn means that the compact manifold must have finite volume.
In the fully noncompactified approach we wish to make no 
{\em a priori\/} assumptions about the nature of the 
extra-dimensional manifold.

We begin with the same five-dimensional metric~(\ref{5dMetric}) 
as before, but choose coordinates such that the four components 
of $A_{\alpha}$ vanish.  Since we are no longer imposing cylindricity 
on our solutions, this entails no loss of algebraic generality; 
it is analogous to the common strategy in electromagnetic theory 
of choosing coordinates such that either the electric or magnetic 
field vanishes.  We also eschew any conformal factor here, 
preferring to treat the fifth dimension on the same footing as 
the other four.  The five-dimensional metric tensor, then, is:
\begin{equation}
\left( \hat{g}_{AB} \right) = \left( \begin{array}{cc}
   g_{\alpha\beta} \; \; & \; \; 0 \\
                 0 \; \; & \; \; \varepsilon\phi^2
   \end{array} \right) \; \; \; ,
\label{Final5dMetric}
\end{equation}
where we have introduced the factor $\varepsilon$ in order to allow
a timelike, as well as spacelike signature for the fifth dimension
(we require only that $\varepsilon^2=1$).

Timelike extra dimensions are rarely considered in compactified 
Kaluza-Klein theory, for several reasons \cite{BL87}:  
(1) they lead to the wrong sign for the Maxwell action in 
eq.~(\ref{4dAction}) relative to the Einstein one; and
(2) they lead to the wrong sign for the mass $m_{n}$ of 
the charged modes in eq.~(\ref{4dPsiAction}); 
ie., to the prediction of tachyons.
The relevance of these two arguments to noncompactified theory 
may be debated.  A third common objection (3) is that 
additional temporal \cite{BL87} or timelike \cite{ACF87,Fre80} 
dimensions would lead to closed timelike curves 
and hence allow causality violation.
One should be careful here to discriminate between temporal dimensions, 
which actually have physical units of time; and timelike ones, 
which merely have timelike signature.  
If the physical nature of the fifth coordinate were actually temporal,
one could certainly imagine problems with causality.  
One can, however, transpose units with the proper combination of 
fundamental constants; changing a temporal one, for instance, into 
a spatial one with $c$.  With regard to timelike extra dimensions,
the situation is also less clear than is sometimes claimed.
It has even been argued \cite{Fri91} that physics might be quite 
compatible with closed timelike curves.
All in all, it is probably prudent to keep an open mind
toward the signature of a physical fifth dimension.

\subsection{The Field Equations} \label{Fields}

One now follows the same approach as Kaluza,
using the same definitions~(\ref{5dChristRicci}) of the 
five-dimensional Christoffel symbols and Ricci tensor.
Now, however, one {\em keeps\/} derivatives with respect to the fifth 
coordinate $x^4$ rather than assuming that they vanish.
The resultant expressions for the $\alpha\beta$-, $\alpha 4$- 
and $44$-parts of the five-dimensional 
Ricci tensor $\hat{R}_{AB}$ are \cite{W92a}:
\begin{eqnarray}
\hat{R}_{\alpha\beta} & = & R_{\alpha\beta} - \frac{\nabla_{\beta} 
   ( \partial_{\alpha} \phi )}{\phi} + \frac{\varepsilon}{2\phi^2} 
   \left( \frac{\partial_4 \phi \, \partial_4 g_{\alpha\beta} }{\phi} -
   \partial_4 g_{\alpha\beta} \nonumber \right. \\
                      &   & + \left. g^{\gamma\delta} \, \partial_4 
   g_{\alpha\gamma} \, \partial_4 g_{\beta\delta} - 
   \frac{g^{\gamma\delta} \, \partial_4 g_{\gamma\delta} \, 
   \partial_4 g_{\alpha\beta}}{2} \right) \; \; \; , \nonumber \\
\hat{R}_{\alpha 4}    & = & \frac{g^{44} g^{\beta\gamma}}{4} 
   (\partial_4 g_{\beta\gamma} \, \partial_{\alpha} g_{44} - 
   \partial_{\gamma} g_{44} \, \partial_4 g_{\alpha\beta} ) + 
   \frac{\partial_{\beta} g^{\beta\gamma} \, \partial_4 
   g_{\gamma\alpha}}{2} \nonumber \\
                      &   & + \frac{g^{\beta\gamma} \, \partial_4 
   (\partial_{\beta} g_{\gamma\alpha})}{2} - \frac{\partial_{\alpha} 
   g^{\beta\gamma} \, \partial_4 g_{\beta\gamma}}{2} - 
   \frac{g^{\beta\gamma} \, \partial_4 (\partial_{\alpha} 
   g_{\beta\gamma})}{2} \nonumber \\
                      &   & + \frac{g^{\beta\gamma} g^{\delta\epsilon}
   \, \partial_4 g_{\gamma\alpha} \, \partial_{\beta} 
   g_{\delta\epsilon}}{4} + \frac{\partial_4 g^{\beta\gamma} \, 
   \partial_{\alpha} g_{\beta\gamma}}{4} \; \; \; , \nonumber \\
\hat{R}_{44}          & = & -\varepsilon \phi \Box \phi - 
   \frac{ \partial_4 g^{\alpha\beta} \, \partial_4 g_{\alpha\beta}}{2}
   - \frac{g^{\alpha\beta} \, \partial_4 (\partial_4 
   g_{\alpha\beta})}{2} \nonumber \\
                      &   & + \frac{ \partial_4 \phi \, 
   g^{\alpha\beta} \, \partial_4 g_{\alpha\beta}} {2\phi} - 
   \frac{g^{\alpha\beta} g^{\gamma\delta} \, \partial_4 
   g_{\gamma\beta} \, \partial_4 g_{\alpha\delta}}{4} \; \; \; ,
\label{5dRicci}
\end{eqnarray}
where ``$\Box$'' is defined as usual (in four dimensions) by
$\Box\phi \equiv g^{\alpha\beta} \nabla_{\beta} 
( \partial_{\alpha} \phi)$.

We assume that there is no ``higher-dimensional matter,''
so the Einstein equations take the form~(\ref{5dEFE2}), 
$\hat{R}_{AB}=0$.
The first of eqs.~(\ref{5dRicci}) then produces the following
expression for the {\em four-dimensional\/} Ricci tensor:
\begin{eqnarray}
R_{\alpha\beta} & = & \frac{\nabla_{\beta} ( \partial_{\alpha}
                      \phi )}{\phi} - \frac{\varepsilon}{2\phi^2} 
                      \left( \frac{\partial_4 \phi \, \partial_4 
                      g_{\alpha\beta} }{\phi} - \partial_4 
                      ( \partial_4 g_{\alpha\beta} ) 
                      \right. \nonumber \\
                &   & + \left. g^{\gamma\delta} \, \partial_4
                      g_{\alpha\gamma} \, \partial_4 g_{\beta\delta} -
                      \frac{g^{\gamma\delta} \, \partial_4 
                      g_{\gamma\delta} \, \partial_4 
                      g_{\alpha\beta}}{2} \right) \; \; \; .
\label{4dRicciTensor}
\end{eqnarray}
The second can be written in the form of a {\em conservation law}:
\begin{equation}
\nabla_{\beta} P^{\beta}_{\alpha} = 0 \; \; \; ,
\label{Pconservation}
\end{equation}
where we have defined a new four-tensor by:
\begin{equation}
P_{\alpha}^{\beta} \equiv \frac{1}{2\sqrt{\hat{g}_{44}}} \left( 
                          g^{\beta\gamma} \, \partial_4 
                          g_{\gamma\alpha} - \delta^{\beta}_{\alpha} 
                          \, g^{\gamma\epsilon} \, \partial_4 
                          g_{\gamma\epsilon} \right) \; \; \; .
\label{Pdefn}
\end{equation}
And the third of eqs.~(\ref{5dRicci}) takes the form of a 
{\em scalar wave equation} for $\phi$:
\begin{equation}
\varepsilon \phi \Box \phi = - \frac{ \partial_4 g^{\alpha\beta} \, 
                             \partial_4 g_{\alpha\beta}}{4} - 
                             \frac{g^{\alpha\beta} \, \partial_4 
                             (\partial_4 g_{\alpha\beta})}{2} + 
                             \frac{ \partial_4 \phi \, g^{\alpha\beta}
                             \, \partial_4 g_{\alpha\beta}}{2\phi} 
                             \; \; \; .
\label{ScalarWaveEqun}
\end{equation}
Eqs.~(\ref{4dRicciTensor}) -- (\ref{ScalarWaveEqun}) form the basis 
of five-dimensional noncompactified Kaluza-Klein theory. 
It only remains to interpret their meaning in four dimensions, 
and then to apply them to any given physical problem by choosing 
the appropriate metric $\hat{g}_{AB}$.  
The rest of \S~\ref{NONCOMPACT} is taken up with interpretation; 
applications to cosmology and astrophysics are the subjects
of \S~\ref{COSMOLOGY} and \S~\ref{ASTROPHYSICS}.
We concentrate in this report on the five-dimensional case.
The extension to arbitrary dimensions has yet to be investigated in 
detail, although some aspects of this have recently been discussed by 
Rippl, Romero \& Tavakol \cite{Rip95}.  (These authors also consider
noncompactified {\em lower\/}-dimensional gravity,
which might be more easily quantized than Einstein's theory).

\subsection{Matter from Geometry} \label{Induced}

The best-understood of eqs.~(\ref{4dRicciTensor}) -- 
(\ref{ScalarWaveEqun}) is the first.  It allows us to interpret
{\em four-dimensional matter as a manifestation of five-dimensional
geometry\/} \cite{W92a}.  One simply requires that the usual
Einstein equations (with matter) hold in four dimensions:
\begin{equation}
8 \pi G \, T_{\alpha\beta} = R_{\alpha\beta} - \frac{1}{2} R \, 
                             g_{\alpha\beta} \; \; \; ,
\label{EFEagain}
\end{equation}
where $T_{\alpha\beta}$ is the matter energy-momentum tensor.
Contracting eq.~(\ref{4dRicciTensor}) with the
metric $g^{\alpha\beta}$ gives 
(with the help of eq.~(\ref{ScalarWaveEqun}))
the following expression for the four-dimensional Ricci scalar:
\begin{equation}
R = \frac{\varepsilon}{4\phi^2} \left[ \partial_4 g^{\alpha\beta} \, 
    \partial_4 g_{\alpha\beta} + ( g^{\alpha\beta} \, 
    \partial_4 g_{\alpha\beta} )^2 \right] \; \; \; .
\label{4dRicciScalar}
\end{equation}
Inserting this result, along with eq.~(\ref{4dRicciTensor}), into
eq.~(\ref{EFEagain}), one finds that:
\begin{eqnarray}
8 \pi G \, T_{\alpha\beta} & = & \frac{\nabla_{\beta} ( 
   \partial_{\alpha} \phi )} {\phi} - \frac{\varepsilon}{2 \phi^2} 
   \left[ \frac{\partial_4 \phi \, \partial_4 g_{\alpha\beta} }{\phi}
   - \partial_4 (\partial_4 g_{\alpha\beta} ) + g^{\gamma\delta} \, 
   \partial_4 g_{\alpha\gamma} \, \partial_4 g_{\beta\delta} 
   \right. \nonumber \\
                           &   & - \left. \frac{g^{\gamma\delta} \, 
   \partial_4 g_{\gamma\delta} \, \partial_4 g_{\alpha\beta}}{2} + 
   \frac{g_{\alpha\beta}}{4} \left( \partial_4 g^{\gamma\delta} \, 
   \partial_4 g_{\gamma\delta} + ( g^{\gamma\delta} \, \partial_4 
   g_{\gamma\delta} )^2 \right) \right] \; \; \; .
\label{EMTinduced}
\end{eqnarray}
Provided we use this expression for $T_{\alpha\beta}$, 
the four-dimensional Einstein equations 
$G_{\alpha\beta} = 8 \pi G \, T_{\alpha\beta}$ are
{\em automatically contained in\/} the five-dimensional vacuum ones
$\hat{G}_{AB} = 0$.  The matter described by $T_{\alpha\beta}$
is a manifestation of pure geometry in the higher-dimensional world.
This has been termed the ``induced-matter interpretation'' of 
Kaluza-Klein theory, and eq.~(\ref{EMTinduced}) is said to define 
the energy-momentum tensor of induced matter.

This tensor satisfies the appropriate requirements:
it is symmetric (the first term is a second derivative, 
while the others are all explicitly symmetric), and reduces 
to the expected limit when the cylinder condition is re-applied 
(ie., when all derivatives $\partial_4$ with respect to 
the fifth dimension are dropped).  
In this case, the scalar wave equation~(\ref{ScalarWaveEqun})
becomes just the Klein-Gordon equation for a massless scalar field:
\begin{equation}
\Box \phi = 0 \; \; \; ;
\end{equation}
and the contracted energy momentum tensor of the induced matter 
vanishes:
\begin{equation}
T = g^{\alpha\beta} \, T_{\alpha\beta} = 0 \; \; \; ,
\end{equation}
which implies a radiationlike equation of state ($p=\rho/3$) for 
the induced matter, in agreement with earlier work \cite{Man85} 
based on the cylinder condition.  
The induced matter in this case consists of photons, 
the gauge bosons of electromagnetism --- 
exactly the same result obtained by Kaluza.
This is the only kind of matter one can obtain in the induced-matter
interpretation as long as the cylinder condition is in place.
To extend Kaluza's approach to other kinds of matter, it is necessary
to do one of two things:
(1) go to higher dimensions and add an explicit energy-momentum tensor 
(or other terms) to the higher-dimensional vacuum field equations 
(compactified theories in practice involve both these things); or
(2) loosen the restriction of cylindricity.
In noncompactified theory, which takes the latter course,
it turns out that matter described by $T_{\alpha\beta}$ --- 
even in five dimensions --- is already general enough to describe 
many physical systems, including in particular those connected
with cosmology and the classical tests of general relativity.

The interpretation of eqs.~(\ref{Pconservation}) and 
(\ref{ScalarWaveEqun}) --- the $\alpha 4$- and $44$- components 
of the five-dimensional field equations ($\hat{R}_{AB}=0$) --- 
is not as straightforward as that of eq.~(\ref{4dRicciTensor}).
The relative simplicity of the conservation 
equation~(\ref{Pconservation}) suggests that there is 
a deeper physical significance to the four-tensor
$P_{\alpha}^{\beta}$, whose fully covariant form is
$P_{\alpha\beta} \equiv (\partial_4 g_{\alpha\beta} - 
g_{\alpha\beta} \, g^{\gamma\delta} \, \partial_4 
g_{\gamma\delta} ) / (2 \sqrt{\hat{g}_{44}} \, )$.
It may be related to more familiar conserved physical quantities, 
or to the Bianchi identities \cite{W92a}.

Alternatively, it has been conjectured \cite{W94a} that, as the
$\alpha\beta$-components of the field equations link geometry with
the {\em macroscopic\/} properties of matter, so the $\alpha 4$-
and $44$-components might describe their {\em microscopic\/} ones.
In particular, if one makes the tentative identification:
\begin{equation}
P_{\alpha\beta} = k ( m_i v_{\alpha} v_{\beta} +
                      m_g g_{\alpha\beta} ) \; \; \; ,
\label{NewPdefn}
\end{equation}
where $k$ is a constant, $m_i$ and $m_g$ are the (suitably defined) 
inertial and gravitational mass of a particle in the induced-matter
fluid, and $v^{\alpha} \equiv dx^{\alpha}/ds$ is its four-velocity, 
then the conservation equation~(\ref{Pconservation}) turns out to be 
the {\em four-dimensional geodesic equation\/} 
(for one class of metrics at least).
This is interesting, since equations of motion are 
usually quite distinct from the field equations.  
Similarly, using appropriate definitions of
particle mass $m$, one can identify the scalar wave
equation~(\ref{ScalarWaveEqun}) with the simplest possible
relativistic quantum wave equation, namely the 
{\em Klein-Gordon equation\/}:
\begin{equation}
\Box \phi = m^2 \phi \; \; \; .
\end{equation}
The relevant expression for particle mass turns out to depend 
explicitly on the components of the metric, which means that 
this variant of noncompactified Kaluza-Klein theory is a 
realization of Mach's Principle \cite{Sci53,Bar95}.  
These are interesting results, but speculative ones, 
and we do not discuss them further here.
Some other Machian aspects of noncompactified theories 
have been explored in \cite{W96a,Ma90b,Mas94,Liu95b,W92e}.

\subsection{The Spherically-Symmetric Case} \label{Sph-Sym}

To appreciate what the induced-matter energy-momentum 
tensor~(\ref{EMTinduced}) means physically, one has to supply a 
five-dimensional metric $\hat{g}_{AB}$ --- 
preferably one specific enough to simplify the mathematics
but general enough to be broadly applicable, eg., to both 
cosmological and one-body problems.
We begin here with the {\em general spherically-symmetric\/}
five-dimensional line element:
\begin{equation}
d\hat{s}^2 = e^{\nu} \, dt^2 - e^{\lambda} \, dr^2 - R^2 \left( 
             d\theta^2 + \sin^2 \theta \, d\phi^2 \right) + 
             \varepsilon e^{\mu} \, d\psi^2 \; \; \; ,
\label{LineElement}
\end{equation}
where $\varepsilon$ serves the same function as before,
$t,r,\theta$ and $\phi$ have their usual meanings, $\psi$ is the 
fifth coordinate, and $\nu,\lambda,R$ and $\mu$ are, for now, 
arbitrary functions of $r,t$ and $\psi$.
Denoting derivatives with respect to $t$ by 
overdots $(\dot{\mbox{ }})$, derivatives with respect to $r$ by 
primes $(^{\prime})$, and derivatives with respect to $\psi$ by 
star superscripts $(^{\ast})$, one finds \cite{PdL93} that 
the energy-momentum tensor~(\ref{EMTinduced}) of induced matter 
has the following nonzero components:
\begin{eqnarray}
8 \pi G \, T_0^0 & = & -e^{-\nu} \left( \frac{\dot{\lambda} 
   \dot{\mu}}{4} + \frac{\dot{R} \dot{\mu}}{R} \right) + e^{-\lambda} 
   \left( \frac{R^{\prime} \mu^{\prime}}{R} - \frac{\lambda^{\prime} 
   \mu^{\prime}}{4} + \frac{\mu^{\prime\prime}}{2} + \frac{\mu^{\prime 
   \, 2}}{4} \right) \nonumber \\
                 &   & -\varepsilon e^{-\mu} \left( 
   \frac{\lambda^{\ast\ast}}{2} + \frac{\lambda^{\ast \, 2}}{4} - 
   \frac{\mu^{\ast} \lambda^{\ast}}{4} + \frac{R^{\ast} 
   \lambda^{\ast}}{R} - \frac{R^{\ast} \mu^{\ast}}{R} + \frac{R^{\ast 
   \, 2}}{R^2} + \frac{2 R^{\ast\ast}}{R} \right) \nonumber \\
8 \pi G \, T_0^1 & = & -e^{-\lambda} \left( 
   \frac{\dot{\mu}^{\prime}}{2} + \frac{\dot{\mu} \mu^{\prime}}{4} - 
   \frac{\nu^{\prime} \dot{\mu}}{4} - \frac{\dot{\lambda} 
   \mu^{\prime}}{4} \right) \nonumber \\
8 \pi G \, T_1^1 & = & -e^{-\nu} \left( \frac{\ddot{\mu}}{2} + 
   \frac{\dot{\mu}^2}{4} - \frac{\dot{\nu} \dot{\mu}}{4} + 
   \frac{\dot{R} \dot{\mu}}{R} \right) + e^{-\lambda} \left( 
   \frac{R^{\prime} \mu^{\prime}}{R} + \frac{\nu^{\prime} 
   \mu^{\prime}}{4} \right) \nonumber \\
                 &   & -\varepsilon e^{-\mu} \left( \frac{R^{\ast 
   \, 2}}{R^2} + \frac{2R^{\ast\ast}}{R} + \frac{R^{\ast} 
   \nu^{\ast}}{R} - \frac{R^{\ast} \mu^{\ast}}{R} + 
   \frac{\nu^{\ast\ast}}{2} + \frac{\nu^{\ast \, 2}}{4} - 
   \frac{\nu^{\ast} \mu^{\ast}}{4} \right) \nonumber \\
8 \pi G \, T_2^2 & = & -e^{-\nu} \left( \frac{\dot{R} \dot{\mu}}{2 R} - 
   \frac{\dot{\nu} \dot{\mu}}{4} + \frac{\dot{\lambda} \dot{\mu}}{4} + 
   \frac{\ddot{\mu}}{2} + \frac{\dot{\mu}^2}{4} \right) + e^{-\lambda} 
   \left( \frac{R^{\prime} \mu^{\prime}}{2R} + 
   \frac{\mu^{\prime\prime}}{2} \right. \nonumber \\
                 &   & \left. + \frac{\mu^{\prime \, 2}}{4} - 
   \frac{\lambda^{\prime} \mu^{\prime}}{4} + \frac{\nu^{\prime} 
   \mu^{\prime}}{4} \right) -\varepsilon e^{-\mu} \left( 
   \frac{R^{\ast\ast}}{R} + \frac{R^{\ast} \nu^{\ast}}{2R} + 
   \frac{R^{\ast} \lambda^{\ast}}{2R} - \frac{R^{\ast} \mu^{\ast}}{2R} 
   \right. \nonumber \\
                 &   & \left. + \frac{\nu^{\ast\ast}}{2} + 
   \frac{\nu^{\ast \, 2}}{4} + \frac{\lambda^{\ast\ast}}{2} + 
   \frac{\lambda^{\ast \, 2}}{4} + \frac{\nu^{\ast} 
   \lambda^{\ast}}{4} - \frac{\nu^{\ast} \mu^{\ast}}{4} - 
   \frac{\mu^{\ast} \lambda^{\ast}}{4} \right) \nonumber \\
           T_3^3 & = & T_2^2 \; \; \; .
\label{SphSymEMT}
\end{eqnarray}
If one then assumes that this induced matter takes the form of 
a perfect fluid:
\begin{equation}
T^{\alpha}_{\beta} = (\rho + p) \, u^{\alpha} u_{\beta} - 
                     p \, \delta^{\alpha}_{\beta} \; \; \; ,
\label{PerfectFluid}
\end{equation}
where $u^{\alpha}$ is the four-velocity of the fluid elements, 
then the density $\rho$ and pressure $p$ can be readily 
identified \cite{PdL93} from the relations 
$\rho = T_0^0 + T_1^1 - T_2^2$ and $p = -T_2^2$.
Inserting the expressions~(\ref{SphSymEMT}), one obtains:
\begin{eqnarray}
8\pi G \rho & = & \frac{3}{2} \left( \frac{e^{-\lambda} \mu^{\prime} 
                  R^{\prime}}{R} - \frac{e^{-\nu} \dot{\mu} \dot{R}}{R} 
                  \right) + \varepsilon e^{-\mu} \left( 
                  \frac{\nu^{\ast} \lambda^{\ast}}{4} + \frac{3R^{\ast}
                  \mu^{\ast}}{2R} - \frac{R^{\ast} \lambda^{\ast}}{2R} 
                  \right. \nonumber \\
            &   & \left. -\frac{R^{\ast} \nu^{\ast}}{2R} -
                  \frac{2R^{\ast \, 2}}{R^2} -
                  \frac{3R^{\ast\ast}}{R} \right) \; \; \; , 
                  \nonumber \\
   8\pi G p & = & \frac{1}{2} \left( \frac{e^{-\lambda} \mu^{\prime} 
                  R^{\prime}}{R} - \frac{e^{-\nu} \dot{\mu} \dot{R}}{R} 
                  \right) + \varepsilon e^{-\mu} \left( 
                  \frac{\nu^{\ast} \lambda^{\ast}}{4} + \frac{R^{\ast} 
                  \mu^{\ast}}{2R} + \frac{R^{\ast} \lambda^{\ast}}{2R}
                  \right. \nonumber \\
            &   & \left. +\frac{R^{\ast} \nu^{\ast}}{2R} -
                  \frac{R^{\ast\ast}}{R} \right) \; \; \; .
\label{SphSymRhoP}
\end{eqnarray}
It is immediately apparent that under the restriction of cylindricity
(all starred quantities vanish), one can obtain only {\em radiation\/}
($p = \rho/3$) from Kaluza's mechanism, as noted already.

With the relaxation of this condition, by contrast, 
one obtains a very {\em general\/} equation of state.  
For example, one can split the density and pressure into four
components ($\rho = \rho_r + \rho_d + \rho_v + \rho_s$ 
and $p = p_r + p_d + p_v + p_s$),
where the radiation component obeys $p_r = \rho_r/3$, 
the dust-like component obeys $p_d = 0$, 
the vacuum component obeys $p_v = -\rho_v$, 
and the stiff component obeys $p_s = \rho_s$.  
One then finds from eqs.~(\ref{SphSymRhoP}) that:
\begin{eqnarray}
\rho_r & = & \frac{3}{16\pi G} \left( \frac{e^{-\lambda} \mu^{\prime} 
             R^{\prime}}{R} - \frac{e^{-\nu} \dot{\mu} \dot{R}}{R} 
             \right) + \frac{3\varepsilon e^{-\mu}}{8\pi G} \left( 
             \frac{R^{\ast} \mu^{\ast}}{2R} - \frac{R^{\ast\ast}}{R} 
             \right) \; \; \; , \nonumber \\
\rho_d & = & -\frac{\varepsilon e^{-\mu} R^{\ast \, 2}}{4\pi G R^2} 
             \; \; \; , \nonumber \\
\rho_v & = & -\frac{\varepsilon e^{-\mu} R^{\ast}}{16\pi G}
             \left( \nu^{\ast} + \lambda^{\ast} \right) \; \; \; , 
             \nonumber \\
\rho_s & = & \frac{\varepsilon e^{-\mu} \nu^{\ast} 
             \lambda^{\ast}}{32\pi G} \; \; \; .
\end{eqnarray}
From the first of these equations, it follows that in a 
radiationlike universe whose metric coefficients depend only on time, 
the fifth dimension must contract with time ($\dot{\mu} < 0$) if one 
is to have spatial expansion ($\dot{R} > 0$) and positive density 
($\rho_r > 0$).  Mechanisms of this sort have been used in 
compactified Kaluza-Klein cosmology to pump entropy into the 
four-dimensional universe, solving the horizon and flatness 
problems \cite{Alv83}; or indeed to explain why the fifth dimension 
is compact in the first place \cite{Cho80}
(see \S~\ref{Dynamical}).
In the noncompactified approach, 
they no longer have to be assumed {\em a priori},
but can be seen to be {\em required by the field equations}.
From the second of the above equations, meanwhile, it follows 
that a dustlike universe must have a spacelike fifth dimension 
($\varepsilon = -1$ in our convention) 
in order for its density to be positive ($\rho_d > 0$).
This agrees with the causality argument (\S~\ref{Metric}).

\subsection{The Isotropic and Homogeneous Case} \label{IsoHom}

One can go farther by making additional assumptions about the metric.
Suppose the line element~(\ref{LineElement}) is rewritten in 
spatially isotropic form:
\begin{equation}
d\hat{s}^2 = e^{\nu} dt^2 - e^{\omega} \left( dr^2 + r^2 d\Omega^2 
             \right) + \varepsilon e^{\mu} d\psi^2 \; \; \; ,
\label{IsoLineElement}
\end{equation}
where $d\Omega^2 \equiv d\theta^2 + \sin^2 \theta \, d\phi^2$.
If one assumes that $\nu,\omega$ and $\mu$ are {\em separable\/} 
functions of the variables $t,r$ and $\psi$, one can obtain 
specialized solutions to the field equations~(\ref{5dEFE2}) 
whose properties of matter, as specified by the energy-momentum 
tensor~(\ref{SphSymEMT}), agree very closely with those
expected from four-dimensional theory.

Consider first the case of dependence on $t$ only.
The metric~(\ref{IsoLineElement}) is then just a five-dimensional
generalization of a flat homogeneous and isotropic 
Friedmann-Robertson-Walker (FRW) cosmology.
But in the context of noncompactified Kaluza-Klein theory, 
one ought also to allow dependence on the extra coordinate $\psi$.  
So the {\em general\/} flat five-dimensional cosmological metric, 
assuming separability, should have:
\begin{equation}
e^{\nu} \equiv T^2(t) \; X^2(\psi) \; \; \; , \; \; \;
e^{\omega} \equiv U^2(t) \; Y^2(\psi) \; \; \; , \; \; \;
e^{\mu} \equiv V^2(t) \; Z^2(\psi) \; \; \; .
\label{SepIsoHom}
\end{equation}
Ponce de Leon \cite{PdL88} was the first to investigate solutions 
of the vacuum Einstein equations~(\ref{5dEFE2}) with this form.
Of his eight solutions, one is of special interest because 
it reduces on hypersurfaces $\psi=$ constant to the spatially
flat {\em four-dimensional\/} FRW metric.  This solution has
$\varepsilon=-1$ and:
\begin{eqnarray}
T(t) = \mbox{constant} \; \; \;                & , & \; \; \; 
       X(\psi) \propto \psi \; \; \; , \nonumber \\
U(t) \propto t^{1/\alpha} \; \; \;             & , & \; \; \; 
       Y(\psi) \propto \psi^{1/(1-\alpha)} \; \; \; , \nonumber \\
V(t) \propto t \; \; \; & , & \; \; \; 
       Z(\psi) = \mbox{constant} \; \; \; ,
\label{TUVXYZ}
\end{eqnarray}
and can be written in the form:
\begin{equation}
d\hat{s}^2 = \psi^2 dt^2 - t^{2/\alpha} \psi^{2/(1-\alpha)}
             \left( dx^2 + dy^2 + dz^2 \right) -
             \frac{\alpha^2}{(1-\alpha)^{2}} \, t^2 d\psi^2 \; \; \; ,
\label{CosmoMetric}
\end{equation}
where $dx^2 + dy^2 + dz^2 = dr^2 + r^2 d\Omega^2$ are the usual 
rectangular coordinates, and $\alpha$ is a free parameter of the 
theory \cite{W92c}.
Because this solution reduces on spacetime sections 
($d\psi=0$) to the familiar $k=0$ FRW metric:
\begin{equation}
ds^2 = dt^2 - R^2(t) \left( dx^2 + dy^2 + dz^2 \right) \; \; \; ,
\end{equation}
it can properly be called the {\em generalization of the flat FRW 
cosmological metric to five dimensions}.

Assuming that cosmological matter behaves like a perfect
fluid, one obtains from eqs.~(\ref{SphSymRhoP}) the following 
expressions for density and pressure \cite{W92b}:
\begin{equation}
\rho = \frac{3}{8\pi G \alpha^2 \psi^2 \, t^2} \; \; \; , \; \; \;
   p = \left( \frac{2\alpha}{3} - 1 \right) \, \rho \; \; \; .
\label{CosmoRhoP}
\end{equation}
These are consistent with a wide variety of equations of state:
a radiation-dominated universe, for example, if $\alpha=2$;
a dust-filled one if $\alpha = 3/2$; or an inflationary one
if $0 < \alpha < 1$.  Physical properties of cosmologies based on
the metric~(\ref{CosmoMetric}) have been explored in
\cite{W92b,W92c,W94b,W95c,Bil96a}, and its implications for the 
equations of motion (eg., of galaxies) are known \cite{W95b}.  
Generalizations to $k\ne0$ cosmologies \cite{Liu94,McM94} and 
extended (eg., Gauss-Bonnet) theories of gravity \cite{Col94,Col95}
have been made, and a connection to Mach's principle
\cite{W96a,W94a,Mas94,Liu95b} has been identified.
These and related issues are reviewed in \S~\ref{COSMOLOGY}.

One other of Ponce de Leon's homogeneous and isotropic 
solutions \cite{PdL88} deserves mention.  It has:
\begin{eqnarray}
T(t) = \mbox{constant} \; \; \;                                 & , &
       \; \; \; X(\psi) \propto \psi \; \; \; , \nonumber \\
U(t) \propto \exp \left( \sqrt{\Lambda/3} \; t \right) \; \; \; & , &
       \; \; \; Y(\psi) \propto \psi \; \; \; , \nonumber \\
V(t) = \mbox{constant} \; \; \;                                 & , &
       \; \; \; Z(\psi) = \mbox{constant} \; \; \; ,
\end{eqnarray}
and looks like:
\begin{equation}
d\hat{s}^2 = \psi^2 dt^2 - \psi^2 e^{2\sqrt{\Lambda/3} \; t} 
             \left( dx^2 + dy^2 + dz^2 \right) - d\psi^2 \; \; \; .
\end{equation}
This reduces on spacetime hypersurfaces ($\psi=$~constant)
to the {\em de Sitter\/} metric, and $\Lambda \equiv 3/\psi^2$ 
is a {\em cosmological constant\/} induced in four-dimensional 
spacetime by the existence of the fifth coordinate $\psi$.
The equation of state of the ``matter'' induced in four dimensions 
is that of the classical de~Sitter vacuum, $p=-\rho$, 
with $\rho=\Lambda/(8 \pi G)$.

Billyard \& Wesson \cite{Bil96a} have considered generalizations 
of this solution:
\begin{equation}
d\hat{s}^2 = \psi^2 dt^2 - \psi^2 e^{i\omega t} \left( e^{ik_1 x} 
             dx^2 + e^{ik_2 y} dy^2 + e^{ik_3 z} dz^2 \right) + 
             \ell^2 d\psi^2 \; \; \; ,
\end{equation}
where $\omega$ is a frequency, $k_i$ a wave vector, 
and $\ell$ measures the size of the extra dimension.  
The induced-matter equation of state is again $p=-\rho$, 
but now with $\rho = -3\omega^2/(32\pi G \psi^2)$.
The field equations~(\ref{5dEFE2}) turn out to require 
$\ell^2=4/\omega^2$ so the vacuum has positive energy density 
if the fifth dimension is spacelike.  The metric coefficients 
of ordinary three-space exhibit wave-like behaviour, 
but the associated medium is an unperturbed de~Sitter vacuum ---
so this solution describes what might be termed ``vacuum waves'' in
Kaluza-Klein theory.  
(They are not gravitational waves of the conventional
sort because three-space is spherically-symmetric.)  
One might apply this to the inflationary universe scenario; 
imagining, for example, that $\omega$ starts out with real values 
(corresponding to a vacuum-dominated universe with oscillating 
three-space coefficients) but later takes on {\em imaginary\/} values 
(for which the universe enters an expanding de~Sitter phase 
of the usual kind).
On this interpretation, the big bang occurs as a 
(presumably quantum-induced) phase change --- 
as has previously been suggested elsewhere 
on other grounds \cite{Vil82,W85b}.

\subsection{The Static Case} \label{Static}

The metric~(\ref{IsoLineElement}) reduces to another well-known form 
when the coefficients $\nu,\omega$ and $\mu$ depend only on the 
radial coordinate $r$.
This is just a five-dimensional generalization of the one-body or 
Schwarzschild metric, and has been variously
interpreted in the literature as describing a 
magnetic monopole \cite{Sor83},
``black hole'' \cite{Dav85b}, and
{\em soliton\/} \cite{Gro83} (see \S~\ref{Solitons} for discussion).
Again, however, from a noncompactified point of view there is 
no {\em a priori\/} reason to suppress dependence 
on $\psi$, so a {\em general\/} static spherically-symmetric
metric, assuming separability, should have:
\begin{equation}
e^{\nu} \equiv A^2(r) \; D^2(\psi) \; \; \; , \; \; \;
e^{\omega} \equiv B^2(r) \; E^2(\psi) \; \; \; , \; \; \;
e^{\mu} \equiv C^2(r) \; F^2(\psi) \; \; \; .
\label{SepStatic}
\end{equation}
Ponce de Leon \& Wesson \cite{PdL93} searched for
two-parameter solutions of the five-dimensional field 
equations~(\ref{5dEFE2}) with this form and found only four.
(Liu \& Wesson \cite{Liu96b,Liu96c} have recently obtained a 
three-parameter generalization of this class).
The most useful is the one which contains the 
ordinary {\em four-dimensional\/} Schwarzschild solution 
as a limiting case.  This is the solution with $D,E$ and $F$ constant 
($=1$ without loss of generality), and is thus identical to the 
soliton metric just mentioned.  The coefficients $A,B$ and $C$ are:
\begin{eqnarray}
A(r) & = & \left( \frac{ar-1}{ar+1} \right)^{\epsilon k} \; \; \; , 
           \; \; \;
B(r) = \frac{1}{a^2 r^2} \frac{(ar+1)^{\epsilon (k-1) +1}}
           {(ar-1)^{\epsilon (k-1) -1}} \; \; \; , \nonumber \\
C(r) & = & \left( \frac{ar+1}{ar-1} \right)^{\epsilon} \; \; \; ,
\label{ABC}
\end{eqnarray}
where $a$ is a constant related to the mass of the central body,
and $\epsilon$ and $k$ are other parameters (in the notation of
\cite{Dav85b}).  Only one of these is strictly a free parameter, 
as they are related by a consistency relation:
\begin{equation}
\epsilon^2 (k^2 - k + 1) = 1 \; \; \; .
\label{Consistency}
\end{equation}
Written out explicitly, the metric is:
\begin{equation}
d\hat{s}^2 = A^2(r) dt^2 - B^2(r) \left( dr^2 + r^2 d\Omega^2 \right) - 
             C^2(r) d\psi^2 \; \; \; ,
\label{DOsoliton}
\end{equation}
where we have assumed a spacelike fifth coordinate ($\varepsilon=-1$)
in agreement with other work.
In the limits $\epsilon \rightarrow 0, k \rightarrow \infty$, and
$\epsilon k \rightarrow 1$ (where $a \equiv 2/GM_{\ast}$ and $M_{\ast}$ 
is the mass of the central body), this metric reduces on spacetime 
sections $d\psi=0$ to the familiar Schwarzschild metric 
(in isotropic coordinates):
\begin{equation}
ds^2 = \left( \frac{1-GM_{\ast}/2r}{1+GM_{\ast}/2r} \right)^2 dt^2 -
       \left( 1 + \frac{GM_{\ast}}{2r} \right)^4 \left( dr^2 + r^2
           d\Omega^2 \right) \; \; \; .
\label{SchwarzMetric}
\end{equation}
It is therefore properly called the {\em generalization of the 
Schwarzschild metric to five dimensions}.  
Elsewhere in \S~\ref{NONCOMPACT} we will refer to the above values 
of $\epsilon$ and $k$ as the ``Schwarzschild limit'' of the theory.

Assuming as usual that the induced matter takes the form of a perfect 
fluid, eqs.~(\ref{SphSymRhoP}) give for both solutions the following
density and pressure \cite{W92d}:
\begin{equation}
\rho = \frac{\epsilon^2 k a^6 r^4}{2\pi G (ar-1)^4 (ar+1)^4} \,
       \left( \frac{ar-1}{ar+1} \right)^{2\epsilon (k-1)} \; \; \; , 
       \; \; \;
   p = \frac{\rho}{3} \; \; \; .
\label{SolitonRhoP}
\end{equation}
The soliton metric~(\ref{DOsoliton}) thus describes a central mass 
surrounded by an inhomogeneous cloud of radiation-like matter 
whose density goes as~$\sim \! \! 1/a^2 r^4$ at large values of $r$.
(The Schwarzschild limit defined above is the special case where
the density and pressure of the cloud are zero; in this case
$p = 0 = -\rho$ which is the usual vacuum solution, with its
attendant classical tests of general relativity.)
The $\epsilon^2 k$-term indicates that this combination of 
the two (related) parameters $\epsilon$ and $k$ may characterize 
the soliton's energy density \cite{Liu92}.  (This is
somewhat different from the traditional interpretation, 
in which these parameters are related to its 
``scalar charge'' \cite{Sok86,Cho82}.)
The equation of state~(\ref{SolitonRhoP}) obtained in the
induced-matter interpretation differs from the one found
by Davidson \& Owen \cite{Dav85b}, who used an approach based on 
Kac-Moody symmetries \cite{Dol84} and concluded that $p=-\rho/3$.
Density shows the same $r$-dependence at large distances in both 
approaches, however, and goes to zero in the same Schwarzschild limit.
Both solutions are invariant under 
$a \rightarrow -a,\epsilon \rightarrow -\epsilon$, 
and require $k>0$ for positive density.  
One can define a {\em pressure three-tensor\/} $p^a_b$ 
in the induced-matter interpretation, 
using the $ab$-components of the vacuum field 
equations~(\ref{5dEFE2}), and this yields a result \cite{W92d} 
very similar to the series expression obtained by Davidson \& Owen.
If one then takes $p \equiv p^a_a/3$ (as in \cite{Dav85b}),
one gets back exactly the result in eq.~(\ref{SolitonRhoP}).

In Cartesian spatial coordinates the pressure tensor in general
contains off-diagonal components, which implies that the matter 
making up the soliton is a sum of both a material (perfect) fluid 
and a free electromagnetic field \cite{W92d}.  
The terms ``density'' and ``pressure'' therefore have to be treated 
with caution.  Solitonic matter, in fact, is best be described as 
a {\em relativistic fluid with anisotropic pressure\/} \cite{Liu92}.  
Anisotropic spherically-symmetric fluids have an energy-momentum 
tensor given by:
\begin{equation}
T_{\alpha\beta} = (\rho + p_{\perp}) \, u_{\alpha} u_{\beta} +
                  (p_{\parallel} - p_{\perp}) \chi_{\alpha} 
                  \chi_{\beta} + p_{\perp} \, g_{\alpha\beta} 
                  \; \; \; ,
\label{AnisoFluid}
\end{equation}
where $\chi_{\alpha}$ is a unit spacelike vector orthogonal 
to $u_{\alpha}$, and $p_{\parallel}, p_{\perp}$ refer to pressure 
parallel and perpendicular to the radial direction.  Assuming that the
induced matter~(\ref{SphSymEMT}) takes this form rather than that of
the perfect fluid~(\ref{PerfectFluid}); and choosing
$u^{\alpha}=(u^0,0,0,0),\chi^{\alpha}=(0,\chi^1,0,0)$, one finds:
\begin{eqnarray}
p_{\parallel} & = & \left[ 1 - \left( \frac{a^2 r^2 - 2 \epsilon  
                    (k-1) ar + 1}{\epsilon a r} \right) \right] \, 
                    \rho \; \; \; , \nonumber \\
    p_{\perp} & = & \left( \frac{a^2 r^2 - 2 \epsilon (k-1) ar + 1}{2 
                    \epsilon a r} \right) \, \rho \; \; \; ,
\end{eqnarray}
with $\rho$ exactly as in eq.~(\ref{SolitonRhoP}).
These expressions satisfy $\rho=p_{\parallel}+2p_{\perp}$, 
confirming that the fluid has the nature of radiation.  
The physical properties of solitons 
based on the metric~(\ref{DOsoliton}) have been studied by several 
authors \cite{W92d,Liu92,W94c,Liu96a,Sor83,Gro83,Dav85b,Cho82}.
Their implications for astrophysics \cite{W92e,W94d},
the classical tests of general relativity \cite{Lim92,Kal95,Lim95},
and the equivalence principle \cite{W96b} have been explored,
and the class has been extended to time-dependent \cite{W93b,Liu93}
and charged solutions \cite{Liu96b,Liu96c}.
These and related issues are reviewed in \S~\ref{ASTROPHYSICS}.

We mention for completeness the other three static solutions
of the form~(\ref{SepStatic})
obtained by Ponce de Leon \& Wesson \cite{PdL93}.
Two of them have $A(r),B(r)$ and $C(r)$ exactly as in eqs.~(\ref{ABC}),
but have $F(\psi)$ an arbitrary function of $\psi$, with
$D^{\ast}(\psi) \propto F(\psi)$ and $E(\psi)=$ constant in
the first case, and $D(\psi) \propto E^{-1}(\psi)$ 
and $E^{\ast}(\psi) \propto F(\psi)/E(\psi)$ in the second one.
The density and pressure for both these solutions is exactly 
as in eqs.~(\ref{SolitonRhoP}) above, except for an added factor
of $E^2(\psi)$ in the denominators.  This is physically innocuous in
the first case ($E(\psi)=$ constant) but means in the second one that
these attributes of the radiation cloud depend on the extra coordinate.
The $\psi$-dependent components of the field equations place an extra
constraint on these both these solutions, restricting the allowed 
values of the parameters $\epsilon$ and $k$.
The final solution is more interesting, 
and can be written in the form:
\begin{equation}
d\hat{s}^2 = \frac{\psi^2}{(\psi^2 + \ell)} \, dt^2 -
             \frac{(\psi^2 + \ell)}{(ar^2 + b)}^2 \left( dr^2 + 
             r^2 d\Omega^2 \right) + \varepsilon d\psi^2\; \; \; ,
\label{FourthSoln}
\end{equation}
where $a$ is related to the mass of the central object as before,
$b=-\varepsilon/4a$, and $\ell$ is one other independent constant of
the system.  Although this solution was found by assuming separability
in $r$ and $\psi$, it also satisfies the field equations~(\ref{5dEFE2})
when $a$ and $b$ are {\em arbitrary\/} functions of $\psi$.  
This is intriguing, as it hints at a relationship between the mass 
of the central object and the fifth coordinate.  
Another interesting feature of the metric~(\ref{FourthSoln}) 
is its induced-matter equation of state, which ---
unlike that of the other soliton solutions found so far --- 
is not radiationlike, but turns out to be the one discussed by 
Davidson \& Owen \cite{Dav85b}: $p = -\rho/3$.
This is an unusual form of matter, but has been considered 
previously in several other contexts \cite{Got87,Kol89,W91,PdL93b}, 
largely because it describes matter that does not disturb other 
objects gravitationally (gravitational or Tolman-Whittaker mass is 
proportional to $3p+\rho$).  Thus it might, for instance, be useful in
reconciling the extremely high energy densities expected for quantum
zero-point fields with the small value observed for Einstein's 
cosmological constant \cite{W91}.

\subsection{General Covariance in Higher Dimensions} \label{Covariance}

We have reviewed a number of solutions to the
spherically-symmetric vacuum field equations in five dimensions.  
In each case the five-dimensional geometry manifests itself in four
dimensions as induced matter, with an associated equation of state.
The equation of state, in fact, {\em follows from\/} the field 
equations in the induced-matter interpretation of Kaluza-Klein theory,
rather than having to be supplied separately as in four-dimensional
general relativity.  In several cases, the physical form of the metric
on spacetime hypersurfaces $d\psi=$ constant, or the equation
of state for the induced matter, is such as to make the solutions
useful for testing the predictions of noncompactified theory.
The theory is so far not in conflict with any experimental data
(see \S~\ref{COSMOLOGY} and \S~\ref{ASTROPHYSICS}).

However, it is important to keep in mind that physical quantities 
such as the scalars $\rho$ and $p$, which are designed to be 
invariant with respect to {\em four-dimensional\/} coordinate changes 
$x^{\alpha} \rightarrow \tilde{x}^{\alpha}$ cannot in general 
stay that way in noncompactified Kaluza-Klein theory, 
which is invariant with respect to {\em five-dimensional\/} ones
$x^A \rightarrow \tilde{x}^A$.
Any quantities --- even those normally thought of as conserved ---
are vulnerable if they depend on the fifth coordinate $x^4$.

What this means in practice is that density, pressure, 
and the equation of state in the induced-matter interpretation are 
to some extent dependent on the coordinates in which one chooses to 
express them.  A search for the correct solution to a 
(four-dimensional) physical problem is also a search for the 
appropriate system of (five-dimensional) {\em coordinates}.
This can perhaps best be illustrated with a series of simple
$x^4$-dependent coordinate transformations \cite{W95a},
beginning with five-dimensional Minkowski space:
\begin{equation}
d\hat{s}^2 = dt^2 - dr^2 - r^2 \, d\Omega^2 - d\psi^2 \; \; \; .
\label{5dMinkowski}
\end{equation}
Spacetime sections of this metric are of course four-dimensional
Minkowski spaces.  If one transforms to primed coordinates:
\begin{equation}
t^{\prime} = t \; \; \; , \; \; \; 
r^{\prime} = \frac{r}{\psi} \left( 1 + \frac{r^2}{\psi^2} 
             \right)^{-1/2} \; \; \; , \; \; \;
\psi^{\prime} = \psi \left( 1 + \frac{r^2}{\psi^2} \right)^{1/2} 
                \; \; \; ,
\end{equation}
this metric becomes:
\begin{equation}
d\hat{s}^{\prime \, 2} = dt^{\prime \, 2} - \psi^{\prime \, 2} \left( 
                         \frac{dr^{\prime \, 2}} {1-r^{\prime \, 2}} + 
                         r^{\prime \, 2} d\Omega^2 \right) - 
                         d\psi^{\prime \, 2} \; \; \; .
\label{PrimedMetric}
\end{equation}
Spacetime sections ($\psi^{\prime}=$ constant) of the new primed 
metric are {\em static Einstein cosmologies\/}; ie., 
four-dimensional FRW metrics:
\begin{equation}
ds^{\prime \, 2} = dt^{\prime \, 2} - R^2(t^{\prime}) \left( 
                   \frac{dr^{\prime \, 2}} {1 - kr^{\prime \, 2}} + 
                   r^{\prime \, 2} d\Omega^2 \right) \; \; \; ,
\end{equation}
with $k=+1$ and a {\em constant\/} scale factor 
$R(t^{\prime})=\psi^{\prime}$.
One can obtain from the Friedmann equations the value of Einstein's 
cosmological constant $\Lambda$, and (assuming a perfect fluid) 
expressions for the density and pressure of matter:
\begin{equation}
\Lambda = \frac{1}{\psi^{\prime \, 2}} \; \; \; , \; \; \;
\rho_m = \frac{1}{4\pi G \psi^{\prime \, 2}} \; \; \; , \; \; \;
p_m = 0 \; \; \; .
\end{equation}
The cosmological constant represents a vacuum energy density
$\rho_v = \Lambda/(8\pi G)$ with associated pressure $p_v = -\rho_v$.
So altogether one has:
\begin{equation}
\rho = \rho_m + \rho_v = \frac{3}{8\pi G \psi^{\prime \, 2}} \; \; \; , 
       \; \; \;
p = p_m + p_v = -\frac{\rho}{3} \; \; \; .
\end{equation}
The effective equation of state in the four-dimensional spacetime 
sections of the primed metric~(\ref{PrimedMetric}) is thus that 
of {\em non-gravitating matter} of the kind discussed 
in \S~\ref{Static}.
(The same result could have been obtained by plugging the metric 
directly into eqs.~(\ref{SphSymRhoP}) for induced-matter
density and pressure.)
A second coordinate transformation to double-primed coordinates:
\begin{equation}
t^{\prime} = \psi^{\prime\prime} \sinh t^{\prime\prime} \; \; \; , 
             \; \; \;
r^{\prime} = r^{\prime\prime} \; \; \; , \; \; \;
\psi^{\prime} = \psi^{\prime\prime} \cosh t^{\prime\prime} \; \; \; ,
\end{equation}
puts the metric into the new form:
\begin{equation}
d\hat{s}^{\prime\prime \, 2} = \psi^{\prime\prime \, 2} 
   dt^{\prime\prime \, 2} - \psi^{\prime\prime \, 2} \cosh^2 
   t^{\prime\prime \, 2} \left( \frac{dr^{\prime\prime \, 
   2}}{1-r^{\prime\prime \, 2}} + r^{\prime\prime \, 2} d\Omega^2 
   \right) - d\psi^{\prime\prime \, 2} \; \; \; .
\label{DoublePrimedMetric}
\end{equation}
Spacetime sections of this double-primed metric 
are {\em expanding FRW cosmologies\/}, with $k=+1$ 
and $R(t^{\prime\prime})=\psi^{\prime\prime} \cosh t^{\prime\prime}$.
The density and pressure of the accompanying perfect fluid, 
as obtained from eqs.~(\ref{SphSymRhoP}), are:
\begin{equation}
\rho = \frac{3}{8\pi G \psi^{\prime\prime \, 2}} \; \; \; , \; \; \;
p = -\rho \; \; \; ,
\end{equation}
so that the effective equation of state is that of a {\em pure vacuum}.  

Of the metrics (\ref{5dMinkowski}), (\ref{PrimedMetric}) and
(\ref{DoublePrimedMetric}), which one is the best choice for a
description of the real universe?  None, of course, since none of them
admits spacetime sections with realistic four-dimensional properties.
A metric which {\em is\/} adequate to this task is the cosmological
one~(\ref{CosmoMetric}).  It can be obtained from
Minkowski space~(\ref{5dMinkowski}) by transforming from
$t,r,\psi$ to $\tilde{t},\tilde{r},\tilde{\psi}$ via:
\begin{eqnarray}
t    & = & \frac{\alpha}{2} \left( 1 + \frac{\tilde{r}^2}{\alpha^2} 
           \right) \tilde{t}^{1/\alpha} \tilde{\psi}^{1/(1-\alpha)} - 
           \frac{\alpha}{2(1-2\alpha)} \left[ \tilde{t}^{-1} 
           \tilde{\psi}^{\alpha/(1-\alpha)} 
           \right]^{(1-2\alpha)/\alpha} \; \; \; , \nonumber \\
r    & = & \tilde{r} \tilde{t}^{1/\alpha} \tilde{\psi}^{1/(1-\alpha)} 
           \; \; \; , \nonumber \\
\psi & = & \frac{\alpha}{2} \left( 1 - \frac{\tilde{r}^2}{\alpha^2} 
           \right) \tilde{t}^{1/\alpha} \tilde{\psi}^{1/(1-\alpha)} + 
           \frac{\alpha}{2(1-2\alpha)} \left[ \tilde{t}^{-1} 
           \tilde{\psi}^{\alpha/(1-\alpha)} 
           \right]^{(1-2\alpha)/\alpha} \; \; \; ,
\end{eqnarray}
and dropping the tildes \cite{W96a}.  The cosmological metric,
as we have seen (\S~\ref{IsoHom}), gives back good models for 
the early (radiation) and late (dust) universe on spacetime 
sections $\psi=$ constant if $\alpha$ is chosen appropriately.

The point of this exercise is that
{\em all four of the metrics~(\ref{CosmoMetric}),(\ref{5dMinkowski}),
(\ref{PrimedMetric}) and (\ref{DoublePrimedMetric})
are flat in five dimensions\/}, although they would be perceived very 
differently by four-dimensional observers 
(as evinced by their expansion factors and equations of state).
The reason for these differences is the $\psi$-dependence of the 
coordinate transformations, and the fact that the theory is covariant 
with respect to {\em five-}, not four-dimensional coordinates.  
To properly describe a given four-dimensional problem in 
noncompactified theory, one needs to choose five-dimensional 
coordinates judiciously.
This is not a reflection of some fundamental ambiguity in the theory, 
but is rather forced on us as long as we insist on retaining 
four-dimensional concepts like density and pressure in a 
five-dimensional theory (see also \S~\ref{MoreCovariance}).

\subsection{Other Exact Solutions} \label{Exact}

Similar remarks apply to astrophysical situations.
One has to choose five-dimensional coordinates appropriate to 
each problem, if one wants to couch the results in terms of 
familiar four-dimensional quantities.  
There is thus a rich field here for future inquiry.
The one-body metric which has received most attention so far is that
of the soliton~(\ref{DOsoliton}), which contains the four-dimensional
Schwarzschild solution on spacetime sections.
As we saw in \S~\ref{Static}, however,
the induced matter associated with this metric
is necessarily radiationlike (except in the Schwarzschild limit), 
and its density falls off with distance rather steeply.
To describe bodies with {\em different\/} properties, one must 
find new static spherically-symmetric solutions of the field equations.
This is possible in Kaluza-Klein theory because Birkhoff's theorem 
(which guarantees the uniqueness of the Schwarzschild solution 
in four dimensions)
no longer holds in higher dimensions \cite{W93a,W92e,Cho91,Gro83}.

One such solution has recently been found by 
Billyard \& Wesson \cite{Bil96b}.  It is actually a modification of 
the cosmological metric~(\ref{CosmoMetric}):
\begin{eqnarray}
d\hat{s}^2 & = & \left( \frac{r}{r_0} \right)^{2(\alpha+1)}
                 \psi^{2(\alpha +3)/\alpha} dt^2 -
                 (3-\alpha^2) \psi^2 dr^2 -
                 \psi^2 r^2 d\Omega^2 \nonumber \\
           &   & + 3(3\alpha^{-2} -1) r^2 d\psi^2 \; \; \; ,
\label{BillMetric}
\end{eqnarray}
where $r_0$ is a constant and $\alpha$ is a parameter related to the
properties of matter.
On spacetime hypersurfaces $d\psi=0$ this metric is very similar to
a four-dimensional one originally used to describe inhomogeneous
spheres of matter in static isothermal equilibrium \cite{Hen78}.
With the aid of eqs.~(\ref{SphSymRhoP}), one finds that the associated
induced matter has:
\begin{equation}
\rho = \frac{(2-\alpha^2)}{8\pi G (3-\alpha^2) \psi^2 r^2} \; \; \; , \; \; \;
   p = \left( \frac{\alpha^2 + 2\alpha + 2/3}
                   {2-\alpha^2} \right) \, \rho \; \; \; .
\label{BillEqState}
\end{equation}
In addition, one can use the standard (Tolman-Whittaker) definition 
\cite{Lan75} of the gravitational mass of a volume of fluid to obtain:
\begin{equation}
M_g(r) = \frac{(1+\alpha)}{G \sqrt{3-\alpha^2}} \; \psi^{2+3/\alpha}
         \left( \frac{r}{r_0} \right)^{2+\alpha} r_0 \; \; \; .
\end{equation}
The object described by this metric has positive density 
for $\alpha^2 \le 2$, and positive mass (assuming $\alpha \ne 0$) 
for $\alpha \ge -1$.
So altogether one has a nonzero $\alpha$ between $-1$ and $\sqrt{2}$, 
which allows for equations of state~(\ref{BillEqState}) anywhere 
in the range $-\rho/3 \le p \le \rho$.
These are potentially relevant to a wide variety of 
astrophysical problems.
But the fact that $\rho$ and $p$ are both proportional to $r^{-2}$, 
rather than $\sim \! \! r^{-4}$ as for solitons, 
indicates that eq.~(\ref{BillMetric}) may be especially useful 
for modelling phenomena such as galaxies, 
or clusters of them \cite{Bah94,Cro94,Gir95}.
To go further one needs to rederive the classical tests of 
general relativity for this metric, just as has been done for 
the soliton one (see \S~\ref{ASTROPHYSICS}).
Some work has been done in this direction in \cite{Bil96b}.

\subsection{The Equations of Motion} \label{EqsMotion}

Like the higher-dimensional field equations, the higher-dimensional 
equations of motion are also modified when dependence is allowed on
extra coordinates.  In this section, in order to explicitly include
electromagnetic effects, we no longer restrict our choice of 
coordinates to those in which $A_{\alpha}=0$.
The metric $\hat{g}_{AB}$ is given by eq.~(\ref{5dMetric}), with the
addition of the $\varepsilon$-factor to allow for timelike, as well
as spacelike $x^4$.
We then obtain the equations of motion by minimizing the 
five-dimensional interval $d\hat{s}^2 = \hat{g}_{AB} \, dx^A dx^B$.
This results in a five-dimensional version of the 
geodesic equation \cite{W95b}:
\begin{equation}
\frac{d^2 x^A}{d\hat{s}^2} + \hat{\Gamma}^A_{BC} 
\frac{dx^B}{d\hat{s}} \frac{dx^C}{d\hat{s}} = 0 \; \; \; ,
\label{5dGeodesic}
\end{equation}
with the five-dimensional Christoffel symbol defined as in 
eq.~(\ref{5dChristRicci}).  The $A=4$ component of 
eq.~(\ref{5dGeodesic}) can be shown \cite{W95b} to take the form:
\begin{equation}
\frac{d{\cal B}}{d\hat{s}} = \frac{1}{2} \frac{\partial 
                             \hat{g}_{CD}}{\partial x^4} 
                             \frac{dx^C}{d\hat{s}} 
                             \frac{dx^D}{d\hat{s}} \; \; \; ,
\label{4geodesic}
\end{equation}
where ${\cal B}$ is a scalar function:
\begin{equation}
{\cal B} \equiv \varepsilon \phi^2 \left( \frac{dx^4}{d\hat{s}} +
         \kappa A_{\alpha} \frac{dx^{\alpha}}{d\hat{s}} \right) 
         \; \; \; .
\label{Bdefn}
\end{equation}
In the case where $\hat{g}_{AB}$ does not depend on $x^4$, 
${\cal B}$ is a constant of the motion (since $d{\cal B}/d\hat{s}=0$), 
but this is not generally so in noncompactified theory.
The definition of ${\cal B}$, 
together with the form of the metric~(\ref{5dMetric}), 
allow us to express the five-dimensional interval in terms of the
four-dimensional one via
$d\hat{s} = (1-\varepsilon {\cal B}^2/\phi^2)^{-1/2} ds$.
Using this relation, the $A=\alpha$ components of 
eq.~(\ref{5dGeodesic}) can be shown \cite{W95b} to take the form:
\begin{eqnarray}
\frac{d^2 x^{\mu}}{ds^2} + \Gamma^{\mu}_{\alpha\beta}
      \frac{dx^{\alpha}}{ds} \frac{dx^{\beta}}{ds} 
& = & \frac{{\cal B}}{\sqrt{1-\varepsilon {\cal B}^2/\phi^2}}
      \left[ F^{\mu}_{\nu} \frac{dx^{\nu}}{ds} - 
      \frac{\kappa A^{\mu}}{{\cal B}} \frac{d{\cal B}}{ds} -
      \kappa g^{\mu \lambda} \frac{\partial A_{\lambda}}{\partial x^4}
      \frac{dx^4}{ds} \right] \nonumber \\
&   & + \frac{\varepsilon {\cal B}^2}{(1-\varepsilon {\cal B}^2 / 
      \phi^2) \phi^3} \left[ \nabla^{\mu} \phi + \left( 
      \frac{\phi}{{\cal B}} \frac{d{\cal B}}{ds} - \frac{d\phi}{ds} 
      \right) \frac{dx^{\mu}}{ds} \right] \nonumber \\
&   & - g^{\mu\lambda} \frac{\partial g_{\lambda\nu}}{\partial x^4}
      \frac{dx^{\nu}}{ds} \frac{dx^4}{ds} \; \; \; .
\label{4dGeodesic}
\end{eqnarray}
This is the fully general equation of motion in Kaluza-Klein theory,
and for $\mu=1,2,3$ shows how a test particle moves in ordinary space.

The left-hand side of eq.~(\ref{4dGeodesic}) is identical to that
in Einstein's theory; the terms on the right-hand side are 
deviations from four-dimensional geodesic motion.
In the case of no dependence on the extra coordinate $x^4$,
the four terms in $d{\cal B}/ds$ and $\partial/\partial x^4$ all vanish
and we correctly recover the same result obtained previously by those
working in compactified Kaluza-Klein theory \cite{Lei73,Geg84,Fer89}.
The terms in the first set of square brackets depend on a nonvanishing
electromagnetic potential $A_{\alpha}$, and
the first of these can be recognized as the Lorentz force 
if the charge-to-mass ratio of the test particle is:
\begin{equation}
\frac{Q}{M} = \frac{{\cal B}}{\sqrt{1-\varepsilon {\cal B}^2 / 
              \phi^2}} \; \; \; .
\end{equation}
This relation, however, is only useful in the limit where the
metric is independent of $x^4$, and its extra-dimensional part is flat
\cite{W95b}.  In coordinate frames where this is not the case, one
cannot readily identify quantities like mass or charge, which after
all are four-dimensional concepts.  The same caution applies to the
``scalar charge-to-mass ratio'' given by:
\begin{equation}
\frac{Q^{\prime}}{M} = \frac{\varepsilon {\cal B}^2}{(1-\varepsilon 
                       {\cal B}^2 / \phi^2)\phi^3} \; \; \; ,
\end{equation}
which can be identified analogously to the electromagnetic one
from the multiplicative factor in front of the {\em second\/}
set of square brackets in eq.~(\ref{4dGeodesic}).

The $0$-component of the geodesic equation~(\ref{4dGeodesic}), 
meanwhile, can be written \cite{W95b} in a form analogous to 
eq.~(\ref{4geodesic}) above:
\begin{equation}
\frac{d{\cal C}}{ds} = \frac{1}{2} \frac{\partial 
                       \hat{g}_{CD}}{\partial x^0} \frac{dx^C}{ds} 
                       \frac{dx^D}{d\hat{s}} \; \; \; ,
\label{0geodesic}
\end{equation}
where ${\cal C}$ is a new scalar function:
\begin{equation}
{\cal C} \equiv \sqrt{ \frac{g_{00}}{1-v^2} \left( 1 - 
                \frac{\varepsilon {\cal B}^2}{\phi^2} \right) } +
                \kappa \, {\cal B} A_0 \; \; \; .
\label{Cdefn}
\end{equation}
Here $v^2=\lambda_{ab} v^a v^b$ is the square of the test particle's
spatial 3-velocity
$v^a=dx^a / (\sqrt{g_{00}} \, [dx^0+(g_{0b}/g_{00})dx^b])$,
with $\lambda_{\alpha\beta} \equiv g_{\alpha 0} g_{\beta 0}/g_{00}
-g_{\alpha\beta}$ a suitable projector.
If the metric $\hat{g}_{AB}$ were independent of time $x^0$
then ${\cal C}$ would be a constant of motion.  
Where this is not the case, as in cosmology,
the geodesic equation~(\ref{4dGeodesic}) could in principle 
be applied to test noncompactified theory.
We return to this question in \S~\ref{MoreEqsMotion}.
The possible physical significance of the quantity ${\cal C}$ is
explored in more detail in \cite{W95b}.

\subsection{Physical Meaning of the Fifth Coordinate} \label{STM}

We have noted that the charge of a test particle 
can be readily identified in the limit as
$\psi \equiv x^4=$ constant.  We have also found that
a variety of realistic four-dimensional cosmological 
models and one-body metrics can be identified as constant-$\psi$ 
hypersurfaces of flat five-dimensional Minkowski space.
So far, then, it appears that useful coordinate 
systems can be specified by the condition $u^4 \equiv d\psi/ds = 0$.  
(This is perfectly legitimate from a mathematical point of view
as the introduction of a fifth coordinate into general relativity 
means an extra degree of freedom that can always be used if one 
wishes to set a condition on $u^4$.)  
However, we have not much improved on Kaluza's cylinder condition 
unless we confront the question:  are there any {\em physical\/} 
reasons why we should expect $d\psi/ds=0$?

In answering this one is obliged to interpret $\psi$ physically.
We review here one such interpretation, which has been advanced by
Wesson and his collaborators \cite{W95a,W96a,W83,W92c}.
Noncompactified theory in general (and elsewhere in this report, 
including the next two parts on experimental constraints)
stands or falls quite independently of this additional work.
The proposal we consider is that the fifth coordinate
$\psi$ might be related to {\em rest mass}.  
The coordinate frame picked out by $u^4=0$ is then just the one 
in which {\em particle rest masses are constant}.
There are at least three independent pieces of evidence 
(besides the empirical fact that rest masses are conserved!) 
in support of this conjecture:
(1) All of mechanics depends on base units of length, time and mass.
So if the former two can be treated as coordinates, then maybe the
last should also.  Dimensionally, $x^4=Gm/c^2$ allows us to treat
the rest mass $m$ of a particle as a length coordinate, in analogy
with $x^0=ct$.
(2) Metrics which do not depend on $x^4$, like the soliton
metric~(\ref{DOsoliton}), can give rise only to induced matter
composed of {\em photons\/}; while those which depend on $x^4$,
like the cosmological metric~(\ref{CosmoMetric}), give back
equations of state for fluids composed of {\em massive particles\/}.
(3) The metrics $d\hat{s}^2=dT^2-d\sigma^2-d\Psi^2$ and
$d\hat{s}^2=\psi^2 \, dt^2 - d\sigma^2 - t^2 \, d\psi^2$ are related
by the coordinate transformations
$T=t^2\psi^2/4+\ln [(t/\psi)^{1/2}]$ and
$\Psi=t^2\psi^2/4-\ln [(t/\psi)^{1/2}]$.
The former metric is flat, while the latter gives an action principle
$\delta\int\psi dt=0$ for particles at rest in ordinary space 
($d\sigma/ds=0$), viewed on hypersurfaces $\psi=$ constant.
This action principle is formally the same as that of particle physics
if $\psi \rightarrow m$ in the local, low-velocity limit.
(The same argument applies to 
cosmological metrics~(\ref{CosmoMetric}).)
This view of the origin of mass is similar to that in some 
quantum field theories \cite{Bro78}, where rest masses are generated 
spontaneously in a conformally invariant theory that includes a scalar 
dilaton field or Nambu-Goldstone boson in Minkowski space.

Several other, more philosophical reasons \cite{W95a,W96a} to consider 
the STM (``Space-Time-Matter'') hypothesis
that $\psi$ might be related to $m$ can perhaps be mentioned here:
(4) A theory in which mass is placed on the same footing 
as space and time will be naturally {\em scale-invariant\/}, simply by 
virtue of being {\em coordinate-invariant\/} (because particle masses 
are a necessary part of any system of units, or ``scales'').  
The idea that nature might be scale-invariant has been considered from 
time to time by such eminent thinkers as Dirac, Hoyle and others
\cite{W95a,Dic62,Dir73,Dir74,Hoy74,Can77a,Can77b,Wes78,Bek79,W92f}.
(The STM approach is, however, otherwise quite distinct from these 
theories, not least in the fact that it predicts a variation in 
rest mass $m$ rather than the dimension-transposing constant $G$.)
(5) There is also a pleasing symmetry in the elevation of $G$ to 
the same status as $c$: as the latter puts {\em distances\/} 
into temporal units, so the former is needed to do the same for 
{\em masses\/}.  The actual conversion factors are $1/c$ 
and $G/c^3$ respectively, and this helps explain why any change in 
mass with time --- a generic feature of scale-invariant theories ---
has been so small as to have escaped detection so far:  
the latter factor is some 43 orders of magnitude smaller 
than the former, and the former is already tiny enough to have 
made special relativistic effects unnoticeable until
the second half of this century.
(6) Finally, we note that if $x^4$ is not restricted to be lengthlike 
(or timelike) in nature, then the extra part of the metric can have 
either sign without running afoul of closed timelike curves and 
causality problems (\S~\ref{Metric}).  
We will not consider the STM theory further in this report, 
noting however that its observational implications have been studied 
over the years by Wesson \cite{W90,W84,W85,W86,W88} and numerous others 
\cite{Cha86a,Cha87,Cha90a,Cha90b,Ban90a,PdL88,Gro88a,Gro88b,Bao89},
\cite{Ma90a,Ma90b,Ma91,Col90,Chi90,Roq91,Car91,Car92,Wag92,Mac93},
\cite{Fuk87,Fuk88a,Fuk88b,Fuk92a,Fuk93}.

\section{Cosmology} \label{COSMOLOGY}

\subsection{Compactified Kaluza-Klein Cosmology} \label{Dynamical}

Cosmological aspects of compactified Kaluza-Klein theory have received
less attention than those related to particle physics \cite{DNP86}.
Where they have been addressed \cite{ACF87,BL87}, 
much of the discussion has focused on the search for exact solutions 
of higher-dimensional general relativity 
(or extended gravity theories) which contain the 
familiar FRW universes on spacetime-like sections.
This was first done in five dimensions 
(with no extra-dimensional matter)
by Chodos \& Detweiler \cite{Cho80} in 1980, and extended to ten- and
eleven-dimensional supergravities (with appropriate higher-dimensional 
matter tensors) by Freund \cite{Fre82}.  
The key feature of these and subsequent models
\cite{Dav85a,Ish86,Gle85,Gle86,Wil87,Mae84,Ran84},
\cite{Tom86,Szy88,Dem90,Mak91,Hen86,Ros87,Cha89}
was that extra dimensions could 
(and in some cases necessarily {\em would\/}) 
shrink as the spatial ones expanded, thus lending support 
to the whole notion of compactification.
(The possibility that compact subspaces could ``bounce back'' 
from a contracting phase was also investigated \cite{Yos84,Sat84}.)
This approach to explaining why the universe appears four-dimensional 
is sometimes referred to as ``dynamical'' or ``cosmological'' 
dimensional reduction.  {\em Non\/}-compactified theory can exhibit 
the same behaviour, as noted in \S~\ref{Sph-Sym} 
and \S~\ref{Precession}.

In more than five dimensions, compactification requires either 
explicit matter terms or modifications to the Einstein equations 
(\S~\ref{Matter}).
All kinds of matter have been invoked to induce cosmological 
compactification (usually in addition to that already required 
for spontaneous compactification; 
eg., as in supergravity \cite{Fre82,Gle85,Gle86,Wil87}).
There are theories with dilaton fields \cite{Wil87},
quantized five-dimensional scalar fields \cite{Mae84},
a $D$-dimensional gas of non-interacting scalar particles \cite{Ran84},
general higher-dimensional perfect 
fluids \cite{Gle85,Tom86,Szy88,Dem90,Mak91},
$D$-dimensional radiation \cite{Hen86},
five-dimensional dust \cite{Ros87}, and
scalar fields in nonlinear sigma models \cite{Cha89}.
Cosmological compactification mechanisms
based on modifications of Einstein's theory of gravity
are just as colorful, employing quadratic \cite{Shi86,Der89},
cubic \cite{Met87},
and even quartic terms \cite{Dem92} in the curvature, both generally
and in special combinations known variously as
Gauss-Bonnet terms \cite{Ish86,Hen86,Pau90,Dem91},
Lanczos terms \cite{Mad86},
Lovelock terms \cite{Dem92,Der90},
Euler-Poincar\'e densities \cite{Ari87}, and
dimensionally continued Euler forms \cite{Mul86,Ari89}.
Even changes of metric signature \cite{Ker93} have been considered
as instruments of compactification.
An exhaustive survey and classification of generalized 
higher-dimensional vacuum cosmologies has recently been carried out 
by Coley \cite{Col94}.

An important step in Kaluza-Klein cosmology was the demonstration 
that shrinking extra dimensions could transfer entropy into 
the four-dimensional universe, providing a new way to solve the 
horizon and flatness problems \cite{Alv83,Sah84},
although many ($\sim$40) extra dimensions were required \cite{Abb84}.
Inflation itself has also been incorporated directly in compactified 
Kaluza-Klein theories \cite{Der83,Sha83}, 
and indeed ``Kaluza-Klein inflation'' has burgeoned into
a sub-field of its own \cite{Oli90}.  
It is difficult to obtain in some supergravity \cite{Gle85} 
and most superstring \cite{Lin90} theories,
and again requires in general that either additional matter 
terms \cite{Der83} or higher-derivative corrections \cite{Sha83} 
be added to Einstein's theory.  Examples of the former include 
higher-dimensional dust \cite{Ish84}, scalar fields 
with conformal transformations of the metric \cite{Cho90,Yoo90}
or non-minimal couplings to the curvature \cite{Sun90},
generalized perfect fluids \cite{Szy90,Bel94},
and others \cite{Bur95}.
Examples of the latter include higher-derivative corrections 
to Einstein's equations \cite{Sha85,Sha87},
Gauss-Bonnet terms \cite{Pau90,Ker93},
and Euler forms \cite{Ari89}.
Inflation has also been obtained with
multiple compact subspaces \cite{Eza91} and
explicit ``chaotic inflaton'' fields \cite{Car96}.
Other inflationary Kaluza-Klein cosmologies include versions of 
extended inflation \cite{Cha93a} and STM theory \cite{Gro88b}.
An exciting recent development is the use of {\em COBE\/}
measurements of microwave background anisotropy
to put experimental limits --- 
surprisingly restrictive ones in some cases ---
on inflationary Kaluza-Klein models \cite{Fab93,Far95,Ben96,Kub95}.

Cosmological constraints on compactified Kaluza-Klein 
theories {\em apart\/} from those relating to inflation 
in the early universe have also received attention, 
beginning with Marciano's observation \cite{Mar84}
that time-variation in the scale of extra dimensions would have
important consequences for the fundamental constants of 
four-dimensional physics.  Implications of the same phenomenon for
primordial nucleosynthesis \cite{Kol86} and
nuclear resonance levels in carbon and oxygen atoms \cite{Bar87}
have also been discussed.
If the extra dimensions are spatial in nature, these arguments 
imply that the present rate of change in their mean radius is 
less than about $10^{-19}$ yr$^{-1}$.  
Another interesting idea is to use observations of 
{\em gravitational waves\/} to constrain Kaluza-Klein cosmologies;
this however turns out to be impractical 
at the present time \cite{Bie94}.
Other issues in compactified Kaluza-Klein cosmology include the
possibility of excessive contributions to the global energy density
from massive Fourier modes \cite{BL87,Kol84} 
and solitons \cite{BL87,Har84} (see also \S~\ref{DarkMatter}),
gravitational effects due to massless scalar components of the
compactified higher-dimensional metric \cite{Kos89},
and the stability of solutions with respect to 
classical perturbations \cite{Gon87},
chaotic behaviour \cite{Szy87},
and quantum effects \cite{Mae87,Yos84,Oso93}.
Inhomogeneous Kaluza-Klein cosmologies have been considered 
in \cite{Cha94}.

\subsection{The Equation of State} \label{EqState}

Compactified Kaluza-Klein cosmology, as described above, 
is characterized by a profusion of competing expressions 
for the energy-momentum tensor $\hat{T}_{AB}$ in higher dimensions, 
reflecting the fact that there is no consensus on how to define 
``higher-dimensional matter.''
In noncompactified cosmology, by contrast, one avoids this ambiguity
with the natural and economical assumption that 
$\hat{T}_{AB} \equiv 0$; that is, that 
the universe in higher dimensions is {\em empty\/}.

This cannot be done in compactified theory because 
the cylinder condition imposes uncomfortable restrictions 
on the resulting equation of state 
(and other properties of matter) in four dimensions.
Consider as a simple example the uniform five-dimensional line
element~(\ref{IsoLineElement}) with $\nu=0,\omega=\ln t$, 
and $\mu=-\ln t$:
\begin{equation}
d\hat{s}^2 = dt^2 - t d\sigma^2 - t^{-1} d\psi^2 \; \; \; ,
\label{RadMetric}
\end{equation}
where $d\sigma^2=dx^2+dy^2+dz^2$ is shorthand for the spatial part
of the metric.  This would be an acceptable solution in compactified
cosmology in that its constant-$\psi$ sections are FRW,
none of the metric coefficients depend on $\psi$, and the extra
coordinate shrinks with time.  Indeed its spatial part grows 
in exactly the same way as that of a four-dimensional FRW model 
with the radiation equation of state, $p=\rho/3$.
And in fact, in the induced-matter interpretation, this metric
literally {\em does\/} describe radiation.  
That is, putting eq.~(\ref{RadMetric}) into the five-dimensional 
vacuum field equations $\hat{G}_{AB}=0$ gives back the 
four-dimensional ones $G_{\alpha\beta}=8\pi T_{\alpha\beta}$, 
where $T_{\alpha\beta}$ is the energy-momentum tensor of 
a perfect fluid with:
\begin{equation}
\rho = \frac{3}{32\pi t^2} \; \; \; , \; \; \; 
   p = \frac{\rho}{3} \; \; \; ,
\label{RadRhoP}
\end{equation}
as can be shown explicitly using eqs.~(\ref{SphSymRhoP}).
(Units are such that $G=c=1$ throughout \S~\ref{COSMOLOGY},
except where otherwise noted.)
These are the same expressions as those used to describe the 
radiation era in (flat) four-dimensional cosmology \cite{W92c}.
In fact, unless five-dimensional matter is put in to begin with,
this metric is incapable of manifesting itself as anything {\em but\/} 
electromagnetic radiation in four dimensions (\S~\ref{Induced}).

In noncompactified cosmology, by contrast, one can describe 
the universe at {\em any\/} stage of its history without 
higher-dimensional matter
(or modifications to the higher-dimensional field equations).
As discussed in \S~\ref{IsoHom}, the best metric for this purpose is 
the ``cosmological metric''~(\ref{CosmoMetric}).
Consider first the case $\alpha=2$, which looks like:
\begin{equation}
d\hat{s}^2 = \psi^2 dt^2 - t \psi^{-2} d\sigma^2 - 4t^2 d\psi^2 
             \; \; \; .
\label{RelMetric}
\end{equation}
This line element again has FRW-like constant-$\psi$ sections, 
and gives exactly the same expressions for induced density 
and pressure as eqs.~(\ref{RadRhoP}) above, provided that the
unphysical coordinate label $t$ is replaced by the
proper time $\psi t$.
So it again describes a radiation-dominated universe, 
or one filled with relativistic particles such as neutrinos.  
This time, however, the metric coefficients depend on $\psi$, 
and the fifth dimension {\em grows\/} with time.  
Solutions of this type tend to be discarded in particle physics, 
where the assumed lengthlike nature of the extra coordinates 
constrains them to be very small at the present time \cite{Kos91}.
Here we make no {\em a priori\/} assumptions about the physical 
nature of extra dimensions.  
This allows us to obtain more general kinds of 
cosmological matter \cite{W92b,W92c}.  In the case $\alpha=3/2$, 
for instance, the same metric~(\ref{CosmoMetric}) reads:
\begin{equation}
d\hat{s}^2 = \psi^2 dt^2 - t^{4/3} \psi^{-4} d\sigma^2 - 9t^2 d\psi^2
             \; \; \; ,
\label{DustMetric}
\end{equation}
which, from eqs.~(\ref{CosmoRhoP}), represents matter with 
induced density and pressure:
\begin{equation}
\rho = \frac{1}{6\pi (\psi t)^2} \; \; \; , \; \; \; p=0 \; \; \; ,
\end{equation}
and therefore describes a {\em dust-filled\/} universe.
One can also model {\em inflation\/} (in flat FRW models)
by choosing $0< \alpha <1$.
Provided one is willing to tolerate a dependence on the extra 
coordinate, then, and a non-lengthlike interpretation of 
its physical nature, one can describe the universe 
at any stage of its history as a manifestation of pure geometry
in five dimensions.  In every case, the parameters $\rho$ and $p$
appear as products of an underlying geometric theory, 
and the equation of state manifests itself as a {\em consequence\/} 
of the field equations.
This is more satisfying than the usual situation in (four-dimensional) 
cosmology, where pressure and density have merely phenomenological
status and the equation of state must be put into the theory by hand.

\subsection{Extension to $k \ne 0$ Cosmologies} \label{Curved}

The cosmological metric~(\ref{CosmoMetric}), 
and the others mentioned in \S~\ref{IsoHom}, 
are all five-dimensional generalizations of {\em spatially flat\/} 
four-dimensional FRW spacetimes.  
One could also consider a {\em curved\/} version of 
the homogeneous and isotropic line element~(\ref{IsoLineElement}):
\begin{equation}
d\hat{s}^2 = e^{\nu} \, dt^2 - e^{\omega} \, 
             \left( \frac{dr^2}{1-kr^2} + r^2 d\Omega^2 \right) -
             e^{\mu} \, d\psi^2 \; \; \; .
\label{CurvedMetric}
\end{equation}
McManus \cite{McM94} has investigated solutions of the vacuum Einstein
equations~(\ref{5dEFE2}) with this form.  
Like Ponce de Leon \cite{PdL88}, he assumed that $\nu,\omega$ 
and $\mu$ were {\em separable\/} functions,
as given by eq.~(\ref{SepIsoHom}), with $T(t)=Z(\psi)=$ constant.
He found four solutions with $k \ne 0$,
each associated with a well-defined induced-matter equation of state.
We list his final results here.  In the first solution, $X(\psi)$ 
and $Y(\psi)$ are constant as well as $Z(\psi)$, 
and the line element reads:
\begin{equation}
d\hat{s}^2 = dt^2 - (-kt^2 + \xi t + \eta) d\chi^2 - \frac{(kt - 
             \xi /2)^2}{-kt^2 + \xi t + \eta} \, d\psi^2 \; \; \; ,
\end{equation}
where $d\chi^2 = [ (1-kr^2)^{-1} dr^2 + r^2 d\Omega^2 ]$ is new 
shorthand for the spatial part of the metric and $\xi$ 
and $\eta$ are arbitrary constants.
Since none of the metric coefficients depend on $\psi$, 
the equation of state is that of radiation:
\begin{equation}
\rho = \frac{3(\xi^2 + 4k \eta)}
            {32\pi (-kt^2 + \xi t + \eta)^2} \; \; \; , \; \; \;
   p = \frac{\rho}{3} \; \; \; .
\end{equation}
This solution was originally discussed by Davidson {\em et al.\/}
\cite{Dav85a}.  The second solution reads:
\begin{equation}
d\hat{s}^2 = \frac{(k\psi + \xi/2)^2}{k\psi^2 + \xi\psi + \eta} \, dt^2 -
             (k\psi^2 + \xi\psi + \eta) d\chi^2 - d\psi^2 \; \; \; ,
\end{equation}
and has:
\begin{equation}
\rho = \frac{3k}{8\pi (k\psi^2 + \xi\psi + \eta)} \; \; \; , \; \; \;
   p = -\frac{\rho}{3} \; \; \; .
\end{equation}
This is the equation of state of ``nongravitating matter'' discussed 
in \S~\ref{Static}.  The same equation of state characterizes 
the third solution:
\begin{equation}
d\hat{s}^2 = dt^2 - \frac{1}{4} t^2 (e^{\psi} - k e^{-\psi})^2 
             d\chi^2 - t^2 d\psi^2 \; \; \; ,
\end{equation}
which has:
\begin{equation}
\rho = \frac{3}{8\pi t^2} \, \frac{1}{(\tanh \psi)^{2k}} \; \; \; .
\end{equation}
Finally, McManus' fourth solution is given by:
\begin{equation}
d\hat{s}^2 = \psi^2 dt^2 - \frac{1}{4} \psi^2 (e^t + ke^{-t})^2 
             d\chi^2 - d\psi^2 \; \; \; ,
\end{equation}
with:
\begin{equation}
\rho = \frac{3}{8\pi \psi^2} \; \; \; , \; \; \; p = -\rho \; \; \; .
\end{equation}
This is the equation of state of a vacuum.

Liu \& Wesson \cite{Liu94} have extended the search 
to {\em non\/}-separable $k \ne 0$ solutions~(\ref{CurvedMetric}).  
Instead of eq.~(\ref{SepIsoHom}),
they assume metric coefficients of the form:
\begin{equation}
e^{\nu} \equiv L^2(t-\lambda\psi) \; \; \; , \; \; \;
e^{\omega} \equiv M^2(t-\lambda\psi) \; \; \; , \; \; \;
e^{\mu} \equiv N^2(t-\lambda\psi) \; \; \; ,
\end{equation}
where $L,M$ and $N$ are {\em wavelike\/} functions of the
argument $(t-\lambda\psi)$, with $\lambda$ acting as a ``wave number.''
Their solutions turn out to be determined by two relations:
\begin{eqnarray}
\dot{L}^2 + kM^2 & = & \zeta^2 M^2 \left( \lambda^2 M^2 + 
                       \frac{\kappa}{L^2} \right) \; \; \; , 
                       \nonumber \\
             N^2 & = & \lambda^2 M^2 - \frac{k}{\zeta^2} +
                       \frac{\kappa}{L^2} \; \; \; ,
\end{eqnarray}
where $L$ (or $M$) must be supplied, and $\zeta$ and $\kappa$ are
integration constants.  With $L,M$ and $N$ specified in this way,
the induced-matter energy-momentum tensor can be calculated, and the
density and pressure of the cosmological induced matter found.
If one supposes, for example, that $\kappa=0$ and
$M=L^{(3\gamma +1)/2}$, with $\gamma$ a new constant,
then one obtains a perfect fluid with:
\begin{equation}
\rho = \frac{3\zeta^2 \lambda^2}{8\pi L^{3+3\gamma}} \; \; \; , 
       \; \; \;
   p = \gamma\rho \; \; \; .
\end{equation}
In this case the metric~(\ref{CurvedMetric}) reads:
\begin{equation}
d\hat{s}^2 = \frac{1}{L^{1+3\gamma}} \, dt^2 - L^2 \chi^2 - \left(
             \frac{\lambda^2}{L^{1+3\gamma}} - \frac{k}{\zeta^2}
             \right) d\psi^2 \; \; \; ,
\end{equation}
where $L(t-\lambda\psi)$ plays the role of the
{\em cosmological scale factor\/}, obeying the field equation:
\begin{equation}
\dot{L}^2 + \frac{k}{L^{1+3\gamma}} = \frac{\zeta^2 \lambda^2}
                                           {L^{2+6\gamma}} \; \; \; .
\end{equation}
This solution can be used to describe, for example, the 
matter-dominated era ($\gamma=0$) or the radiation-dominated one 
($\gamma=1/3$).
The properties of this model are discussed in more detail 
in \cite{Liu94}.
The fact that the scale factor depends on $\psi$ as well as $t$ is 
particularly interesting, and implies that observers with different 
values of $\psi$ would disagree on the time elapsed since the big bang;
that is, {\em on the age of the universe}.  Rather than being a single
event, in fact, the big bang in this picture resembles a sort of shock
wave propagating along the fifth dimension.
This effect could in principle allow one to constrain
the theory using observational data on the age spread of objects
such as globular clusters \cite{Ste89,Bol89,Van90}.
This has yet to be investigated in detail.

\subsection{Newton's Law, the Continuity Equation, and Horizon Size}
\label{Laws}

Besides the equation of state, there are two
other important laws relating $\rho$ and $p$ in cosmology; and it is
natural to ask whether or not they are automatically satisfied
by the induced-matter fluid.  These are the continuity
(or mass conservation) equation, or equivalently the first law of 
thermodynamics $dE + p \, dV = 0$ 
(where $E$ is energy and $V$ is the three-dimensional volume); 
and the equation of motion (or geodesic equation).  
The former can be written:
\begin{equation}
\frac{\partial}{\partial T} (\rho R^3) +
p \, \frac{\partial}{\partial T} (R^3) = 0 \; \; \; .
\label{Continuity}
\end{equation}
The latter is in general quite complicated (see \S~\ref{EqsMotion}), 
and we defer discussion of the noncomoving case to the next section.  
For matter which is comoving with a uniform fluid, however, 
only the radial direction is of interest and
the equation of motion is just Newton's law:
\begin{equation}
\frac{\partial^2 R}{\partial T^2} = -\frac{M}{R^2} \; \; \; , 
                                    \; \; \;
M \equiv \frac{4}{3} \pi R^3 (\rho + 3p) \; \; \; ,
\label{Newton}
\end{equation}
where we have used gravitational, rather than inertial mass,
as pressure can be significant in cosmological problems \cite{Bon89}.
In these equations one must be careful to use {\em proper\/} time
$T=\int e^{\nu/2} dt$ and distance $R=\int e^{\omega/2} dr$ rather 
than the ``raw coordinates'' $t$ and $r$ (where $r^2=x^2+y^2+z^2$ is
comoving radial distance).
For the cosmological metric~(\ref{CosmoMetric}) one has just
$T = \psi t$ and $R = t^{1/\alpha} \, \psi^{1/(1-\alpha)} \, r$,
and it is straightforward to show, using eqs.~(\ref{CosmoRhoP})
for $\rho$ and $p$, that the conservation equation~(\ref{Continuity})
and equation of motion~(\ref{Newton}) are both satisfied \cite{W92b}.
In fact, the same thing is true for {\em any\/} spatially-flat
perfect-fluid cosmology induced in this way by five-dimensional 
geometry\cite{W92c}.  As far as Newton's law and the continuity
equation are concerned, then, noncompactified Kaluza-Klein cosmology
is {\em indistinguishable from standard cosmology.}  To the extent that
these laws depend on the field equations, this is not surprising,
since the five-dimensional field equations $\hat{G}_{\alpha\beta}=0$
contain exactly the same information as the usual four-dimensional
ones $G_{\alpha\beta}=8\pi T_{\alpha\beta}$.

There are, however, effects which depend on the metric but not
the field equations, and the noncompactified versions of these will 
in general show departures from standard cosmology.  
The size of the particle horizon, for example, 
can be computed directly from the line element (assuming a
null geodesic, $d\hat{s}^2=0$).  For the above ``dust-like'' 
metric~(\ref{DustMetric}), it reads:
\begin{equation}
d = t_0^{2/3} \int_0^{t_0} \; \; \left[ \psi_0^2 - 9t^2 \left(
              \frac{d\psi}{dt} \right)^2 \right]^{1/2} 
              \frac{dt}{t^{2/3}} \; \; \; .
\end{equation}
This is just the usual (four-dimensional) expression, plus a
term in $(d\psi/dt)$.  (This term necessarily acts to {\em reduce\/}
the size of the particle horizon because the extra dimension of the 
cosmological metric~(\ref{CosmoMetric}) is spacelike.)
Similar results are found for the ``radiation'' 
metrics~(\ref{RelMetric}) and (\ref{RadMetric}) above \cite{W92c}.  
The value of the derivative $(d\psi/dt)$ can be evaluated with 
the help of the full geodesic equation, to which we turn next.

\subsection{The Equation of Motion} \label{MoreEqsMotion}

The general equation of motion, or geodesic 
equation~(\ref{4dGeodesic}), is also a metric-based relation 
and will contain nonstandard terms if the fifth dimension is real.  
Since the cosmological fluid is neutral, we disregard 
electromagnetic terms.  The spatial components
($\mu=i$, with $i=1,2,3$) of eq.~(\ref{4dGeodesic}) then read:
\begin{eqnarray}
\frac{d^2 x^i}{ds^2} + \Gamma^i_{\alpha\beta}
                       \frac{dx^{\alpha}}{ds} \frac{dx^{\beta}}{ds}
& = & \frac{\varepsilon {\cal B}^2}{(1-\varepsilon {\cal B}^2/\phi^2) 
      \phi^3} \left[ \nabla^i \phi + \left( \frac{\phi}{{\cal B}} 
      \frac{d{\cal B}}{ds} - \frac{d\phi}{ds} \right) 
      \frac{dx^i}{ds} \right] \nonumber \\
&   & - g^{i\lambda} \frac{\partial g_{\lambda\nu}}{\partial x^4}
      \frac{dx^{\nu}}{ds} \frac{dx^4}{dx} \; \; \; ,
\label{SpatialGeodesic}
\end{eqnarray}
where ${\cal B}$ is as given in eq.~(\ref{Bdefn}).
We then define a five-velocity $\hat{v}^A \equiv dx^A/d\hat{s}$, 
which is related to the usual four-velocity 
$v^{\alpha} \equiv dx^{\alpha}/ds$ by
$v^{\alpha} = ( d\hat{s}/ds ) \hat{v}^{\alpha}$.
Using the cosmological metric~(\ref{CosmoMetric}),
one can show that, for objects which are {\em comoving\/}
with the cosmological fluid ($\hat{v}^i=0$), all terms on the 
right-hand side of eq.~(\ref{SpatialGeodesic}) vanish \cite{W95b}.  
Comoving objects, in other words, satisfy the spatial components 
of the five-dimensional geodesic equation in {\em exactly the same 
way as in the standard four-dimensional theory.}  
This echoes the result obtained above for Newton's law.

For {\em noncomoving\/} objects, however, the right-hand side of 
eq.~(\ref{SpatialGeodesic}) will in general contain nonzero terms
involving the spatial velocities, the extra part of the metric, and
derivatives of the metric coefficients with respect to the extra
coordinate.  From the viewpoint of four-dimensional general relativity
such terms would appear as violations of the weak equivalence principle
or manifestations of a ``fifth force'' \cite{Kal95,Cho91,Gro83}.
To put this in practical terms, eq.~(\ref{SpatialGeodesic}) tells us 
that {\em galaxies with large peculiar velocities will not 
necessarily travel along four-dimensional geodesics}.
Observations of the peculiar motions of galaxies (and groups and
clusters of them) are now becoming available
\cite{Dre87,Lyn88,Luc88,Cal93,Mou93,Goi93}, and in principle 
these can be used to discriminate between noncompactified Kaluza-Klein
theory and ordinary general relativity \cite{W95b}, although this has
yet to be investigated in detail.  
Similar considerations will apply to the dynamics of {\em charged\/} 
test particles such as cosmic rays, for which the electromagnetic
terms in eq.~(\ref{4dGeodesic}) would need to be included.

We turn next to the $0$- and $4$-components of the geodesic
equation~(\ref{4dGeodesic}).  With comoving spatial coordinates
($\hat{v}^i=0$) one finds, using the metric~(\ref{CosmoMetric}):
\begin{eqnarray}
\frac{d\hat{v}^0}{d\hat{s}} + 
   \frac{2}{\psi} \, \hat{v}^0 \hat{v}^4 +
   \frac{\alpha^2}{(1-\alpha)^2} 
   \frac{t}{\psi^2} \, \hat{v}^4 \hat{v}^4 & = & 0 \; \; \; , 
   \nonumber \\
\frac{d\hat{v}^4}{d\hat{s}} + 
   \frac{(1-\alpha)^2}{\alpha^2} 
   \frac{\psi}{t^2} \, \hat{v}^0 \hat{v}^0 +
   \frac{2}{t} \, \hat{v}^0 \hat{v}^4      & = & 0 \; \; \; .
\label{04geodesic}
\end{eqnarray}
A solution of these must be compatible with the metric itself, 
which imposes the condition:
\begin{equation}
\psi^2 \hat{v}^0 \hat{v}^0 - \frac{\alpha^2}{(1-\alpha)^2} \; t^2 
       \hat{v}^4 \hat{v}^4 = 1 \; \; \; .
\label{Compatible}
\end{equation}
From eqs.~(\ref{04geodesic}) and (\ref{Compatible}) the $0$- 
and $4$-components of the five-velocity are found \cite{W95c} to be:
\begin{equation}
\hat{v}^0 = \mp \frac{\alpha}{\sqrt{2\alpha-1}} \, 
            \frac{1}{\psi} \; \; \; , \; \; \;
\hat{v}^4 = \pm \frac{(1-\alpha)^2}{\alpha\sqrt{2\alpha-1}} \, 
            \frac{1}{t} \; \; \; .
\end{equation}
The ratio of these gives us the rate of change with time of the extra
coordinate ($d\psi/dt=\hat{v}^4/\hat{v}^0$), 
and this is easily integrated to yield:
\begin{equation}
\psi(t)=\left( \frac{t_0}{t} \right)^{\cal A} \; \; \; ,
\end{equation}
where $t_0$ is an integration constant 
and ${\cal A} \equiv (1-\alpha)^2/\alpha^2$.
For $\alpha=2$ (the radiation-dominated era), ${\cal A}=0.25$; while
for $\alpha=3/2$ (the matter-dominated era), ${\cal A} \approx 0.11$.
The relative rate of change of the extra coordinate is:
\begin{equation}
\frac{d\psi/dt}{\psi} = -\frac{{\cal A}}{t} \; \; \; ,
\label{RelRateChange}
\end{equation}
and this is comfortably small in either case at late times.
(This is also true in the STM interpretation, where $\psi$
is related to rest mass; we return to this in the next section.)
The small size of $d\psi/dt$ means that the horizon sizes discussed
in the last section will be close to those in standard cosmology.
The discrepancies are, however, necessarily {\em nonzero\/} if the
spatial coordinates are chosen to be comoving.

\subsection{Cosmological Implications of General Covariance} 
\label{MoreCovariance}

In the previous section we worked entirely in coordinates defined
by the cosmological metric~(\ref{CosmoMetric}).
We are of course free to transform to coordinates in which the spatial
components of the five-velocity are {\em not\/} comoving.
For example, we can switch from $t,r,\psi$ to $t^{\prime},r^{\prime}$
and $\psi^{\prime}$, where:
\begin{equation}
t^{\prime} = t\psi \; \; \; , \; \; \;
r^{\prime} = t^{1/\alpha}r \; \; \; , \; \; \;
\psi^{\prime} = {\cal A} t^{\pm {\cal A}}\psi \; \; \; .
\label{NewCoords}
\end{equation}
In terms of these new coordinates, the density and pressure of the
cosmological fluid are no longer given by eqs.~(\ref{CosmoRhoP}) 
but by:
\begin{equation}
\rho = \frac{3}{8\pi \alpha^2 t^{\prime \, 2}} \; \; \; , \; \; \;
   p = \left( \frac{2\alpha}{3} - 1 \right) \, \rho \; \; \; .
\end{equation}
These are identical to the expressions in standard early 
($\alpha=2$) and late ($\alpha=3/2$) cosmology.
Also, since the metric transforms as a tensor,
$\hat{g}^{\prime \, AB} = (\partial x^{\prime \, A}/\partial x^C) 
                          (\partial x^{\prime \, B}/\partial x^D) 
                          \hat{g}^{CD}$, we have the result that
$\hat{g}^{\prime \, 00}=(2\alpha-1)/\alpha^2=$ constant,
which implies that in the new coordinates~(\ref{NewCoords})
there is a universal or cosmic time.
Similarly, using the vector transformation law
$\hat{v}^{\prime \, A}=(\partial x^{\prime \, A}/\partial x^B) 
\hat{v}^B$, we find that the new components of the five-velocity are:
\begin{equation}
\hat{v}^{\prime \, 0} = \mp \frac{\sqrt{2\alpha-1}}{\alpha} \; \; , 
                        \; \;
\hat{v}^{\prime \, 1} = \mp \frac{1}{\sqrt{2\alpha-1}} \,
                        \frac{r^{\prime}}{t^{\prime}} \; \; , \; \;
\hat{v}^{\prime \, 2} = \hat{v}^{\prime \, 3} = \hat{v}^{\prime 
                        \, 4} = 0 \; \; .
\end{equation}
The $0$-component of the test particle velocity
(which in four-dimensional theory is related to its energy) 
is {\em constant}.
The first component is proportional to $r^{\prime}$, 
which represents a version of {\em Hubble's law}.
And the fourth component vanishes.
Taken together, the above observations tell us that
the new coordinates defined by eq.~(\ref{NewCoords}) ---
the ones in which $\hat{v}^4 = d\psi/d\hat{s} = 0$ --- 
are {\em just the ones which give back standard cosmology}.
In a fully covariant five-dimensional theory,
there can be no {\em a priori\/} reason to prefer coordinates
in which $\hat{v}^4=0$ over those in which $\hat{v}^4 \ne 0$.  
It is a matter for experiment to decide.  
As emphasized in \S~\ref{Covariance},
the choice between the coordinates defined by the
metric~(\ref{CosmoMetric}) and those defined by~(\ref{NewCoords})
is {\em not arbitrary\/} --- not as long as the laws of physics
are written in terms of four-dimensional concepts like
density, pressure, and comoving four-velocity \cite{W95a,W95c}.
To decide whether or not the coordinates of the last section
are appropriate to describe the ``real world,'' one must look
for the effects associated with them, like
the nongeodesic motion of galaxies with
large peculiar velocities.

Another promising possibility arises if one can interpret the
fifth dimension physically, since eq.~(\ref{RelRateChange})
shows explicitly how it will change with time.  In particular,
in the context of STM theory, where $\psi$ is related to
particle rest mass, this equation implies a {\em slow variation
in rest mass\/} with time:
\begin{equation}
\frac{\dot{m}}{m} = -\frac{{\cal A}}{t} \; \; \; .
\end{equation}
Putting ${\cal A} \approx 0.11$ for the matter-dominated era and
$t \approx 15 \times 10^9$ yr for the present epoch, we obtain a value
of $\dot{m}/m = -7 \times 10^{-12} \; \mbox{yr}^{-1}$.
This is marginally consistent with ranging data from the {\em Viking\/}
space probe to Mars, where errors are reported as
$\pm 4 \times 10^{-12} \; \mbox{yr}^{-1}$, and
$\pm 10 \times 10^{-12} \; \mbox{yr}^{-1}$; and quite consistent
with timing data for the binary pulsar 1913+16,
where errors are reported as
$\pm 11 \times 10^{-12} \; \mbox{yr}^{-1}$ \cite{W95c,Wil81}.
If the STM hypothesis is valid, these data tell us that observation
is close to settling the question of whether cosmology is using 
coordinates with $\hat{v}^4 = 0$ or ones with $\hat{v}^4 \ne 0$.

It may seem unusual that physical effects can depend on the 
reference frame in which one observes them.  In fully covariant 
Kaluza-Klein theory this is a necessary consequence of trying to 
measure a higher-dimensional universe with four-dimensional tools.
Perhaps the most graphic example of this is the big bang itself.
As demonstrated in \S~\ref{Covariance}, the cosmological 
metric~(\ref{CosmoMetric}) is five-dimensionally {\em flat\/}.
The universe may therefore be far simpler than previously suspected,
in that it may have zero curvature.
What then of the big bang singularity, the Hubble expansion, the 
microwave background, and primordial nucleosynthesis?
In noncompactified Kaluza-Klein cosmology, these phenomena, 
which are all defined in four-dimensional terms, 
are in a sense recognized as geometrical illusions --- artifacts of a 
choice of coordinates in the higher-dimensional world \cite{W94b}.
Something like this occurs even in four-dimensional general relativity
when one works with {\em comoving spatial coordinates\/}, 
in which galaxies remain forever apart and there is no initial 
singularity \cite{W95c}.
Relativity is founded on the idea that there should be no preferred 
coordinate systems; yet in spatially comoving frames 
there is no big bang.
This paradox has no resolution within Einstein's theory,
which must consequently be seen as incomplete.
In practice, one usually regards the comoving coordinates as useful 
but ``not real.''  
Noncompactified Kaluza-Klein theory gives us a new way to think about 
these issues in terms of general covariance in higher dimensions.

\section{Astrophysics} \label{ASTROPHYSICS}

\subsection{Kaluza-Klein Solitons} \label{Solitons}

To model astrophysical phenomena like the Sun or other stars in 
Kaluza-Klein theory, one must extend the spherically-symmetric 
Schwarzschild solution of general relativity to higher dimensions.  
Birkhoff's theorem guarantees that the four-dimensional 
Schwarzschild metric is both static and {\em unique\/} to within 
its single free parameter (the mass of the central object).
This theorem, however, does not hold in higher dimensions, 
where solutions that are spherically-symmetric 
(in three or more spatial dimensions) 
depend in general on a number of parameters 
(such as electric and scalar charge) 
besides mass, and can in some cases be time-dependent as well.  
Unlike four-dimensional stationary solutions, some can also be
nonsingular \cite{Gro83,Gib82,Mat87}.
Such localized solutions of finite energy can legitimately be called
``solitons'' in the same broad sense used elsewhere 
in physics \cite{Fad87}.
In fact, some workers \cite{W92d,W94c,W94d,W93b,Liu93}
have found it convenient to apply this term to the {\em entire class\/} 
of higher-dimensional generalizations of the Schwarzschild metric
with finite energy (including those which, technically speaking, 
do contain geometrical singularities).  
We follow this convention here.

Kaluza-Klein solitons (in this general sense) were noted 
as early as 1951 by Heckmann, Jordan \& Fricke \cite{Hec51}, 
who found several solutions of the five-dimensional vacuum 
Einstein equations that were stationary and 
spherically-symmetric in three-space.
K\"uhnel \& Schmutzer \cite{Kuh61} carried the problem further in 1961,
studying for the first time the motion of test particles in the
field of the central mass.  (Tangherlini \cite{Tan63} used the alleged
instability of such ``generalized Keplerian orbits'' to argue
that there were only three spatial dimensions.)  
This crucial aspect of Kaluza-Klein theory has been re-examined 
over the years by several other authors \cite{Gro83,Geg84,Fer89,Cas87},
and provides one of the most promising ways to constrain it 
observationally.  We will return to it below.

The first {\em systematic\/} studies of stationary Kaluza-Klein
solutions with spherical symmetry appeared in 1982 with the work of
Chodos \& Detweiler \cite{Cho82} and Dobiasch \& Maison \cite{Dob82}.
The former authors obtained a class of five-dimensional solutions
characterized by three parameters 
(mass plus electric and scalar charge)
and emphasized the important point that {\em solitons are generic to
Kaluza-Klein theory in the same way that black holes are to
ordinary general relativity\/}.  This is what makes them so
important in confronting the theory with experiment.
The latter authors worked in $D$ dimensions 
(although the internal space was restricted to be flat) 
and their solutions accordingly possess four or more parameters.  
Various aspects of Chodos-Detweiler and Dobiasch-Maison 
solitons have been studied in \cite{Sok86,Pol83}.

The physical properties of five-dimensional solitons 
with zero electric charge were first described in detail 
by Sorkin \cite{Sor83}, Gross \& Perry \cite{Gro83}, 
and Davidson \& Owen \cite{Dav85b}, 
whose solutions (given by eq.~(\ref{DOsoliton}) 
in the notation of \cite{Dav85b}) are characterized by two parameters.  
Although these latter authors 
(along with many others) describe their solutions as ``black holes,'' 
it is important to note that in some cases the objects 
being considered are naked singularities \cite{Pol83}, or
have singular event horizons \cite{Sok86,Der85}.
The term ``monopole'' is also potentially misleading 
since more complicated solitons can, for example, take the form of 
dipoles \cite{Gro83}.  For these reasons we prefer to stay with 
the broader term ``solitons'' in this report.

The Chodos-Detweiler metric was generalized by 
Gibbons \& Wiltshire \cite{Gib86} to include extra nondiagonal terms, 
introducing a fourth parameter (associated with magnetic charge).  
These authors also considered the thermodynamics of Kaluza-Klein 
solitons for the first time.
Myers \& Perry \cite{Mye86} then extended the discussion 
to $D$-dimensional solitons with spherical symmetry in $(D-1)$, 
rather than three spatial dimensions, 
which allowed them to obtain Kaluza-Klein versions of the 
Reissner-N\"ordstrom and Kerr metrics, 
as well as the Schwarzschild one.
The thermodynamical properties of these objects, 
especially in six and ten dimensions, 
were examined by Accetta \& Gleiser \cite{Acc87}.
Myers \cite{Mye87} considered solitons which were {\em not\/} 
asymptotically Minkowskian.  
And Yoshimura \cite{Yos86} took the bold step of allowing
dependence of his solutions (albeit only the $(D-4)$-dimensional part)
on extra dimensions.
Others have studied the stability of soliton solutions 
with respect to classical perturbations \cite{Tom85,Gre88,Azr93} 
and quantum effects \cite{Iwa84}.  

All this work was done in a higher-dimensional {\em vacuum\/}; 
that is, with no explicit higher-dimensional matter.  
But most compactified Kaluza-Klein theories, as we have seen, 
operate in curved higher-dimensional spaces and require such matter 
(or other modifications of the Einstein equations) 
to ensure proper compactification, among other things.
This is just as true for soliton solutions as cosmological ones.
{\em Non-Abelian\/} solitons have accordingly been constructed 
by many authors using, for example, the Freund-Rubin fields 
of $D=11$ supergravity \cite{van84},
suitably defined six-dimensional \cite{Per84} or
seven-dimensional matter fields \cite{Ang86}, 
and various $D$-dimensional scalar 
fields \cite{Koi87,Gib88,Lid90,Liu91,Kro94,Ble94}.
Others have preferred to stay in a higher-dimensional vacuum, 
opting for higher-derivative corrections to the Einstein equations, 
including (quadratic) Gauss-Bonnet \cite{Wil86} 
and cubic \cite{Wur87} curvature terms; 
or for modifications of the Kaluza-Klein mechanism 
such as ``local compactification'' \cite{Bai85}.

\subsection{Are Solitons Black Holes?} \label{BlackHoles}

The rest of this report is concerned with solitons of 
the five-dimensional Gross-Perry-Davidson-Owen-Sorkin (GPDOS) 
type \cite{Sor83,Gro83,Dav85b}, 
with the line element~(\ref{DOsoliton}) in the notation of 
Davidson \& Owen.
Other spherically-symmetric static solutions, 
like the class found by Billyard \& Wesson \cite{Bil96b},
and those with more than two independent 
parameters \cite{Liu96b,Liu96c},
are subjects for future research.
Insofar as the metric coefficients of eq.~(\ref{DOsoliton}) do not
depend on the fifth coordinate, the distinction between 
compactified and noncompactified approaches is not an issue here.  
It would, however, become crucial in higher-dimensional 
generalizations of what follows.
We will interpret the four-dimensional properties of 
Kaluza-Klein solitons as induced by the geometry of 
{\em empty five-dimensional space\/} \cite{W92a}
in the manner of \S~\ref{Induced}.  
When $D>5$ this requires either a noncompactified approach, 
or modifications to the field equations, as 
described at the end of the last section.

The first question to address is whether GPDOS solitons in the 
induced-matter interpretation can rightly be considered black holes.
The two classes of object are alike in one important respect:
they contain a curvature singularity at the center of 
ordinary three-space.  However:
(1) solitons do not have an event horizon (not as understood
in ordinary general relativity, at any rate); and
(2) they have an {\em extended\/} matter distribution, rather than
having all their mass compressed into the central singularity.
In this section we try to clarify these properties, which
make the term ``black hole'' an inappropriate one
in the context of induced-matter Kaluza-Klein theory.

To begin with, it is apparent from the spatial components of the
metric~(\ref{DOsoliton}) that {\em the center of the 3-geometry is
at $r=1/a$ and not $r=0$\/}.  The surface area of 2-shells varies as 
$(ar-1)^{1-\epsilon (k-1)}$, and this shrinks to zero at $r=1/a$,
given that $k>0$ (as required above for positive density), 
and that the consistency relation~(\ref{Consistency}) holds.
The point $r=0$ is, in fact, not even part of the manifold, 
which ends at $r=1/a$.  
That this spatial center marks the location of a
bona fide curvature singularity, and not merely
a coordinate one, may be verified by evaluating the appropriate
invariant geometric scalars.  The square of the five-dimensional
Riemann-Christoffel tensor (or Kretschmann scalar
$\hat{K} \equiv \hat{R}_{ABCD} \hat{R}^{ABCD}$), for example,
reads in isotropic coordinates \cite{W94c}:
\begin{eqnarray}
\hat{K} & = & \frac{192 a^{10} r^6}{(a^2 r^2 -1)^8} \left( 
              \frac{ar-1}{ar+1} \right)^{4\epsilon (k-1)} \left[ 1 - 
              2 \epsilon (k-1) (2 + \epsilon^2 k) ar \nonumber 
              \right. \\
        &   & + \left. 2(3-\epsilon^4 k^2) a^2 r^2 - 2 \epsilon (k-1) 
              (2 + \epsilon^2 k) a^3 r^3 + a^4 r^4 \right] \; \; \; ,
\end{eqnarray}
and this is manifestly divergent at $r=1/a$ (with $k>0$).
(In the Schwarzschild limit this expression simplifies to
$\hat{K}=192a^{10} r^6 (ar+1)^{-12}$, 
which is formally the same as that in four-dimensional Einstein theory.
This however has little significance from the Kaluza-Klein point 
of view since the point $r=-1/a$ is not in the manifold.)  
The relevant {\em four\/}-dimensional curvature invariant is 
the square of the Ricci tensor,
$C \equiv R_{\alpha\beta} R^{\alpha\beta}$, 
and this comes out as \cite{W94c}:
\begin{eqnarray}
C & = & \frac{8 \epsilon^2 a^{10} r^6}{(a^2 r^2 -1)^8} \left( 
        \frac{ar-1}{ar+1} \right)^{2\epsilon (k-1)} \left[ 3 + 
        4\epsilon (3-2k) ar + 2(3 + 6\epsilon^2 + 4\epsilon^2 k^2 
        \nonumber \right. \\
  &   & \left. - 8\epsilon^2 k) a^2 r^2 +
        12\epsilon a^3 r^3 + 3a^4 r^4 \right] \; \; \; ,
\end{eqnarray}
which is also manifestly divergent at the center of the
soliton, $r=1/a$.

For black holes in general relativity,
the event horizon is commonly defined in general coordinates as the
surface where the norm of the timelike Killing vector vanishes.
In our case the Killing vector is just $(1,0,0,0)$ so its
norm vanishes where $g_{00}$ does.  
For the soliton metric~(\ref{DOsoliton}) this happens 
at $r \rightarrow 1/a$, given that $k>0$ and $\epsilon > 0$
(we will find below that physicality requires both these conditions).
For physical solitons, in other words, 
the event horizon shrinks to a point at the center of ordinary space.
{\em Kaluza-Klein solitons must therefore be classified as 
naked singularities\/}, as noticed previously by several 
authors \cite{Cha90a,Sok86,Der85,Liu91}.
According to the cosmic censorship hypothesis, such objects should
not be realized in nature.  
The relevance of this (essentially four-dimensional) postulate
to five-dimensional objects may, however, be debated.
In any case we will show below that if they exist, they could be 
detectable by conventional astrophysical techniques.

What of the soliton's mass distribution?  Applying the standard
definition \cite{Lan75} and using the soliton metric~(\ref{DOsoliton}),
one finds \cite{W94c}:
\begin{equation}
M_g(r) = \frac{2 \epsilon k}{a} \left( \frac{ar-1}{ar+1} 
         \right)^{\epsilon} \; \; \; .
\label{GravMass}
\end{equation}
($G=c=1$ throughout \S~\ref{ASTROPHYSICS} unless otherwise noted.)
This is the gravitational (or Tolman-Whittaker) mass of a 
Kaluza-Klein soliton as a function of (isotropic) radius $r$.  
Other commonly-used definitions of mass can be evaluated \cite{W94c} 
but do not lend themselves readily to physical interpretation.
For positive mass (as measured at infinity) 
one must have $\epsilon k > 0$.
Since positive density requires in addition that $k > 0$, 
it is apparent that {\em both\/} $k$ {\em and\/} $\epsilon$ 
must be positive for realistic solitons.
Eq.~(\ref{GravMass}) therefore implies that {\em the gravitational 
mass of the soliton goes to zero at the center\/} --- 
behaviour which differs radically from that exhibited by black holes.
Rather than being concentrated into a pointlike singularity,
the mass of the soliton is distributed in an {\em extended\/} fashion 
(although the $\sim \! \! 1/r^4$-dependence of density noted above 
means that this distribution is still a sharply peaked one).

The soliton defined by the Schwarzschild limit is, however, special
in this regard.  If one simply takes the Schwarzschild values
$\epsilon = 0, \epsilon k = 1$
and puts them directly into eq.~(\ref{GravMass}), one finds
that $M_g(r) = 2/a =$ constant for all $r$.  Replacing the parameter
$a$ via $M_{\ast} \equiv 2/a$ and putting this into the metric, 
eq.~(\ref{DOsoliton}), one recovers on spacetime sections
the four-dimensional Schwarzschild solution~(\ref{SchwarzMetric}), 
with ``Schwarzschild mass'' $M_{\ast}$.  Alternatively, however, 
one might keep $\epsilon$ {\em arbitrarily small\/} and allow 
$r \rightarrow 1/a$.  
In this case one finds that $M_g(r) \rightarrow 0$ 
{\em irrespective\/} of $\epsilon$.  
In other words, there is an ambiguity in the limit by which 
one is supposed to recover the Schwarzschild solution
from the soliton metric.  The problem is reminiscent of one
investigated by Janis, Newman \& Winicour \cite{Jan68,Jan69} 
and others \cite{Lid90,Bek74,Cha92}, 
in which the presence of a scalar field in four-dimensional 
general relativity led to ambiguity in defining 
the center of the geometry.  
In their case perturbation analysis led to a satisfactory 
resolution of the problem (in which the Schwarzschild ``horizon''
at $r=2M_{\ast}$ turned out to be a point).  
Adopting the same approach, Wesson \& Ponce de Leon \cite{W94c} 
have conducted a numerical study of eq.~(\ref{GravMass}), 
and this leads unambiguously to the conclusion that in 
the Schwarzschild limit (as defined by $k \rightarrow \infty$
and $\epsilon \rightarrow 0$) the mass does go to {\em zero\/} 
at $r=1/a$.
The picture that emerges from this numerical work is of 
an extended cloud of matter whose mass distribution becomes 
more and more compressed {\em near\/} its center as the 
parameters $\epsilon$ and $k$ approach their Schwarzschild values.  
Due to the nature of the geometry, however, 
the enclosed gravitational mass {\em at\/} the center is always zero.

\subsection{Extension to the Time-Dependent Case} \label{Tdependent}

The results of the last section make it clear that Kaluza-Klein 
solitons, although they contain singularities at their centers, 
are not black holes, since they have neither 
pointlike mass distributions nor event horizons of 
the conventional type.
A third crucial difference between these two classes of objects,
which follows from the fact that Birkhoff's theorem 
does not hold in five dimensions, is that
{\em soliton metrics can be generalized to include time-dependence}.
This goes somewhat against the idea
of a soliton as a {\em static\/} solution of the field equations.
However, it is reasonable to suppose that solitons, if they exist,
must have been formed in some astrophysical or cosmological process
during which they could not have been entirely static.  So it is of
physical, as well as mathematical interest to study the
extension to time-dependent solutions.

Liu, Wesson \& Ponce de Leon \cite{Liu93} have considered 
the case in which the coefficients $\nu,\omega$ and $\mu$  
of the general spherically-symmetric metric~(\ref{IsoLineElement}) 
depend not only on the radial coordinate $r$
(as in the GPDOS solution~(\ref{DOsoliton})), but on $t$ as well.
The metric coefficients are still assumed to be {\em separable\/}
functions, so that eqs.~(\ref{Static}) are in effect replaced by:
\begin{equation}
e^{\nu} \equiv A^2(r) \; T^2(t) \; \; \; , \; \; \;
e^{\omega} \equiv B^2(r) \; U^2(t) \; \; \; , \; \; \;
e^{\mu} \equiv C^2(r) \; V^2(t) \; \; \; .
\end{equation}
The field equations then produce two sets of differential equations,
for which four classes of solutions have been identified.  
We list these here, with brief comments.  
All the solutions have $T(t)=$ constant.
The first class has $U(t)=$ constant as well, 
along with $A(r)=C(r)$, and looks like:
\begin{equation}
d\hat{s}^2 = A^2(r) dt^2 - B^2(r) d\sigma^2 -
             A^2(r) V^2(t) d\psi^2 \; \; \; ,
\end{equation}
where $d\sigma^2 = dr^2+r^2 d\Omega^2$ as usual, 
and $V(t)$ can have either an oscillating form
$V(t)=\cos(\omega t + \varphi)$ or an exponentially varying one
$V(t)=\exp(\pm Ht)$ (the parameters $\varphi$ and $H$ are arbitrary
constants).  The four-dimensional parts of these solutions are static, 
and only the extra-dimensional part varies with time.  In the case of
the decaying exponential solution, the time-dependent soliton
tends toward a static one as $t \gg H^{-1}$.

The second class of solutions has $V(t)=U^{-1}(t)$ 
and $C(r)=A^{-1/2}(r)$, and can be written in the form:
\begin{equation}
d\hat{s}^2 = A^2(r) dt^2 - U^2(t) B^2(r) d\sigma^2 -
             U^{-2}(t) A^{-1}(r) d\psi^2 \; \; \; ,
\label{SecondSoln}
\end{equation}
where $U(t)$ satisfies a differential equation exactly analogous 
to one in standard FRW {\em cosmology\/}, and is given by
$U(t)=\sqrt{\varphi + Ht - \kappa t^2}$ 
(with $\kappa = \pm 1,0$ playing the role of a curvature constant).
This is the most interesting of the time-dependent soliton solutions,
and has been looked at separately by Wesson, Liu \& Lim \cite{W93b}.
The functions $A(r)$ and $B(r)$ can, for instance, 
be taken to be the same as those of the static soliton, 
eqs.~(\ref{ABC}).  The parameters $\epsilon$ and $k$ obey 
the consistency relation~(\ref{Consistency}) as before, 
and here take the values $1/\sqrt{3}$ and $2$ respectively.
Choosing in addition $\varphi=1$ and $\kappa=0$ for convenience,
the metric~(\ref{SecondSoln}) becomes:
\begin{eqnarray}
d\hat{s}^2 & = & \left( \frac{ar-1}{ar+1} \right)^{4/\sqrt{3}} \; 
                 dt^2 - \left( \frac{a^2 r^2 -1}{a^2 r^2} \right)^2 \,
                 \left( \frac{ar+1}{ar-1} \right)^{2/\sqrt{3}} \; 
                 (1+Ht) \, d\sigma^2 \nonumber \\
           &   & - \left( \frac{ar+1}{ar-1} \right)^{2/\sqrt{3}} \;
                 (1+Ht)^{-1} \, d\psi^2 \; \; \; .
\label{CosmoSoliton}
\end{eqnarray}
In the induced-matter interpretation this geometry manifests itself
in four dimensions as matter with anisotropic pressure.  
Using the same technique as in \S~\ref{Static} 
(identifying the pressure three-tensor $p^j_i$ and 
defining $p \equiv p^i_i/3$), one can nevertheless derive 
a unique equation of state.  This turns out to be \cite{W93b}:
\begin{eqnarray}
\rho & = & \frac{a^6 r^4}{3\pi (1+Ht)(a^2 r^2 -1)^4} \,
           \left( \frac{ar-1}{ar+1} \right)^{2/\sqrt{3}} \, +
           \frac{3H^2}{32\pi (1+Ht)^2} \,
           \left( \frac{ar+1}{ar-1} \right)^{4/\sqrt{3}} \nonumber \\
   p & = & \rho/3 \; \; \; .
\end{eqnarray}
The matter comprising this time-dependent soliton satisfies the
the relativistic equation of state, as expected since the 
metric coefficients are all independent of $\psi$.
What is interesting about this solution~(\ref{CosmoSoliton}) is that
{\em it reduces to the radiation-dominated cosmological 
metric\/}~(\ref{RadMetric}) in the limit of zero central mass 
$a \rightarrow \infty$ (ie., $M_{\ast} \rightarrow 0$).
So what began as a metric suitable for astrophysical problems 
may have cosmological applications, perhaps for modelling solitons 
in the early universe \cite{W93b}.

The third class of solutions found in \cite{Liu93} has $A(r)=$ constant 
and $U(t)=V(t)=1+Ht$, and reads:
\begin{equation}
d\hat{s}^2 = dt^2 - (1+Ht)^2 \left[ B^2(r) d\sigma^2 +
             C^2(r) d\psi^2 \right] \; \; \; .
\end{equation}
For these solitons the fifth dimension is expanding along with
the three-dimensional spatial part.  The fourth class of solutions,
finally, also has $U(t)=1+Ht$, but uses $V(t)=U^{\ell}(t)$ and
$C(r)=[A(r)]^{(\ell +2)/(\ell -1)}$, where $\ell$ is another 
arbitrary constant.  The associated line element looks like:
\begin{eqnarray}
d\hat{s}^2 & = & A^2(r) dt^2 - (1+Ht)^2 B^2(r) d\sigma^2
                 \nonumber \\
           &   & - (1+Ht)^{2\ell} [A(r)]^{2(\ell +2)/(\ell -1)} 
                 d\psi^2 \; \; \; .
\end{eqnarray}
For these solitons the three-dimensional space expands, but the
fifth dimension can either expand, contract, or remain static 
accordingly as $\ell >0$, $\ell <0$, or $\ell =0$ respectively.

\subsection{Solitons as Dark Matter Candidates} \label{DarkMatter}

Viewed in four dimensions via the induced-matter mechanism, 
the soliton resembles a hole in the geometry surrounded by a 
spherically-symmetric ball of ultra-relativistic matter 
whose density falls off at large distances as $1/r^4$.
If the universe does have more than four dimensions, 
these objects should be quite common, being generic to Kaluza-Klein 
gravity in exactly the same way black holes are to 
general relativity \cite{W92d,Cho82}.
It is therefore natural to ask whether they could supply the as-yet
undetected {\em dark matter\/} which according to many estimates 
makes up more than 90\% of the matter in the universe.  
Other dark matter candidates,
like massive neutrinos or axions, primordial black holes, and a
finite-energy vacuum, encounter problems with excessive contributions 
to the extragalactic background light (EBL) and the cosmic microwave
background radiation (CMB), among other things \cite{Ove92}.
In view of this we consider here the possibility that the so-called
``missing mass'' consists of solitons.

Adopting the same approach that has led to strong constraints 
on some of these other dark matter candidates \cite{Ove92}, 
one can begin by attempting to assess the effects of solitons 
on {\em background radiation\/} \cite{W94d},
assuming that the fluid making up the soliton is in fact composed 
of photons (although there are no {\em a priori\/}
reasons to rule out, say, ultra-relativistic neutrinos or gravitons).
Rather than guessing at their spectral distribution 
we restrict ourselves to bolometric calculations for the time being.  
The soliton density $\rho_s$ at large distances comes from 
eq.~(\ref{SolitonRhoP}), and in the same regime 
eq.~(\ref{GravMass}) gives $M_g \sim 2\epsilon k/a$.
Therefore, for solitons of asymptotic mass $M_s$:
\begin{equation}
\rho_s \sim \frac{M_s^2}{8 \pi c^2 k r^4} \; \; \; ,
\label{Rho1}
\end{equation}
where we have restored conventional units.  
Because this goes as $1/r^4$ while volume (in the uniform case) 
increases as only $r^3$, local density will be overwhelmingly 
due to just one soliton --- the {\em nearest\/} one ---
and we do not need to know about the global distribution 
of these objects in space.  
The average separation between solitons of mass $M_s$ in terms 
of their mean density $\bar{\rho_s}$ is $r=(M_s/\bar{\rho_s})^{1/3}$,
and we can use this as the distance to the nearest one.  Writing the
mean soliton density as a fraction 
$\Omega_s \equiv \bar{\rho_s}/\rho_{crit}$
of the critical density 
$\rho_{crit}=1.88 \times 10^{-26} h^2 \mbox{kg m}^{-3}$ \cite{Pee93} 
(where $h$ is the usual Hubble parameter in units of
100~km~s$^{-1}$~Mpc$^{-1}$), we then find that the effective
local density~(\ref{Rho1}) of solitonic radiation,
expressed as a fraction of the CMB density, is:
\begin{equation}
\frac{\rho_s}{\rho_{CMB}} \sim 5.0 \times 10^{-13} h^{8/3} k^{-1} 
                          \Omega_s^{4/3} \left( \frac{M_s}{M_{\odot}} 
                          \right)^{2/3} \; \; \; ,
\label{Rho2}
\end{equation}
where $M_{\odot}$ stands for one solar mass and 
$\rho_{CMB}=2.5 \times 10^{-5} h^{-2} \rho_{crit}$ is the 
equivalent mass density of the CMB at zero redshift \cite{Pee93}.
The quantity $k$ is a free parameter, subject only to the consistency 
relation~(\ref{Consistency}).  A particularly convenient choice for
illustrative purposes is $k=1$ (which implies $\epsilon =1$ also).  
This class of solutions was discovered independently by 
Chatterjee \cite{Cha90a} and has the special property that 
gravitational mass $M_g(r)$ at large distances $r \gg 1/a$ is 
equal to the Schwarzschild mass $M_{\ast}$.
If we suppose that individual solitons have galactic mass
($M_s = 5 \times 10^{11} M_{\odot}$) and that they collectively
make up all the dark matter required to close the universe 
($\Omega_s = 0.9$), then eq.~(\ref{Rho2}) tells us that 
distortions in the CMB will be of the order:
\begin{equation}
\frac{\rho_s}{\rho_{CMB}} \sim 1.0 \times 10^{-5} \; \; \; ,
\end{equation}
where we have used $h=0.7$ for the Hubble parameter.
This is precisely the upper limit set by {\em COBE\/} and other
experiments on anomalous contributions to the CMB.
So we can conclude that solitons, if they are to provide a significant
part of the missing mass, are probably less massive than galaxies.

A similar argument can be made on the basis of {\em tidal effects\/}.
It is known that conventional dark matter candidates such as 
black holes can be ruled out if they exceed 
$\sim \! \! 10^8 M_{\odot}$ in mass, since such objects 
would excessively distort the shapes of nearby galaxies.
The same thing would apply to solitons.
However, one has to keep in mind that there is no reason for the
parameters $k$ and $\epsilon$ to be equal to one for all solitons.
They are not universal constants like $c$ or $G$, but can in
principle vary from soliton to soliton.
Those with $k<1$ will have effective gravitational masses below
the corresponding Schwarzschild ones, and will consequently be
less strongly constrained.  A soliton with $k=0.1$, 
for example, will have $\epsilon = 1.05$ from eq.~(\ref{Consistency}),
and its gravitational mass~(\ref{GravMass}) at large $r$ will
be $M_g = 0.105 M_{\ast}$ --- only one-tenth the conventional value.
And in the extreme case $\epsilon k \rightarrow 0$, its
gravitational mass will vanish altogether.  So while these are
promising ways to look for Kaluza-Klein solitons, caution must
be taken in interpreting the results.  
Ideally one would like to be able to apply one or 
more {\em independent\/} tests to a given astrophysical
system.  We therefore devote the rest of this report to outlining the 
implications of Kaluza-Klein gravity for the classical tests of 
general relativity, and for related phenomena such as those 
having to do with the principle of equivalence.

\subsection{The Classical Tests} \label{ClassicalTests}

It is convenient to switch from the notation of 
Davidson \& Owen \cite{Dav85b} to that of Gross \& Perry \cite{Gro83}, 
and to convert from isotropic coordinates to nonisotropic ones 
(in which $r^{\prime} = r(1+GM_{\ast}/2r)^2$).
The soliton metric~(\ref{DOsoliton}) then takes the form:
\begin{eqnarray}
d\hat{s}^2 & = & \left( 1 - \frac{2M_{\ast}}{r^{\prime}} 
                 \right)^{1/\alpha} \, dt^2 - \frac{(1-2M_{\ast}/
                 r^{\prime})^{(\alpha-\beta-1)/\alpha}}
                 {(1-2M_{\ast}/r^{\prime})} \, dr^{\prime \, 2} 
                 \nonumber \\
           &   & - \left( 1 - \frac{2M_{\ast}}{r^{\prime}} 
                 \right)^{(\alpha-\beta-1)/\alpha} 
                 \, r^{\prime \, 2} d\Omega^2 - \left( 1 - 
                 \frac{2M_{\ast}}{r^{\prime}} \right)^{\beta/\alpha}
                 \, d\psi^2 \; \; \; .
\label{NewSol}
\end{eqnarray}
where $\alpha$ and $\beta$ are related to $\epsilon$ and $k$ by 
$\epsilon = -\beta/\alpha$ and $k=-1/\beta$, and where 
we have replaced the Gross-Perry parameter $m$ by $M_{\ast}/2$.
Eq.~(\ref{NewSol}) clearly reduces to the familiar Schwarzschild 
solution on hypersurfaces $\psi=$ constant as $\alpha \rightarrow 1$ 
and $\beta \rightarrow 0$.
Defining two new parameters 
$a \equiv 1/\alpha$ and $b \equiv \beta/\alpha$,
together with the function  $A(r) \equiv 1-2M_{\ast}/r$, 
eq.~(\ref{NewSol}) becomes:
\begin{equation}
d\hat{s}^2 = A^a dt^2 - A^{-(a+b)} dr^2 - A^{(1-a-b)} r^2 d\Omega^2 - 
             A^b d\psi^2 \; \; \; ,
\label{GPsoliton}
\end{equation}
where we have dropped the primes on $r$ for convenience.
The consistency relation~(\ref{Consistency}) takes the form:
\begin{equation}
a^2 + ab + b^2 = 1 \; \; \; .
\label{GPconsist}
\end{equation}
We wish to analyze the motion of photons and massive test particles
in the field described by the metric~(\ref{GPsoliton}).
The Lagrangian density ${\cal L}$ can be obtained from 
${\cal L}^2 = g_{ik} \dot{x}^i \dot{x}^k$, where the $x^i$ are 
generalized coordinates and the overdot denotes differentiation with
respect to an affine parameter (such as proper time in the case
of massive test particles) along the particle's geodesic trajectory.
For the metric~(\ref{GPsoliton}) this gives:
\begin{equation}
{\cal L}^2 = A^a \dot{t}^2 - A^{-(a+b)} \dot{r}^2 - A^{1-a-b} r^2 
             (\dot{\theta}^2 + \sin^2\theta \dot{\phi}^2) - A^b 
             \dot{\psi}^2 \; \; \; .
\label{LagDensity}
\end{equation}
From symmetry we can assume that $\dot{\theta}=0$, so $\theta=\pi/2$ 
without loss of generality.  Application of the Euler-Lagrange
equations to the Lagrangian~(\ref{LagDensity}) immediately produces
three constants of the motion:
\begin{equation}
l \equiv A^a \dot{t} \; \; \; , \; \; \;
h \equiv A^{(1-a-b)} r^2 \dot{\phi} \; \; \; , \; \; \;
k \equiv A^b \dot{\psi} \; \; \; .
\label{Constants}
\end{equation}
The third of these quantities, $k$, is related to the velocity of
the test particle along the fifth dimension.
The ``Schwarzschild limit'' of the theory hereafter refers to
the values $a=1,b=0$ and $k=0$.
With eqs.~(\ref{LagDensity}) and (\ref{Constants}) we are in a 
position to describe the motion of photons and test bodies in 
the weak-field approximation 
(ie., neglecting terms in $(M_{\ast}/r)^2$ and higher orders).
The procedure is exactly analogous to that in ordinary general 
relativity \cite{Adl75}, and since details have been given elsewhere
\cite{Bao89,Lim92,Kal95,Lim95,W96b} we confine ourselves in what
follows to summarizing only the main assumptions and conclusions.

\subsection{Gravitational Redshift} \label{GravRedshift}

This test depends only on the coefficients of the 
metric~(\ref{GPsoliton}) and, since the latter is static, 
one can consider emitters and receivers of light signals 
with fixed spatial coordinates.  The ratio of frequencies of
the received and emitted signals is simply:
\begin{equation}
\frac{\nu_r}{\nu_e} = \frac{g_{00} (r_e)}{g_{00} (r_r)} \; \; \; ,
\end{equation}
where $r_e$ and $r_r$ are the positions of the emitter and receiver 
respectively.  Using the metric~(\ref{GPsoliton}) and 
discarding terms of second and higher orders in $M_{\ast}/r$, 
one finds \cite{Kal95} that:
\begin{equation}
\frac{\nu_r - \nu_e}{\nu_e} = aM_{\ast} \left( \frac{1}{r_r} - 
                              \frac{1}{r_e} \right) \; \; \; .
\end{equation}
From this result it is clear that the gravitational redshift in 
Kaluza-Klein theory is in perfect agreement with that of 
four-dimensional general relativity, as long as one defines 
the gravitational mass $M_g$ of the soliton by $M_g \equiv aM_{\ast}$.

\subsection{Light Deflection} \label{LightDeflection}

The light deflection test is more interesting.  Noting that
$d\hat{s}^2=0$ for photons, and substituting the 
expressions~(\ref{Constants}) into eq.~(\ref{GPsoliton}), 
one finds the following equation of motion:
\begin{equation}
\left( \frac{dr}{d\phi} \right)^2 - \left( A^{(2-2a-b)} l^2 -
   A^{(2-a-2b)} k^2 \right) \frac{r^4}{h^2} + Ar^2 = 0 \; \; \; .
\label{EqMotion}
\end{equation}
For weak fields this can be solved \cite{Kal95} to yield a hyperbolic 
orbit $r(\phi)$ in which the photon approaches the central mass from 
infinity at $\phi=0$ and escapes to infinity along $\phi=\pi+\omega$.
The total deflection angle $\omega$ is given by:
\begin{equation}
\omega = \frac{4M_{\ast}}{r_0} + 2M_{\ast}pr_0 \; \; \; ,
\label{Deflect1}
\end{equation}
where $p \equiv (2-a-2b)(k/h)^2 - (2-2a-b)(l/h)^2$
and $r_0$ is the impact parameter 
(distance of closest approach to the central mass).
The first term in eq.~(\ref{Deflect1}) is the familiar
Einstein light deflection angle.  The second term represents a
correction due to the presence of the fifth dimension,
and is in principle measurable.
(Note that the apparent linear dependence of this term on $r_0$ is 
illusory as $p$ involves the square of the ``angular momentum'' 
constant $h \propto r_0$ in its denominator.)

The physical meaning of this result can be clarified by using 
the metric~(\ref{GPsoliton}) and the definitions~(\ref{Constants}) 
to recast eq.~(\ref{Deflect1}) in the form \cite{Lim95}:
\begin{equation}
\omega = \frac{4M_{\ast}}{r_0} \left[ 1 - \left( 
         \frac{f-m(d\psi/dt)^2}{1-n(d\psi/dt)^2} \right) 
         \right] \; \; \; ,
\label{Deflect2}
\end{equation}
where:
\begin{eqnarray}
f & \equiv & ( 1-a-b/2 ) A^{-(1-2a-b)} \; \; \; , \nonumber \\
m & \equiv & ( 1-a/2-b ) A^{-(1-3b)} \; \; \; , \; \; \;
n \equiv A^{-(a-b)} \; \; \; .
\end{eqnarray}
The $m$- and $n$-terms can be ignored when the velocity $d\psi/dt$ of
the test body along the fifth dimension is negligible.  
This is certainly true for photons (whose velocity is constant 
in four dimensions).
In addition one can go to the weak-field limit and neglect terms of 
first order in $M_{\ast}/r$ compared to one, so that $A=1$.
In that case $f=1-(a+b/2)$ and eq.~(\ref{Deflect2}) becomes:
\begin{equation}
\omega = \frac{4M_{\ast}}{r_0} \left( a + \frac{b}{2} \right) 
         \; \; \; .
\label{Deflect3}
\end{equation}
This reduces to the general relativistic result in the Schwarzschild 
limit.  For other values of $a$ and $b$, the Kaluza-Klein 
light-bending angle will depart from Einstein's prediction, 
and it is natural to inquire how big such a departure could be.  
The consistency relation~(\ref{GPconsist}) implies 
that $(a+b/2)=\sqrt{1-3b^2/4}$, so in principle eq.~(\ref{Deflect3})
is compatible with a range of angles 
$-\omega_{GR} \le \omega \le \omega_{GR}$,
where $\omega_{GR}$ is the general relativistic value.
This would allow for {\em null deflection\/} (for $b^2=4/3$) 
and even {\em light repulsion\/} (for negative roots).
These possibilities are, however, unphysical to the extent that 
they imply negative values for the (four-dimensional) mass of 
the soliton.  Inertial mass $M_i$, for example, 
can be obtained from the Landau pseudo energy-momentum 
tensor \cite{W96b,Cho91,Gro83,Des88}, and turns out to be 
$M_i=(a+b/2)M_{\ast}$.  
Therefore if one requires positivity of inertial mass,
then $(a+b/2) \ge 0$, which is incompatible with light repulsion.
Similarly, gravitational mass $M_g$ is found from the 
asymptotic behaviour of $\hat{g}_{00}$ \cite{W96b,Cho91,Gro83,Des88} 
to be given by $M_g=aM_{\ast}$ (see also \S~\ref{GravRedshift}).
(As discussed in these references, and in \S~\ref{Equivalence}, the 
fact that $M_g \ne M_i$ for $b \ne 0$ need not necessarily constitute 
a violation of the equivalence principle in Kaluza-Klein theory.)
Combining the requirements that $M_g \ge 0$ and $M_i \ge 0$ with the
consistency relation~(\ref{GPconsist}), one finds that $0 \le b \le 1$.
Therefore if one requires positivity of both inertial {\em and\/} 
gravitational mass, then the Kaluza-Klein light deflection 
angle~(\ref{Deflect3}) must lie in the range 
$0.5 \omega_{GR} \le \omega \le \omega_{GR}$, which
rules out null deflection as well as light repulsion.

This however still leaves room for significant departures from 
general relativity.  Why have these not been observed?
Most tests to date have been carried out in the solar system which, 
considered as a soliton, is very close to the limiting Schwarzschild
case since nearly all its mass is concentrated near the center.  
From this perspective the fact that long-baseline interferometric 
measurements of solar light-bending \cite{Wil81} have confirmed 
Einstein's prediction to within a factor of $\pm 10^{-3}$ 
merely tell us --- via eq.~(\ref{Deflect3}) ---
that the Sun must have $b<0.05$.  Larger values of this parameter, and 
hence larger deviations from the predictions of general relativity, 
might be looked for in the halos of large elliptical galaxies, or in 
clusters of galaxies, where mass is more evenly distributed.
Much of the dark matter is widely believed to be in these places,
and if some or all of it is made up of Kaluza-Klein solitons then one
could hope to find evidence of anomalous deflection angles in 
observations of {\em gravitational lensing\/} by 
elliptical galaxies \cite{Bre93},
galaxy clusters \cite{Loe94,Wu94,Fah94,Car94},
and perhaps in observations of {\em microlensing\/} by rich
clusters \cite{Wal95}.

Just as in four-dimensional general relativity, one can also 
solve the equation of motion~(\ref{EqMotion}) for circular, 
as well as hyperbolic photon orbits.
Putting $\dot{r}=0$ gives \cite{Kal95}:
\begin{equation}
(1-2a-b) r^2 + \frac{A}{M_{\ast}} \, r^3 + (b-a)A^{(1-a-2b)} \,
\frac{k^2 r^4}{h^2} = 0 \; \; \; .
\end{equation}
For negligible motion along the fifth dimension ($k=0$) this leads to:
\begin{equation}
r = (1+2a+b) M_{\ast} \; \; \; .
\end{equation}
In the Schwarzschild limit this gives back the general relativistic 
result.  For other values of $a$ and $b$, circular photon orbits 
can occur at other radii.  
However, prospects for distinguishing between alternative theories
of gravity based on this phenomenon are slim \cite{Wil81}, so we do
not consider it further.

\subsection{Perihelion Advance} \label{Perihelion}

The elliptical orbits of massive test bodies in orbit around the 
central mass are of greater interest \cite{Lim92}.
Using $d\hat{s}^2 \ne 0$ leads to a slightly more complicated version 
of the equation of motion~(\ref{EqMotion}).
This can be solved for the orbit of the test body, 
which is nearly periodic.  The departure from periodicity per orbit, 
or perihelion shift, is found \cite{Kal95} to be:
\begin{equation}
\delta \phi = \frac{6\pi M_{\ast}^2}{h^2} \left( d + 
              \frac{e}{6} \right) \; \; \; ,
\label{PerShift}
\end{equation}
where:
\begin{eqnarray}
d & \equiv & (1+k^2) + (a-1)(-1+2l^2-k^2) + b(-1+l^2-2k^2) \; \; \; , 
             \nonumber \\
e & \equiv & 2(2-a-b)(-1+a+b) + 2l^2(-2+2a+b)(-1+2a+b) \nonumber \\
  &        & + 2k^2(2-a-2b)(-1+a+2b) \; \; \; .
\end{eqnarray}
This gives back the usual general relativistic result
in the Schwarzschild limit.  If the orbit is nearly circular then 
eq.~(\ref{PerShift}) can be simplified to read:
\begin{equation}
\delta \phi = \frac{6\pi M_{\ast}}{r} \left( a + 
              \frac{2b}{3} \right) \; \; \; ,
\end{equation}
where $r$ is the orbit's coordinate radius.
As with the light deflection test, solar system experiments
(precession of Mercury's orbit) imply that the Sun, if modelled as
a soliton, must have values of $a$ and $b$ very close to 
the Schwarzschild ones.
Extrasolar systems, however, might show nonstandard periastron shifts.
Candidate systems could include DI Herculis \cite{Gui85} and
AS Camelopardalis \cite{Mal89}, as well as binary pulsars \cite{Tay89},
x-ray binaries \cite{Whi89}, and possibly pulsars with planetary
companions \cite{Wol92,Tho92}.  (Eq.~(\ref{PerShift}) would require
modifications for systems with significant mass ratios.)

\subsection{Time Delay} \label{RadarDelay}

A similar procedure gives the proper time taken by a photon on a 
return trip between any two points in the field of the central mass.
The definitions~(\ref{Constants}) and equation of 
motion~(\ref{EqMotion}) lead to the following result \cite{Kal95}:
\begin{eqnarray}
\Delta \tau & = & 2 \left( 1 - \frac{2M_{\ast}}{r} \right)^a \left\{ 
                  \left[ 1 + \frac{1}{2} \left( \frac{k}{l} \right)^2 
                  \right] \left( \sqrt{r_p^2 - r_o^2} + 
                  \sqrt{r_e^2 - r_o^2} \right) \nonumber \right. \\
            &   & - M_{\ast} \left[ 1 + \frac{1}{2} 
                  \left( \frac{k}{l} \right)^2 \right]
                  \left( \frac{\sqrt{r_p^2 - r_o^2}}{r_p} +
                  \frac{\sqrt{r_e^2 - r_o^2}}{r_e} \right) \nonumber \\
            &   & + M_{\ast} \left[ (2a+b) + \frac{3b}{2} 
                  \left( \frac{k}{l} \right)^2 \right]
                  \left[ \ln \left( \frac{r_p + \sqrt{r_p^2 - 
                  r_o^2}}{r_o} \right) \nonumber \right. \\
            &   & + \left. \left. \ln \left( \frac{r_e + 
                  \sqrt{r_e^2 - r_o^2}}{r_o} \right) \right] 
                  \right\} \; \; \; ,
\label{TimeDelay}
\end{eqnarray}
where $r_o,r_e$ and $r_p$ are the photon's distance of 
closest approach to the central mass and the radius measures to the 
emitting planet (usually Earth) and reflecting planet respectively, 
and $r$ is the coordinate radius at which measurement is made 
(usually the same as $r_e$).
In the Schwarzschild limit, eq.~(\ref{TimeDelay}) gives back 
the usual result of four-dimensional general relativity.  
Experimental data such as that from the Viking spacecraft \cite{Wil81}
tell us that our solar system is close to this limit.

\subsection{Geodetic Precession} \label{Precession}

The motion of a {\em spinning\/} object in five dimensions is 
more complicated, but can be usefully studied in at least 
two important special cases:
(1) the case in which the $4$ component of the spin 
vector $\hat{S}^A$ is {\em zero\/} \cite{Kal95}; and
(2) the case in which the $4$-component of spacetime is {\em flat\/}
\cite{Liu96a}.  We review these in turn.  

The object in both cases is to solve for $\hat{S}^A$ as a function of
proper time $\hat{s}$.  The requirement of parallel transport implies:
\begin{equation}
\frac{d\hat{S}^M}{d\hat{s}} + 
\hat{\Gamma}^M_{AB} \hat{S}^A \hat{v}^B = 0 \; \; \; ,
\label{ParaTrans}
\end{equation}
where $\hat{v}^A \equiv dx^A/d\hat{s}$ is the five-velocity.
Since $\hat{S}^A$ is spacelike whereas $\hat{v}^B$ is timelike, 
their inner product can be made to vanish:
\begin{equation}
\hat{g}_{AB} \hat{S}^A \hat{v}^B = 0 \; \; \; .
\label{InnerProd}
\end{equation}
Eqs.~(\ref{ParaTrans}) and (\ref{InnerProd}), together with the metric,
can be solved for the components of the spin vector $\hat{S}^A$ 
if some simplifying assumptions are made.

To evaluate case (1) we use the Gross-Perry soliton 
metric~(\ref{GPsoliton}), we restrict ourselves to circular 
orbits $x^A(\hat{s})$, 
(for which the velocity vector $\hat{v}^A$ can be written
$\hat{v}^A=(t,0,0,td\phi/dt,0)$), and assume that $\hat{S}^4=0$.
The resulting expressions for the components of $\hat{S}^A$ are 
lengthy \cite{Kal95} and not particularly illuminating.  
The important thing about them is that the spatial 
components $\hat{S}^i$ show a {\em rotation relative to 
the radial direction\/}, with proper angular speed \cite{Kal95}:
\begin{equation}
\Omega = \left( \frac{[r-(1+a+b) M_{\ast}] r^{(a-1)/2}}
                     {(r-2M_{\ast})^{(a+1)/2}} \right) \, 
         \Omega_0 \; \; \; ,
\end{equation}
where $\Omega_0 \equiv d\phi/dt$ is given by:
\begin{eqnarray}
\Omega_0 & = & \left[ \frac{(1-a-b)}{a} \, \left( \frac{1-2M_{\ast}}{r}
               \right)^{1-2a-b} \, r^2 \nonumber \right. \\
         &   & \left. + \frac{1}{aM_{\ast}} \, \left( 
               \frac{1-2M_{\ast}}{r} \right)^{2-2a-b} \, r^3 
               \right]^{1/2} \; \; \; .
\end{eqnarray}
A spin vector $\hat{S}^A$ whose initial orientation is along the 
radial direction will, after one revolution along $x^A(\hat{s})$, 
undergo a {\em geodetic precession}:
\begin{equation}
\delta \phi = 2\pi \left[ 1 - \frac{ \sqrt{r-(1+a+b)M_{\ast}} \;
              \sqrt{r-(1+2a+b)M_{\ast}} }{ \sqrt{r(r-2M_{\ast})} } 
              \right] \; \; \; .
\end{equation}
Going to the weak-field limit and using the consistency 
relation~(\ref{GPconsist}), one can reduce this expression to:
\begin{equation}
\delta \phi = \frac{3\pi M_{\ast}}{r} \left( a + \frac{2b}{3} 
              \right) \; \; \; ,
\label{Precess}
\end{equation}
which gives back the usual general relativistic result in the 
Schwarzschild limit.  In general there will be deviations from 
Einstein's theory which are in principle measurable.  
One way to detect them would be with orbiting gyroscopes 
like those aboard the Gravity Probe-B (GP-B) satellite \cite{Tur86},
designed to orbit the earth at an altitude of 650 km.  
Assuming for the sake of illustration the same value of $b=0.05$ 
mentioned in \S~\ref{LightDeflection} as the largest one compatible 
with solar light-bending experiments, one finds from 
eq.~(\ref{Precess}) that the geodetic precession in Kaluza-Klein 
theory would be 1.238 milliarcseconds per revolution, 
or an angular rate of 6.674 arcsec~yr$^{-1}$.  
This exceeds the general relativistic prediction 
(6.625 arcsec~yr$^{-1}$) by 49 milliarcsec~yr$^{-1}$ --- 
a difference that would easily be detected by GP-B.  
In fact this satellite is expected to measure angular rates 
as small as 0.1 milliarcsec~yr$^{-1}$, 
which would allow it to probe $b$-values as small as $10^{-4}$.

We turn now to case (2), in which the spin vector $\hat{S}^A$ is 
{\em arbitrary\/} but the fifth dimension of spacetime is flat.  
Instead of the soliton metric~(\ref{GPsoliton}) we introduce 
a simpler five-dimensional line element \cite{Liu96a} which is also 
spherically-symmetric in three-dimensional space:
\begin{eqnarray}
d\hat{s}^2 & = & \frac{\psi^2}{L^2} \left[ \left( 1 - 
                 \frac{2M_{\ast}}{r} - \frac{r^2}{L^2} \right) \, 
                 dt^2 \nonumber \right. \\
           &   & - \left. \left( 1 - \frac{2M_{\ast}}{r} - 
                 \frac{r^2}{L^2} \right)^{-1} \, dr^2 - r^2 d\Omega^2 
                 \right] - d\psi^2 \; \; \; .
\label{Spin5dMetric}
\end{eqnarray}
This reduces to the four-dimensional Schwarzschild-{\em de Sitter\/}
line element on surfaces $\psi=$ constant:
\begin{equation}
ds^2 = \left( 1 - \frac{2M_{\ast}}{r} - \frac{r^2}{L^2} \right) \, 
       dt^2 - \left( 1 - \frac{2M_{\ast}}{r} - \frac{r^2}{L^2} 
       \right)^{-1} \, dr^2 - r^2 d\Omega^2 \; \; \; ,
\label{Spin4dMetric}
\end{equation}
and the constant $L$ (which has units of length) gives rise to an
effective four-dimensional cosmological constant $\Lambda=3/L^2$ 
\cite{Mas94,Liu95b}.
The four-dimensional universe is characterized by induced matter 
whose density and pressure are found from eqs.~(\ref{SphSymRhoP}) 
to be given by $\rho=3/\psi^2$ and $p=-\rho$.

Consider first of all an object which is not spinning.
Its orbit $x^A(\hat{s})$ is found using the metric 
relation~(\ref{Spin5dMetric}), which satisfies a consistency relation:
\begin{equation}
\hat{g}_{AB} \hat{v}^A \hat{v}^B = 1 \; \; \; ,
\label{SpinConsist}
\end{equation}
together with the geodesic equation~(\ref{5dGeodesic}).
The $A=5$ component of this latter equation 
turns out \cite{Liu96a} to be:
\begin{equation}
\frac{d^2\psi}{d\hat{s}^2} + \frac{1}{\psi} \left( 
\frac{d\psi}{d\hat{s}} \right)^2 + \frac{1}{\psi} = 0 \; \; \; .
\end{equation}
This has the simple solution $\psi^2 = \psi_m^2 - \hat{s}^2$,
where $\psi_m= $ constant.  Since eqs.~(\ref{Spin5dMetric}) and
(\ref{Spin4dMetric}) are related by 
$d\hat{s}^2=(\psi/L)^2 ds^2 - d\psi^2$, one finds that:
\begin{equation}
\psi = \frac{\psi_m}{\cosh \left[ (s-s_m)/L \right]} \; \; \; ,
\end{equation}
where $s_m$ is some fiducial value of the four-dimensional proper time
at which $\psi=\psi_m$.  Physically, the fifth coordinate in this
spacetime expands from zero size to a maximum value of $\psi_m$, 
and then contracts back to zero.  
We are living in the period $s>s_m$, when $\psi$ is decreasing 
({\em cf.\/} \S~\ref{Dynamical}) on cosmological timescales.
(The length $L$ is large if the cosmological constant $\Lambda$ 
is small.)

The spatial components of the geodesic equation~(\ref{5dGeodesic})
for this five-dimensional metric turn out to be identical with 
the usual four-dimensional ones \cite{Mas94,Liu96a}.  
This is somewhat surprising since the four-dimensional 
metric~(\ref{Spin4dMetric}) depends on $\psi$.
It means that the classical tests of relativity discussed in 
\S~\ref{GravRedshift} -- \S~\ref{RadarDelay} are by themselves 
insufficient to distinguish between Einstein's theory and its 
five-dimensional counterpart.  {\em When spin is included\/}, 
however, the two theories lead to very different predictions.  
For this we require the full machinery of eqs.~(\ref{ParaTrans}) 
and (\ref{InnerProd}) as well as eqs.~(\ref{5dGeodesic}) 
and (\ref{SpinConsist}).
Consider for simplicity a circular orbit as before, and assume
that the spin vector lies in the plane of the orbit (so $\hat{S}^2=0$),
and that one can arrange mechanically to satisfy the inequality
$r\hat{S}^3 \ll \hat{S}^1$.  
(These conditions are close to those in the GP-B experiment, 
or alternatively might be used to model the Sun--Uranus system, 
since the spin axis of Uranus lies near its orbital plane.)
In this case the four equations noted above allow one to solve for
all the components of the spin vector (including $\hat{S}^4$).
Its precession from the radial direction after one orbit turns out
in the weak-field limit ($r^2/L^2 \ll M_{\ast}/r \ll 1$) 
to be \cite{Liu96a}:
\begin{equation}
\delta \phi = \frac{3\pi M_{\ast}}{r} - \frac{2\pi H_4 r}
              {H_1 L \cosh \left[ (s_0 - s_m)/L \right]} \; \; \; ,
\end{equation}
where $s_0$ is the value of $s$ at the beginning of the orbit,
and $H_1$ and $H_4$ are normalized amplitudes of the spin vector
along the $x^1$ and $x^4$ axes respectively.
The first term is the usual geodetic precession of
four-dimensional general relativity.  The extra term depends on
the size $H_4/H_1$ of the spin component along the fifth dimension, 
the mass $M_{\ast}$ of the central body, 
and cosmological factors like the elapsed four-dimensional proper time.
It also involves {\em radius\/} in a manner quite different from that
of the first term, which suggests that the two terms could be
separated experimentally.  Whether this is practical or not has
yet to be established, but further investigation is warranted insofar
as geodetic precession is the {\em only\/} test of relativity
which can in principle allow us to distinguish between the 
five-dimensional metric~(\ref{Spin5dMetric}) and the four-dimensional
one~(\ref{Spin4dMetric}).

\subsection{The Equivalence Principle} \label{Equivalence}

Many of the above tests show departures from four-dimensional geodesic
motion.  These could be interpreted as violations of the weak
equivalence principle (WEP) by the curvature of the fifth dimension.
However, Gross \& Perry \cite{Gro83} have argued that they should 
more appropriately be attributed to a breakdown of 
{\em Birkhoff's theorem\/}, since the underlying theory is fully
covariant in five dimensions and involves only gravitational effects.
Cho \& Park \cite{Cho91} have made similar comments, arguing that
the extra dimension acts like a fifth force which is, however,
indistinguishable from gravity for an uncharged particle.
The nature of the fifth force in noncompactified theory has recently
been treated in depth by Mashhoon {\em et al.\/} \cite{Mas96}.
Here we consider only the simple but dramatic illustration afforded
by a test body in radial free fall near a soliton.
This is the analog of Galileo's experiments with
objects dropped vertically in the Earth's gravitational field.
And while this case is somewhat impractical in the context
of modern tests of gravity, we will see that it leads to several
simple and instructive results.

For vertical free-fall, $d\theta=d\phi=0$, and the equation
of motion~(\ref{EqMotion}) leads directly to the following
result in terms of the constants~(\ref{Constants}):
\begin{equation}
\dot{r}^2 = A^b l^2 - A^a k^2 - A^{(a+b)} \; \; \; .
\label{FreeFall}
\end{equation}
For a particle which begins at rest ($\dot{r}=0$) at $r=r_0$,
this equation gives:
\begin{equation}
l^2 = [ A(r_0) ]^a + [ A(r_0) ]^{(a-b)} \, k^2 \; \; \; .
\label{InitCond}
\end{equation}
Combining eqs.~(\ref{FreeFall}) and (\ref{InitCond}), one obtains
the ``energy condition'' \cite{Kal95}:
\begin{eqnarray}
\dot{r}^2 & = & \left[ \left( 1 - \frac{2M_{\ast}}{r} \right)^b
                \left( 1 - \frac{2M_{\ast}}{r_0} \right)^{(a-b)} -
                \left( 1 - \frac{2M_{\ast}}{r} \right)^a 
                \right] \, k^2 \nonumber \\
          &   & + \left( 1 - \frac{2M_{\ast}}{r} \right)^b
                \left( 1 - \frac{2M_{\ast}}{r_0} \right)^a -
                \left( 1 - \frac{2M_{\ast}}{r} \right)^{(a+b)} 
                \; \; \; .
\label{EnergyCond}
\end{eqnarray}
In the Schwarzschild limit this gives back the familiar 
four-dimensional formula $\dot{r}^2=2M_{\ast}(1/r - 1/r_0)$, 
which has the same form as the energy equation 
(for vertical free-fall) in classical Newtonian theory.

The particle's {\em coordinate velocity\/} in the $r$-direction 
is given by $u_r \equiv dr/dt = \dot{r} ds/dt$, and can be calculated
from eq.~(\ref{EnergyCond}) and the metric~(\ref{GPsoliton}).
It turns out (for $r_0 \rightarrow \infty$) to be \cite{Kal95}:
\begin{eqnarray}
u_r & = & -\frac{1}{\sqrt{1+k^2}} \;
           \left( 1 - \frac{2M_{\ast}}{r} \right)^a 
           \left\{ \left[
           \left( 1 - \frac{2M_{\ast}}{r} \right)^b -
           \left( 1 - \frac{2M_{\ast}}{r} \right)^a 
           \right] \, k^2 \nonumber \right. \\
    &   &  \left. + \left( 1 - \frac{2M_{\ast}}{r} 
           \right)^b \left[ 1 -
           \left( 1 - \frac{2M_{\ast}}{r} \right)^a 
           \right] \right\}^{1/2} \; \; \; .
\end{eqnarray}
This explicitly depends on velocity along the fifth dimension
through $k$.  Test particles with nonzero values for this parameter
will deviate from geodesic trajectories (in four dimensions) and
appear to violate the WEP.  The $a$ and $b$ parameters also
produce discrepancies with four-dimensional theory.  For example,
the radius where $u_r$ begins to decrease (as the test particle 
nears the Schwarzschild surface) differs from the simple value
of $r^{\ast}=6M_{\ast}$ predicted in Einstein's theory.
In the case where $k^2 \ll 1$ one finds instead \cite{Kal95}:
\begin{equation}
r^{\ast} = 2M_{\ast} \left[ 1 - \left( \frac{2a+b}{3a+b} 
           \right)^{1/a} \right]^{-1} \; \; \; .
\end{equation}
This reduces to the general relativistic result in the 
Schwarzschild limit.

The effects of the fifth dimension can perhaps be most readily
appreciated in the particle's {\em acceleration\/}, which comes from
differentiating eq.~(\ref{FreeFall}):
\begin{equation}
\ddot{r} = -\frac{M_{\ast}}{r^2} \left[ (a+b) A^{(a+b-1)} 
           -bl^2 A^{(b-1)} + ak^2 A^{(a-1)} \right]
\end{equation}
In the Schwarzschild limit ($a=1,b=0$) this reduces to:
\begin{equation}
\ddot{r} = -\frac{(1+k^2)M_{\ast}}{r^2} \; \; \; ,
\end{equation}
which gives back the familiar four-dimensional result when $k=0$. 
In general, though, the particle's hidden velocity in the fifth 
dimension affects its rate of fall towards the central body
in a very significant way.  For completeness we note that
a particle which has $k=0$ and starts from rest at infinity 
(in which case eq.~(\ref{InitCond}) implies $l^2=1+k^2$) will have:
\begin{equation}
\ddot{r} = -\frac{aM_{\ast}}{r^2} \; \; \; ,
\end{equation}
at large distances ($r \gg 2M_{\ast}$).
This confirms that a particle accelerates in the field of
the soliton at a rate governed by $M_g = a M_{\ast}$
(the gravitational mass) and not $M_{\ast}$.
As mentioned in \S~\ref{LightDeflection}, neither $M_g$ nor $M_{\ast}$
is necessarily the same as the soliton's {\em inertial\/} 
mass $M_i$ in Kaluza-Klein theory, the two quantities being 
related \cite{W96b,Cho91,Gro83,Des88} by:
\begin{equation}
M_i = \left( 1 + \frac{b}{2a} \right) \, M_g \; \; \; .
\label{2masses}
\end{equation}
These are strictly identical only in the Schwarzschild limit $b=0$,
and in other cases there will be apparent violations of the WEP.
(Note that the factor of two is missing in ref. \cite{Gro83}.)
Experimentally, one can focus on the quantity:
\begin{equation}
\Delta \equiv \left| \frac{ \left( M_g/M_i \right)_A -
              \left( M_g/M_i \right)_B } { \frac{1}{2} \, 
              \left[ \left( M_g/M_i \right)_A +
              \left( M_g/M_i \right)_B \right] } \right| \; \; \; ,
\end{equation}
where the subscripts $A$ and $B$ stand for two objects with
different compositions.  This is known from experiments on the
Earth to be less than about $2 \times 10^{-11}$, and would be 
measured to as little as $10^{-17}$ by the proposed Satellite
Test of the Equivalence Principle (STEP) \cite{Rei93}.  
If eq.~(\ref{2masses}) is valid, then one expects two different
solitons to have:
\begin{equation}
\Delta \approx \frac{1}{2} \left| \left( \frac{b}{a} \right)_B -
               \left( \frac{b}{a} \right)_A \right| \; \; \; ,
\end{equation}
which vanishes in the Schwarzschild limit $b=0$.
This relation provides yet another way to probe experimentally for
the possible existence of extra dimensions.

\section{Conclusions} \label{CONC}

Kaluza unified Einstein's theory of gravity and Maxwell's theory
of electromagnetism by the simple device of letting the indices
run over five values instead of four.  Other interactions can be
included by letting the indices take on even larger values, 
but in our review we have concentrated on the prototype theory
viewed as an extension of general relativity.
Klein's contribution was to explain the apparently unobserved
nature of the extra dimension by assuming it was rolled up to a
small size, and compactified Kaluza-Klein theory remains one of
three principal approaches to the subject.  Another is to use the
extra dimension as an algebraic aid, as in the projective approach.
A third version of Kaluza-Klein theory, on which we have spent
considerable time since it is the newest, regards the fifth dimension
as real but not necessarily a simple length or time.  In the 
space-time-matter theory, it is responsible for mass.

All three versions of Kaluza-Klein theory are viable as judged
by experiment and observation.  In particular, they cannot be
ruled out by the classical tests of relativity or results from
astrophysics and cosmology.  Indeed, it can be difficult to 
distinguish between the three main versions of Kaluza-Klein theory 
at the present time because their observational consequences are often 
similar.  To help differentiate between its variants and bring the 
whole subject closer to critical test, we suggest several things:
(1) A search for exact solutions of new types.  New Kerr-like 
solutions, for example, would help to model spinning elementary 
particles.
(2) More work on quantization.  This is a perennial problem, 
of course, but the richness of Kaluza-Klein theory may offer new
routes to its resolution.
(3) An investigation of the physical nature of the fifth dimension.
While it is merely a construct in the projective approach, it is
real and may become large in certain regimes of exact solutions
in the compactified approach.  In the noncompactified approach,
it is not only real but in principle always observable provided
one chooses a coordinate system or gauge that properly brings it out.

We do not wish to prejudge the issue of which if any version of
Kaluza-Klein gravity will emerge as superior.  However, the progress
of physics lies in explaining more phenomena on the basis of theories
that are constrained by standards of logic, conciseness and elegance.
In this regard, we venture the opinion that the fifth dimension
will be needed.

\begin{ack}
One of us (J. O.) would like to acknowledge the early encouragement
of three inspirational teachers, A. Klassen, H. Janzen and J. Harder.
For comments over the years on higher-dimensional gravity
we also thank (in alphabetical order) A. Billyard, B. J. Carr,
A. A. Coley, C. W. F. Everitt, T. Fukui, D. Kalligas,
G. Lessner, P. Lim, H. Liu, B. Mashhoon, D. J. McManus,
J. Ponce de Leon, E. Schmutzer and R. Tavakol.
This work was supported financially by the National Science and 
Engineering Research Council (NSERC) and the National Aeronautics
and Space Administration (NASA).
\end{ack}

\end{document}